\documentclass[a4paper]{article}
\pdfoutput=1

\usepackage{jheppub}

\usepackage{graphicx}   % need for figures
\usepackage{subcaption}
\usepackage{color}      % use if color is used in text
\usepackage[dvipsnames]{xcolor}      % use if color is used in text
\usepackage{hyperref}   % use for hypertext links, including those to external documents and URLs
\usepackage{slashed}    % for feynmann slashed notation
\usepackage[normalem]{ulem}  % for strike out
\usepackage{textcomp}
\usepackage{tikz}
\usepackage{comment}
\usepackage[utf8]{inputenc} 
\usepackage{bm}
\usepackage{ifthen} % if
\usepackage{soul}
\usepackage{calc}
\usepackage[inline]{enumitem} % inline enumeration

\usepackage{booktabs}
\usepackage{multirow}
\usepackage{siunitx}
\usetikzlibrary{decorations.shapes}
\usetikzlibrary{decorations.pathreplacing}
\usetikzlibrary{calc}
\usetikzlibrary{graphs}

\usepackage{Style.FeynGraphUtils}
\usepackage{Style.Notation}

\newcommand{\Figref}[1]{Figure \ref{#1}}
\newcommand{\figref}[1]{figure \ref{#1}}

\newcommand{\sectionref}[1]{section \ref{#1}}

\newcommand{\appendixref}[1]{appendix \ref{#1}}

\renewcommand{\eqref}[1]{eq.~(\ref{#1})}
\newcommand{\tableref}[1]{table \ref{#1}}

\newcommand{\logit}{{\mathrm{logit}}}

\let\origfootnote\footnote
\renewcommand{\footnote}[1]{\kern.06em\origfootnote{#1}}
\newcommand{\punctfootnote}[1]{\kern-.06em\origfootnote{#1}}

\begin{document}

\title{Neural Network-based Top Tagger with Two-Point Energy Correlations and Geometry of Soft Emissions}

\author[a]{Amit Chakraborty,}
\author[b]{Sung Hak Lim,}
\author[b,c,d]{Mihoko M. Nojiri}
\author[e]{and Michihisa Takeuchi}
\affiliation[a]{Centre for High Energy Physics, Indian Institute of Science, Bengaluru, Karnataka 560012, India}
\affiliation[b]{Theory Center, IPNS, KEK,
1-1 Oho, Tsukuba, Ibaraki 305-0801, Japan}
\affiliation[c]{The Graduate University of Advanced Studies (Sokendai),
1-1 Oho, Tsukuba, Ibaraki 305-0801, Japan}
\affiliation[d]{Kavli IPMU (WPI), University of Tokyo,
5-1-5 Kashiwanoha, Kashiwa, Chiba 277-8583, Japan}
\affiliation[e]{Kobayashi-Maskawa Institute for the Origin of Particles and the Universe,
Nagoya University, Furo-cho, Chikusa-ku, Nagoya, Aichi 464-8602, Japan
}


\emailAdd{achakraborty@iisc.ac.in}
\emailAdd{sunghak.lim@kek.jp}
\emailAdd{nojiri@post.kek.jp}
\emailAdd{takeuchi@kmi.nagoya-u.ac.jp}

\preprint{KEK-TH-2202}

\arxivnumber{2003.11787}

\abstract{
Deep neural networks trained on jet images have been successful in classifying different kinds of jets.
In this paper, we identify the crucial physics features that could reproduce the classification performance of the convolutional neural network in the top jet vs.~QCD jet classification.
We design a neural network that considers two types of substructural features: two-point energy correlations, and the IRC unsafe counting variables of a morphological analysis of jet images.
The new set of IRC unsafe variables can be described by Minkowski functionals from integral geometry.
To integrate these features into a single framework, we reintroduce two-point energy correlations in terms of a graph neural network and provide the other features to the network afterward.
The network shows a comparable classification performance to the convolutional neural network.
Since both networks are using IRC unsafe features at some level, the results based on simulations are often dependent on the event generator choice.
We compare the classification results of \texttt{Pythia 8} and \texttt{Herwig 7}, and a simple reweighting on the distribution of IRC unsafe features reduces the difference between the results from the two simulations.
}

\maketitle
\setcounter{page}{2}
\flushbottom

\section{Introduction}

Interest in deep learning in collider physics \cite{Larkoski:2017jix,Asquith:2018igt,Guest:2018yhq,Radovic:2018dip,Abdughani:2019wuv} has been growing in recent years. 
Many applications of deep learning have appeared in jet classification \cite{Almeida:2015jua,
deOliveira:2015xxd,
Komiske:2016rsd,
Butter:2017cot,
Dery:2017fap,
Kasieczka:2017nvn,
Louppe:2017ipp,
Cheng:2017rdo,
Egan:2017ojy,
Metodiev:2017vrx,
Komiske:2018cqr,
Macaluso:2018tck,
Andreassen:2018apy,
Lim:2018toa,
Qu:2019gqs,
Chakraborty:2019imr,
Andreassen:2019txo,
Chen:2019uar,
Cheng:2019isq,
Chen:2019apv,
Kasieczka:2020yyl}, anomaly detection \cite{Heimel:2018mkt,Farina:2018fyg,Hajer:2018kqm,Dillon:2019cqt,Diefenbacher:2019ezd,Blance:2019ibf,Roy:2019jae,Collins:2019jip,Amram:2020ykb,Nachman:2020lpy,Andreassen:2020nkr}, particle identification \cite{Guest:2016iqz,ATL-PHYS-PUB-2017-003,CMS-DP-2017-005}, pileup mitigation \cite{Martinez:2018fwc,Komiske:2017ubm,Komiske:2018lor}, 
event generation \cite{Bendavid:2017zhk,
Klimek:2018mza,
Otten:2019hhl,
Hashemi:2019fkn,
DiSipio:2019imz,
Butter:2019cae,
Carrazza:2019cnt,
SHiP:2019gcl,
Butter:2019eyo,
Bishara:2019iwh,
Bothmann:2020ywa,
Gao:2020vdv,
Gao:2020zvv,
Matchev:2020tbw,
Badger:2020uow}, unfolding \cite{Andreassen:2019cjw,Bellagente:2019uyp},
and parton distribution functions \cite{Forte:2002fg,
Forte:2002us,
Rojo:2004iq,
DelDebbio:2004xtd,
DelDebbio:2007ee,
Ball:2008by,
Ball:2009mk,
Ball:2010de,
Ball:2010gb,
Ball:2011mu,
Lionetti:2011pw,
Ball:2012cx,Ball:2013hta,Carrazza:2013bra,Carrazza:2013wua,
Ball:2014uwa,
Ball:2017nwa,
Bertone:2017tyb}. 
Deep learning will be used more in the analysis of LHC run III data.
Among those, jet classification using neural networks is one of the well-established areas. 
Several approaches have been proposed, and the performance of different models has been compared \cite{Kasieczka:2019dbj}.
For the classification between top jets and QCD jets, neural networks trained on low-level inputs showed a significant improvement in the classification performance compared to the previous methods \cite{Aaboud:2018psm}.

Before the deep learning in jet classification, the classification using the jet substructure information achieved remarkable success.
The particles coming from the decay of a boosted heavy particle give clear substructures inside the reconstructed jet.
The substructure maybe characterized by various manners; for example, by going through the jet clustering sequence
\cite{Butterworth:2008iy,
Thaler:2008ju,
Kaplan:2008ie,
Ellis:2009su,
CMS:2009lxa,
Plehn:2009rk,
Plehn:2010st,
Dasgupta:2013ihk,
Larkoski:2014wba}, reclustering jet constituents into the jets with smaller radius to identify subjets 
\cite{Butterworth:2008iy,
Krohn:2009th,
Soper:2011cr,
Soper:2012pb,
Soper:2014rya}, or the energy correlations 
\cite{Tkachov:1995kk,
Thaler:2010tr,
Jankowiak:2011qa,
Jankowiak:2012na,
Larkoski:2012eh,
Gallicchio:2012ez,
Larkoski:2013eya,
Larkoski:2014gra,
Moult:2016cvt,
Komiske:2017aww,
Chen:2019bpb}.
Note that such substructures are often defined by infrared and collinear (IRC) safe algorithms or observables which are theoretically more predictable. 
The IRC unsafe quantities are also used in the jet classification. 
For example, the number of charged tracks \cite{Gallicchio:2012ez} is very useful quantity for the quark jet vs.~gluon jet classification.
In some cases, the IRC unsafe counting variable has an IRC safe counterpart such as soft drop multiplicity \cite{Frye:2017yrw}.

The pattern of soft radiation is also important for the classification.
For example, a color singlet boosted heavy particle has emission isolated in terms of soft activity unlike quark and gluon jets.
The related substructure quantity has been incorporated in Higgs taggers \cite{Gallicchio:2010sw,Gallicchio:2010dq} and top taggers \cite{Hook:2011cq}.
Such soft particle distribution may also contribute to the jet classification using neural networks in order to improve the performance.

While the improvement using deep learning is impressive, the physics behind it has not been addressed.  
So far, the classifier based on a convolutional neural network (CNN) trained on the jet image performs well for selecting the top jets.  
It is numerically shown that the CNN uses IRC safe features mostly \cite{Choi:2018dag}, but
it is not easy to make an estimate of systematic uncertainties from various sources without knowing what kind of features of the jet is used in the model. 
Bayesian networks are capable of tracking those uncertainties \cite{Bollweg:2019skg,Kasieczka:2020vlh}, but it is also useful to identify the features in order to interpret the network outputs and uncertainties.
The aim of this paper is to provide a convenient parametrization of the jet feature contributing to the classification using jet images.

In this paper, we address the question in the following steps.
In \sectionref{sec:relation_net_and_s2}, we first introduce a graph neural network \cite{1555942,4700287,DBLP:journals/corr/RaposoSBPLB17,NIPS2017_7082,DBLP:journals/corr/abs-1806-01261} with constraints, and the network is more restrictive than CNN.
Graph networks are flexible enough for analyzing multiple objects appears at the LHC, and have been studied in various contexts \cite{Henrion2017NeuralMP,
Martinez:2018fwc,
Komiske:2018cqr,
Qu:2019gqs,
Chen:2019apv,
Qasim:2019otl,
Abdughani:2018wrw,
Moreno:2019bmu,
Moreno:2019neq,
Ren:2019xhp,
Mikuni:2020wpr,
Bister:2020rfv}.
The graph network in this paper has access to only IRC safe two-point energy correlations \cite{Basham:1978bw,Basham:1978zq,Basham:1979zw,Basham:1979gh,Jankowiak:2011qa,Jankowiak:2012na,Larkoski:2012eh,Lim:2018toa,Chakraborty:2019imr}. 
It was shown that the network has comparable performance to the CNN in the Higgs jet vs.~QCD jet classification \cite{Chakraborty:2019imr}.
We use this network for top jet vs.~QCD jet classification, and it is a good starting point toward the network whose top tagging performance is comparable to the CNN.

To integrate the IRC unsafe quantities to this framework, we formulate a sequence of novel morphological measures based on Minkowski functionals, in \sectionref{sec:minkowski_functionals}.
The sequence includes the number of pixels with finite energy deposit (active pixels), $N^{(0)}$, the number of pixels that touch the active pixels, $N^{(1)}$.  
These numbers can be considered as a discretized version of Minkowski functionals.
They are formulated in a mathematical theory called integral geometry and describes geometric measures to the point distributions.
The application of the Minkowski functionals has already been considered in the astrophysical analysis \cite{Mecke:1994ax,Schmalzing:1995qn,Schmalzing:1997aj,Schmalzing:1997uc,WINITZKI199875,Kerscher:2001ec,Beisbart:2001gk,Matsubara_2003,Hikage:2006fe,G_ring_2013,klatt2017morphometric2,klatt2017morphometric3,Chingangbam:2017sap,Pranav:2018pnu}, and statistical mechanics \cite{PhysRevE.53.4794,10.1007/3-540-45043-2_6,Mantz_2008}.
We perform a morphological analysis to the distribution of soft activity in the jet.

When the first few elements of the Minkowski sequence are included in the graph network inputs, the new classifier has the same performance as the jet image CNN classifier, as shown in \sectionref{sec:s2_top_tagger}.
This means that the improvement of the CNN classifier comes from the geometric quantities of the pixels, and also it is summarized by just a few numbers of additional variables. 
Our result suggests that the CNN output is correlated to a few numbers of geometric quantities derived from the jet image.

In the collider study, event simulators are used extensively to estimate the signal and background distributions. 
The sequence of Minkowski functionals calculated from a jet image is IRC unsafe quantities, and the simulated data need to be calibrated by the experimental data.
We propose an event reweighting method based on the IRC unsafe quantities for the calibration in \sectionref{sec:reweighting}. 
We conclude in \sectionref{sec:conclusion}.

\section{IRC Safe Two-Point Energy Correlations and Relation Network}
\label{sec:relation_net_and_s2}

The jet classifier using a deep learning model trained on the jet image has achieved better performance compared with the other statistical methods.
Still, it is not straightforward to identify the key physical features that contributed to the improvement, other than looking for the hidden data representations of the CNN \cite{deOliveira:2015xxd,Kasieczka:2017nvn,Lin:2018cin,Farina:2018fyg}, or checking the response of the network after perturbing the inputs \cite{Choi:2018dag}.
Note that organized networks whose hidden representations have physical interpretations \cite{Butter:2017cot,Andreassen:2018apy,Komiske:2018cqr,Chakraborty:2019imr,Cheng:2019isq} allow us to interpret the results in terms of physics.
For this purpose, we consider flexible and interpretable quantities derived from the jet image and use them as inputs to a jet classifier modeled by a multilayer perceptron (MLP).
Additional inputs are considered until the performance of the classifier is equivalent to that of the best classifiers using the jet image.

We first introduce two-point energy correlation spectra $S_2$ \cite{Lim:2018toa,Chakraborty:2019imr} as a function of the distance between the jet constituents $R$,
\begin{eqnarray}
S_{2,\jet_a \jet_b}(R) & = &
S_{2,ab}(R) =
\int d \vec{R}_1 d\vec{R}_2 \, P_{T,\jet_a}(\vec{R}_1) P_{T,\jet_b}(\vec{R}_2) \, \delta( R - R_{12} ),
\\
S_{2}(R) 
& = &
S_{2,\jet \jet} (R),
\end{eqnarray}
where $S_{2,ab}$ is a shorthand notation of $S_{2,\jet_a \jet_b}$;
\begin{equation}
P_{T,\jet_a}(\vec{R}) 
 = 
\sum_{i \in \jet_a} p_{T,i} \, \delta(\vec{R} - \vec{R}_i ),
\end{equation}
is an energy flow of a subjet $\jet_a$ of a jet $\jet$; $a$ and $b$ are indices of the subjet. The  $R_{ij}$ is the relative angular distance between two constituents, $\sqrt{(\eta_i - \eta_j)^2 + (\phi_i - \phi_j)^2}$. 
The $S_{2,ab}$ is an IRC safe quantity.
For the Higgs jet vs.~QCD jet classification, an MLP trained on the transverse momenta, masses, and $S_2$'s of the jet and trimmed jet performs nearly as good as a CNN trained on jet images.
In \cite{Chakraborty:2019imr}, we relate $S_{2,ab}$ to the generic jet classifiers through its formal expansion with respect to the energy flow.
This shows that $S_{2,ab}$ is flexible enough to describe many quantities for the classification of jets.

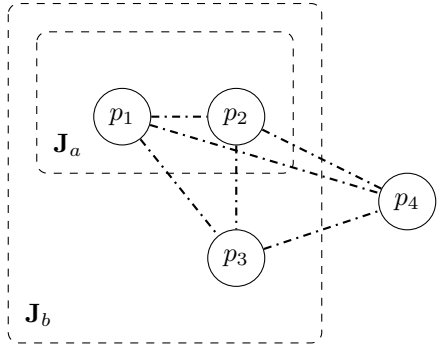
\begin{figure}
\begin{center}
\begin{tikzpicture}[scale=1.5,>=stealth, every node/.style={circle, draw, minimum size=0.75cm}]
% points
\draw (0,0) node(p1) {$p_1$};
\draw (1,0) node(p2) {$p_2$};
\draw (1.0,-1.25) node(p3) {$p_3$};
\draw (2.5,-0.75) node(p4) {$p_4$};
% edges
\draw[thick, dash dot] (p1) -- (p2);
\draw[thick, dash dot] (p1) -- (p3);
\draw[thick, dash dot] (p1) -- (p4);
\draw[thick, dash dot] (p2) -- (p3);
\draw[thick, dash dot] (p2) -- (p4);
\draw[thick, dash dot] (p3) -- (p4);
% loops
%\path (p1) edge [thick, dash dot,out=120,in=150,looseness=15] (p1);
%\path (p2) edge [thick, dash dot,out=75,in=105,looseness=15] (p2);
%\path (p3) edge [thick, dash dot,out=255,in=285,looseness=15] (p3);
%\path (p4) edge [thick, dash dot,out=300,in=330,looseness=15] (p4);
% label box
\draw [dashed, rounded corners] (-1.0,1.0) rectangle ++(2.75,-3.0);
\draw [dashed, rounded corners] (-0.75,0.75) rectangle ++(2.25,-1.25);
% label
\node [draw=none,anchor=west] at (-1.0,-1.75) {$\jet_b$};
\node [draw=none,anchor=west] at (-0.75,-0.25) {$\jet_a$};
\end{tikzpicture}
\end{center}
\caption{
\label{fig:jet_graph}
A schematic diagram of the graph representation of a jet used in this paper. 
Each vertex corresponds to a jet constituent, and a line between two circles represents the variable calculated from the two vertices.
Each dashed rectangle represents a subjet that contains the enclosed jet constituents.
}
\end{figure}

In this section, we first derive $S_{2,ab}$ in terms of a vertex-labeled fully-connected graph to integrate them into a framework of the graph network and extend it for further ML analysis. 
A graph is a set of the points and the lines connecting them, which are called vertices and edges, respectively. 
In our setup, each vertex of the graph corresponds to a jet constituent, and the inputs to the $i$-th vertex are the jet constituent momentum $p_i$. 
The labels of a vertex denote the subjets to which the constituent $i$ belongs. 
The graph network also has the other inputs $\bm{u}$ calculated from the given jet, for example, (sub)jet transverse momentum and mass.
A schematic diagram of the graph is in \figref{fig:jet_graph}. 
Each circle represents the jet constituent assigned to the corresponding vertex.
The dot-dashed lines are the edges.

\begin{figure}
\begin{center}
\begin{subfigure}{0.32\textwidth}
\includegraphics[scale=0.40]{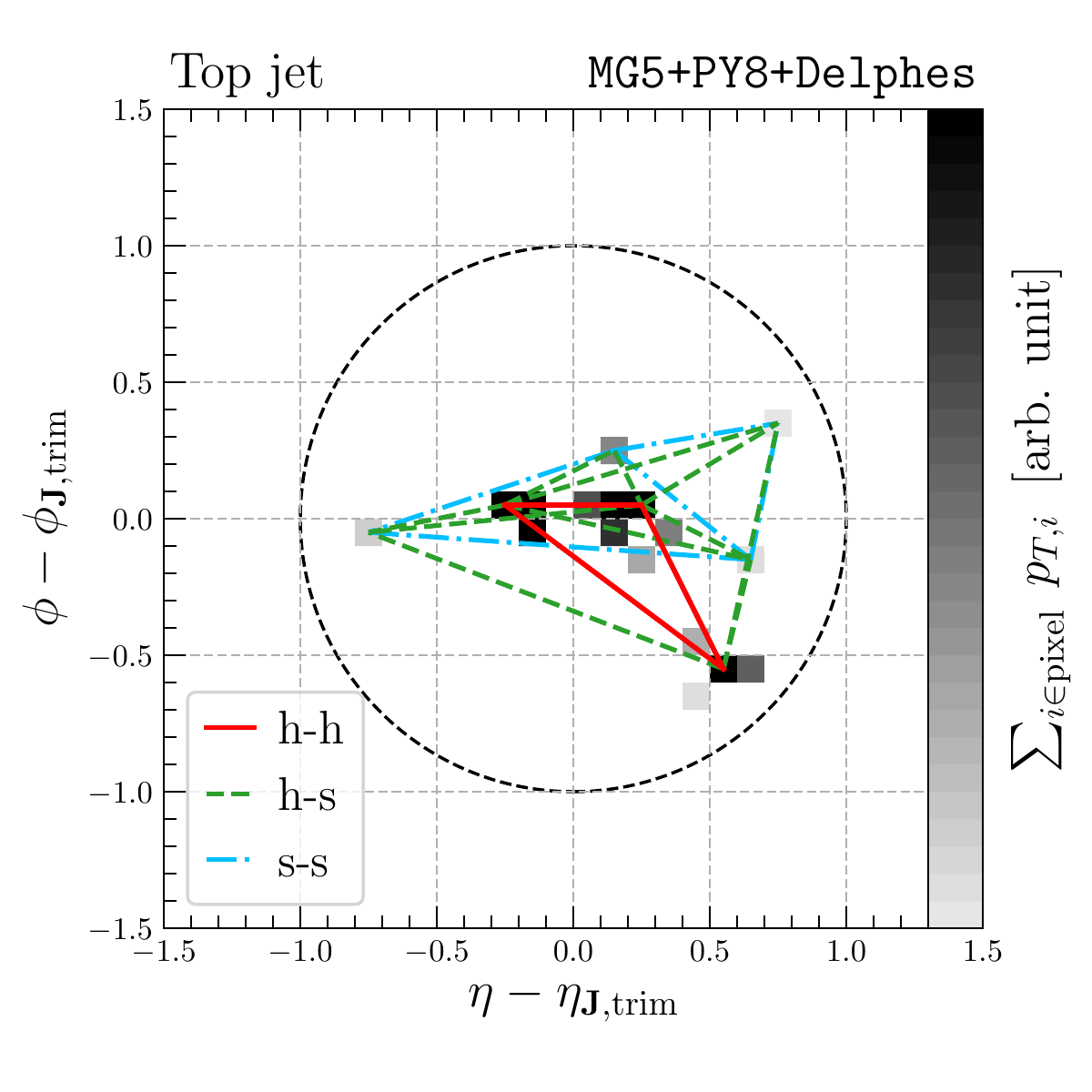}
\caption{
\label{fig:jet_image_graph_on_jet_image_trimsoft_top1}}
\end{subfigure}
\begin{subfigure}{0.32\textwidth}
\includegraphics[scale=0.40]{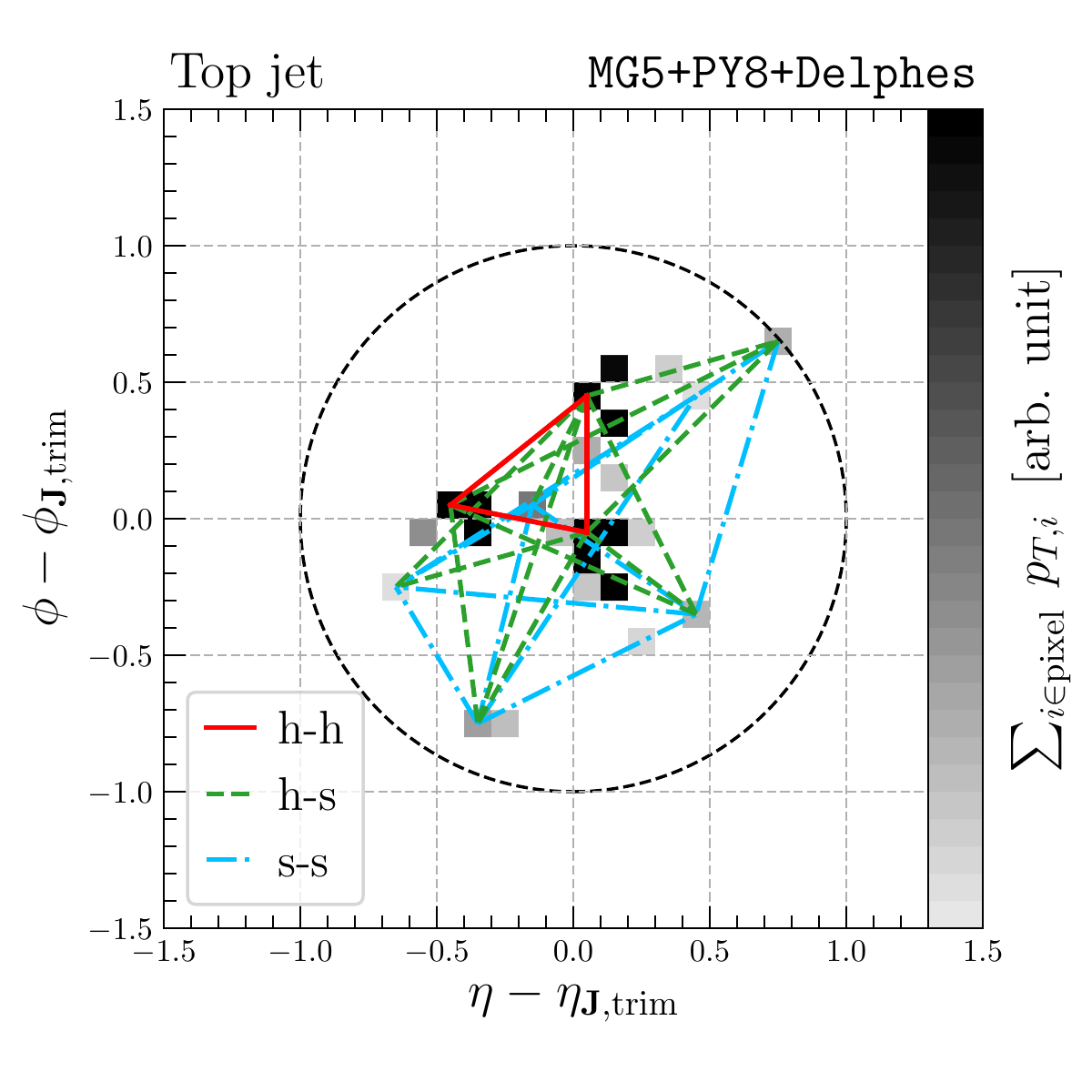}
\caption{
\label{fig:jet_image_graph_on_jet_image_trimsoft_top2}}
\end{subfigure}
\begin{subfigure}{0.32\textwidth}
\includegraphics[scale=0.40]{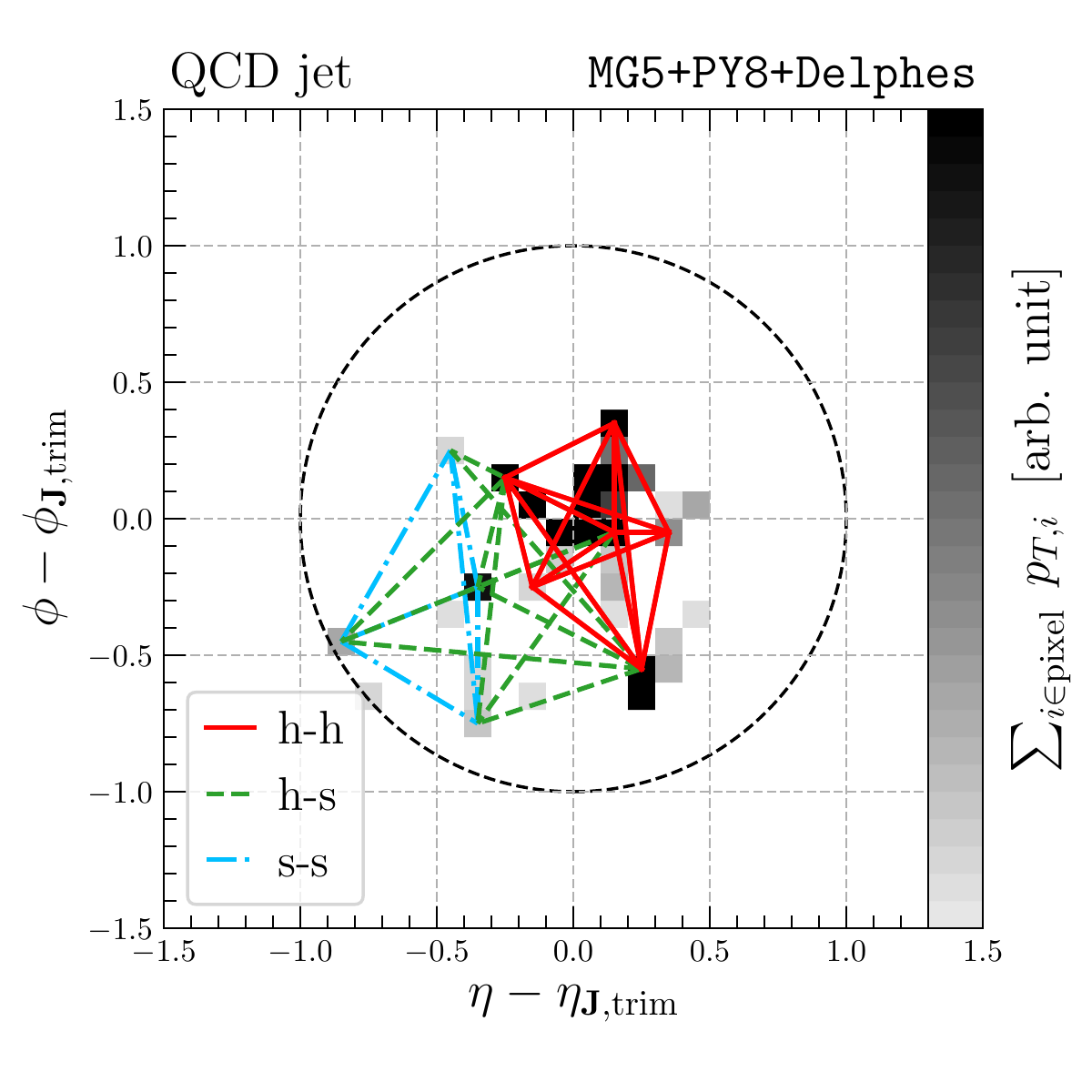}
\caption{
\label{fig:jet_image_graph_on_jet_image_trimsoft_qcd1}}
\end{subfigure}
\end{center}
\caption{
\label{fig:jet_image_graph_on_jet_image_trimsoft}
Schematic diagrams of the graph representations of jets.
(a) and (b) are top jet images, and (c) is a QCD jet image. 
Lines represent  the graphs on the jet images.
The red solid lines are edges between the constituents of the trimmed jet $\jet_\trim$. The green dashed lines are the edges between the constituents of $\jet_\trim$ and the constituents of  $\jet \setminus \jet_\trim$. 
The blue dot-dashed lines are edges between the constituents of  $\jet \setminus \jet_\trim$. 
Note that we omitted some edges for readability.
}
\end{figure}

We use a kind of graph network called a relation network (RN) \cite{DBLP:journals/corr/RaposoSBPLB17,NIPS2017_7082} that mainly utilizes correlations between two vertices.
The reason for using this network is that the kernel of the parton shower model is 1 $\rightarrow$ 2 splitting of partons.
The classifier can focus on the two-point correlations by using the relation network as a functional model.
The classifier output $\bm{u}'$ is the value of a functional model $\phi^u$ applied to the edge outputs $\bar{\bm{e}}_{ab}$, the vertex outputs $\bar{p}_{a}$, and the predefined inputs $\bm{u}$. 
\begin{equation}
\bm{u}'
=
\phi^u \left(
\bar{\bm{e}}_{ab},
\bar{\bm{p}}_a,
\bm{u}
\right).
\end{equation}
The edge output $\bar{\bm{e}}_{ab}$ is the aggregated two-point correlation between $\jet_a$ and $\jet_b$,
\begin{eqnarray}
\bar{\bm{e}}_{ab}
& = &
\sum_{\substack{i\in \jet_a \\ j\in \jet_b}} 
	\phi^e_{ab}(p_i, p_j, \bm{u}),
\end{eqnarray}
where $\phi^e_{ab}(p_i, p_j, \bm{u})$ is a functional model of a two-point correlation assigned on edge linking two jet constituents $i$ and $j$. 
The vertex output $\bar{\bm{p}}_a$ is the aggregated one-point correlation of $\jet_a$,
\begin{equation}
\quad
\bar{\bm{p}}_a
=
\sum_{i \in \jet_a} 
	\phi^v_a (p_i, \bm{u}),
\end{equation}
where $\phi^v_a (p_i, \bm{u})$ is a functional model of a one-point correlation assigned to a vertex that corresponds to a jet constituent $i$.  
The correlations $\bar{\bm{p}}_a$ and $\bar{\bm{e}}_{ab}$ are symmetric for the permutation of the jet constituents. 
We train $\bm{u}'$ to be the logits for the classification.\punctfootnote{See \cite{DBLP:journals/corr/abs-1806-01261} for other formalism of the graph neural network. }

We use energy correlators \cite{Tkachov:1995kk,Komiske:2017aww} for $\phi^e$ and $\phi^v$ to restrict $\bm{u}'$ to be IRC safe. 
Namely, we consider the following IRC safe $C$-correlators for $\bar{\bm{p}}_a$ and $\bar{\bm{e}}_{ab}$,
\begin{eqnarray}
\bar{\bm{p}}_a &= &\sum_{i \in \jet_a} p_{T,i} w_a(\vec{R}_i; \bm{u})\rightarrow p_{T,\jet_a} w_a (\bm{u}),
\\ 
\label{eqn:two_point_corr}
\bar{\bm{e}}_{ab} &=& \sum_{\substack{i \in \jet_a \\ j \in \jet_b}} p_{T,i} p_{T,j} w_{ab}(\vec{R}_i, \vec{R}_j; \bm{u})  \rightarrow   \sum_{i \in \jet_a} \sum_{j \in \jet_b} p_{T,i} p_{T,j} w_{ab}(R_{ij}; \bm{u}),
\end{eqnarray}
where $p_{T,i}$ is the transverse momentum of the $i$-th constituent, $\vec{R}_i = (\eta_i,\phi_i)$ is the pseudorapidity-azimuthal  coordinate of the $i$-th constituent.
The functions $w_a$ and $w_{ab}$ are the angular weighting functions of one-point and two-point energy correlators, respectively. 
The last step of the equation comes from the assumption that 
the classifier does not depend on the absolute angular coordinates of the (sub)jet constituents but uses the relative angular distances.

The last expression in \eqref{eqn:two_point_corr} can be written in terms of an integral \cite{Lim:2018toa}, 
\begin{equation}
\bar{\bm{e}}_{ab} 
 = 
\int dR \, S_{2,ab}(R) w_{ab}(R, \bm{u}).
\end{equation}
We may absorb the angular weighting functions $w_{ab}$ to $\phi^u$ so that the $S_{2,ab}$ and $p_{T,\jet_a}$ can be considered as effective inputs to the network.
\begin{equation}
\label{eqn:model:s2_mlp}
\bm{u}' = \phi^u \left( S_{2,ab}(R), p_{T,\jet_a}, \bm{u} \right).
\end{equation}
This setup is equivalent to the one using $S_{2,ab}$ as input, discussed in \cite{Lim:2018toa}.

\begin{figure}
\begin{center}
\begin{subfigure}{0.32\textwidth}
\includegraphics[scale=0.4]{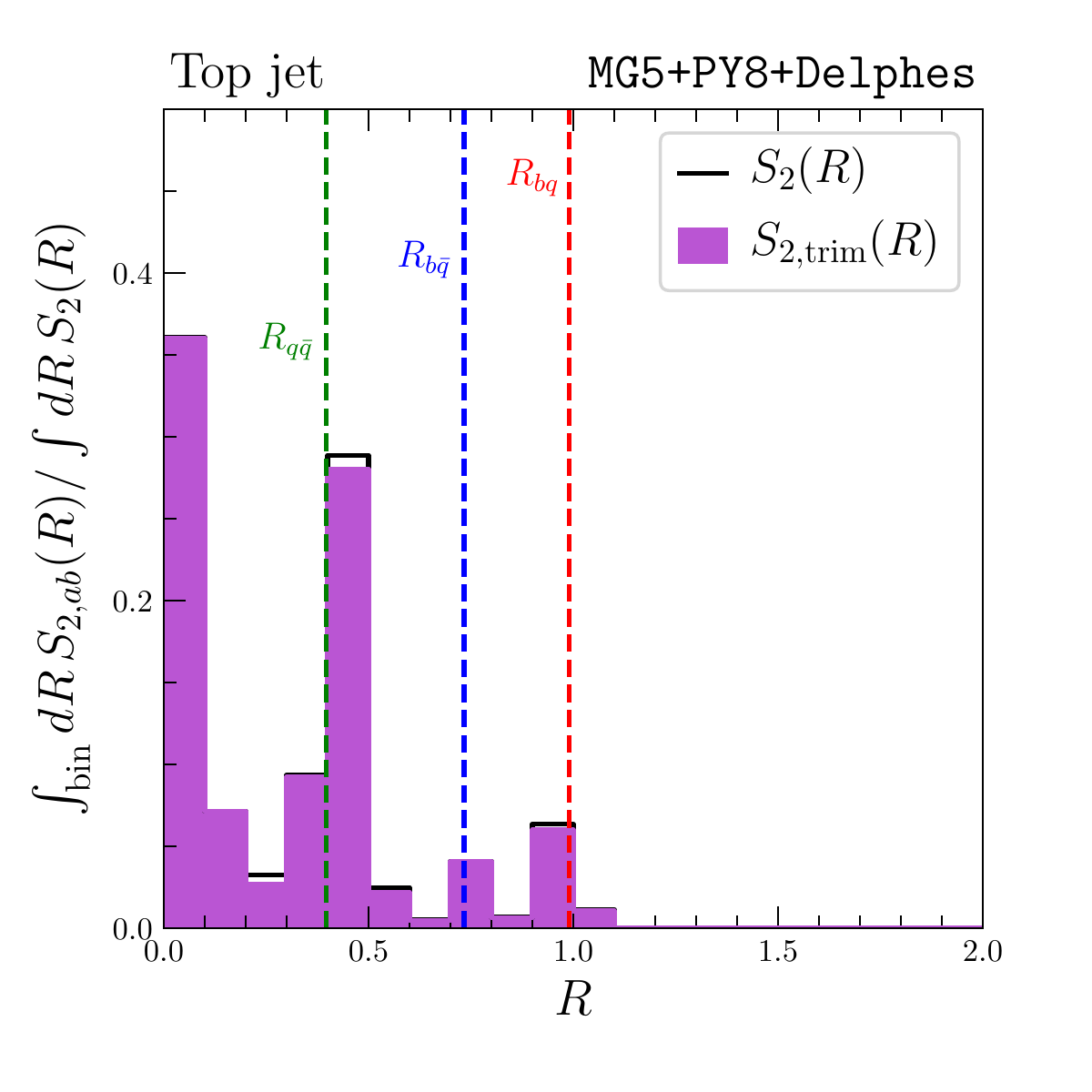}
\caption{\label{fig:jet_spectra_trimsoft_top1}}
\end{subfigure}
\begin{subfigure}{0.32\textwidth}
\includegraphics[scale=0.4]{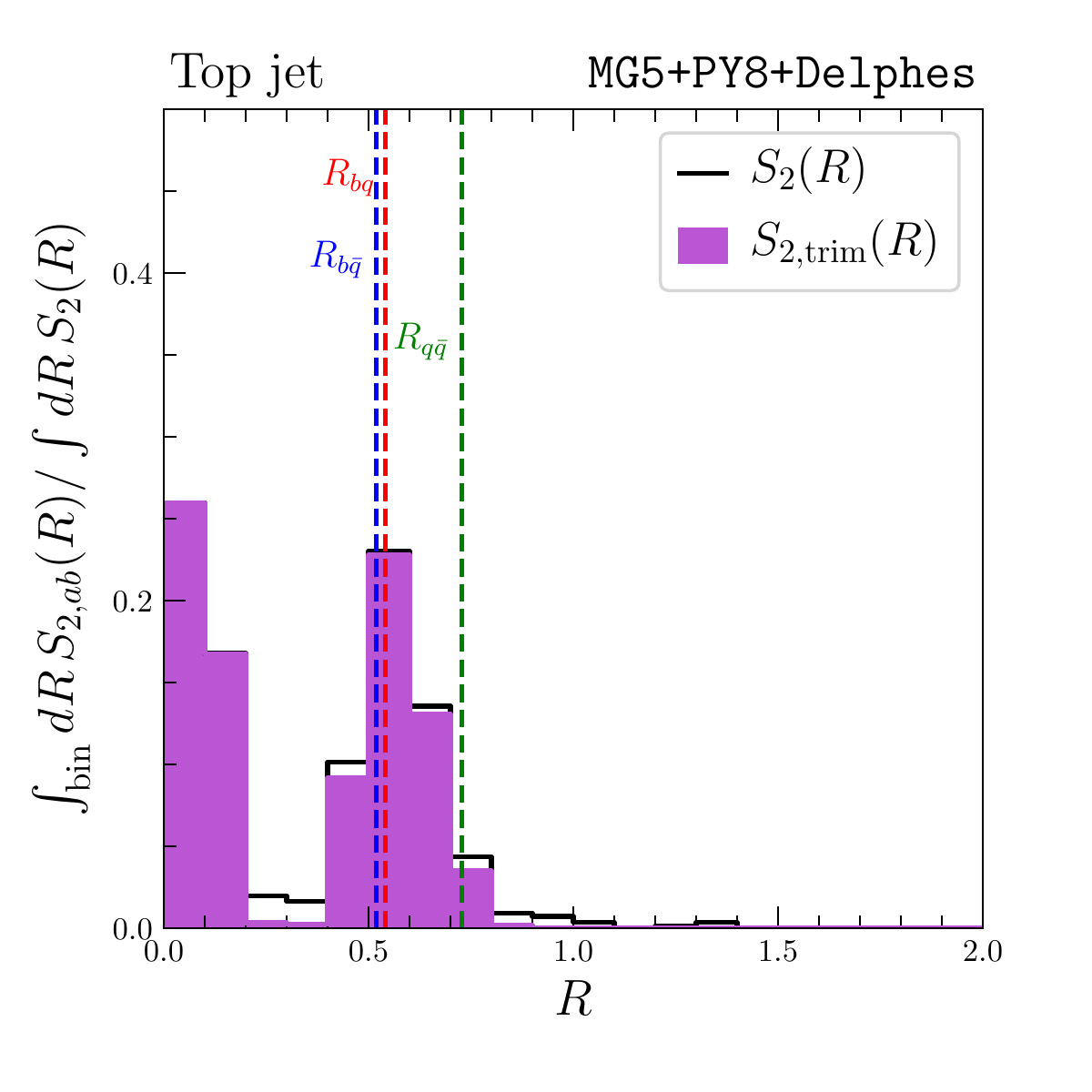}
\caption{\label{fig:jet_spectra_trimsoft_top2}}
\end{subfigure}
\begin{subfigure}{0.32\textwidth}
\includegraphics[scale=0.4]{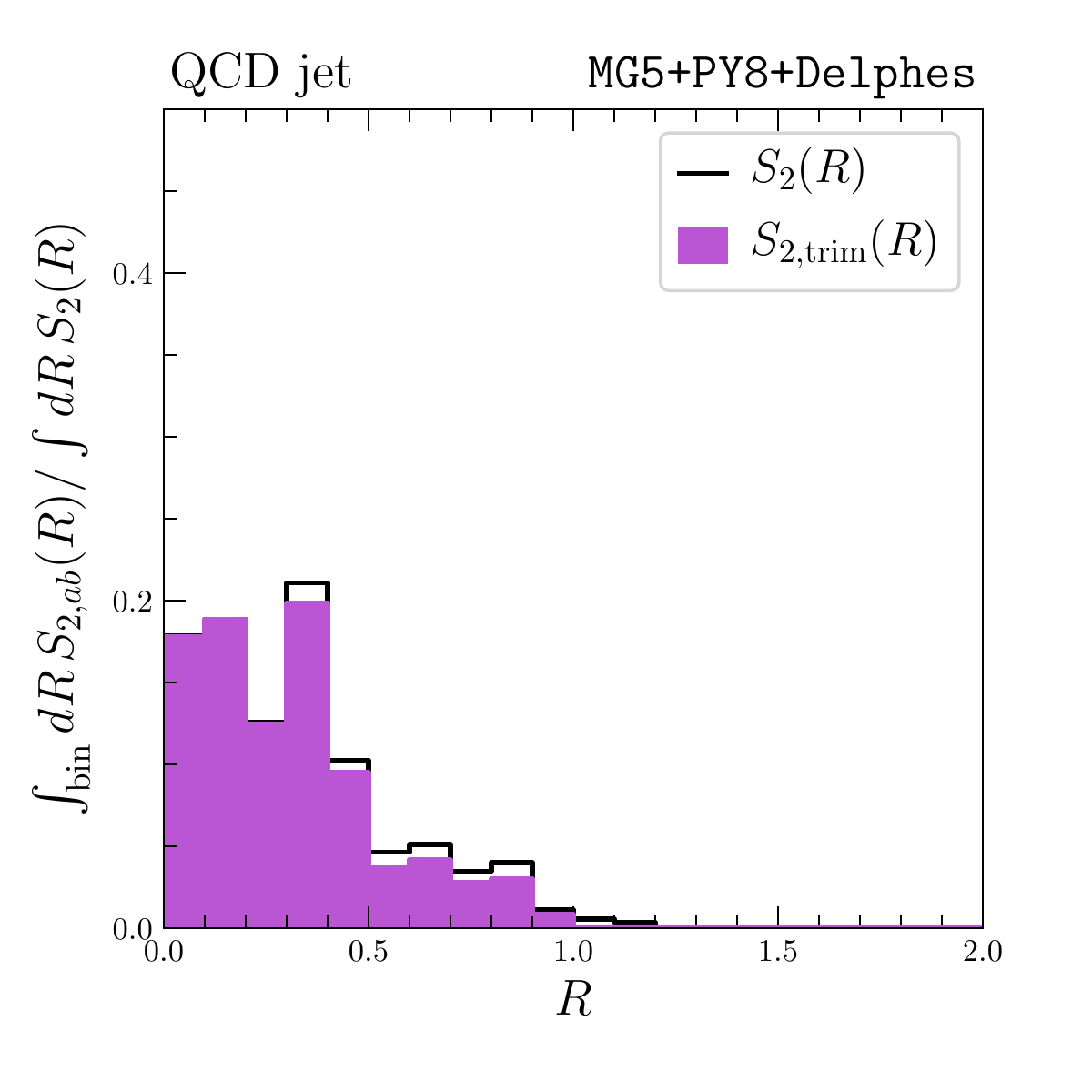}
\caption{\label{fig:jet_spectra_trimsoft_qcd1}}
\end{subfigure}
\end{center}
\caption{
\label{fig:jet_spectra_trimsoft}
The $S_2$ and $S_{2,\trim}$ distributions of the top jets and the QCD jet in \figref{fig:jet_image_graph_on_jet_image_trimsoft}.
The dashed lines are the characteristic angular scales of the top jets in the parton level.
}
\end{figure}

We now design a top tagger based on \eqref{eqn:model:s2_mlp}.  
The structure of the graph is specified by the subjet label $a$ and $b$ of $S_{2,ab}$.
We consider the following subjet labels for the top jet vs.~QCD jet classification.
\begin{itemize}
\item the trimmed jet, $\jet_{\mathrm{trim}}$, denoted by $h$, 
\item the compliment set of  $\jet_{\mathrm{trim}}$,  $\jet \setminus \jet_{\mathrm{trim}}$, denoted by $s$,
\item the leading $p_T$ subjet, $\jet_{\mathrm{1}}$, denoted by $1$,
\item the compliment set of  $\jet_{\mathrm{1}}$,  $\jet \setminus \jet_{\mathrm{1}}$, denoted by $c$,
\end{itemize}
Examples of the vertex-labeled graphs are in \figref{fig:jet_image_graph_on_jet_image_trimsoft}.
Note that the following relations hold for $S_{2}$ and $S_{2,ab}$, 
\begin{eqnarray}
S_{2}(R) 
& = &
S_{2,hh}(R) + 2 S_{2,hs}(R) + S_{2,ss}(R),
\\
& = &
S_{2,11}(R) + 2 S_{2,1c}(R) + S_{2,cc}(R).
\end{eqnarray}
Because $S_{2,ss}$ contains only the correlations between soft constituents, which is theoretically unpredictable and less reliable experimentally, 
we define the following combinations as in \cite{Lim:2018toa}.
\begin{eqnarray}
\label{eqn:s2_trim}
S_{2,\trim}(R) 
& = &
S_{2,hh}(R),
\\
\label{eqn:s2_soft}
S_{2,\soft}(R) 
& = &
2 S_{2,hs}(R) + S_{2,ss}(R).
\end{eqnarray}
The $S_{2,\trim}$ and $S_{2,\soft}$ distributions of the top jets and QCD jets in \figref{fig:jet_image_graph_on_jet_image_trimsoft} are shown in \figref{fig:jet_spectra_trimsoft}.

In parton level, $S_{2,\trim}$ and $S_{2,\soft}$ of a top quark have up to four peaks of delta functions and written as follows if all partons are sufficiently high $p_T$.
\begin{eqnarray}
\nonumber
S_{2,\trim}(R) 
& = &
(p_{T,b}^2 + p_{T,q}^2 + p_{T,\bar{q}}^2) \, \delta(R)
\\ & &
\vphantom{1}
+
2 p_{T,b} p_{T,q} \delta( R - R_{bq} )
+
2 p_{T,b} p_{T,\bar{q}} \delta( R - R_{b\bar{q}} )
+
2 p_{T,q} p_{T,\bar{q}} \delta( R - R_{q\bar{q}} ),
\\
S_{2,\soft}(R) 
& = &
0.
\end{eqnarray} 
Here, $b$ is a bottom quark from a top quark decay, and $q$ and $\bar{q}$ are quarks from the subsequent $W$ boson decay.
\Figref{fig:jet_spectra_trimsoft_top1} is the $S_{2,\trim}$ of the top jet that has those four peaks clearly.
This pattern is relatively rare for QCD jets.
\Figref{fig:jet_spectra_trimsoft_qcd1} is the $S_{2,\trim}$ of a typical QCD jet.

In the case where the characteristic angular scales of the top quark, $R_{bq}$, $R_{b\bar{q}}$, and $R_{q\bar{q}}$, are close to each other, it is not possible to see all peak structures in the $S_{2,\trim}(R)$ distributions.
Such an example is shown in \figref{fig:jet_spectra_trimsoft_top2},
 although the relative strength of the peaks in the $S_{2,\trim}$ distribution contains partial information of the three-prong structures.\punctfootnote{
For example, if all the partons from three-prong decay carry an equal fraction of momenta and their angular distances are the same, the ratio between the intensity of the two peaks is 1:2 in the parton level, while it is 1:1 for a two-prong decay \cite{Lim:2018toa}.
}

\begin{figure}
\begin{center}
\includegraphics[scale=0.5]{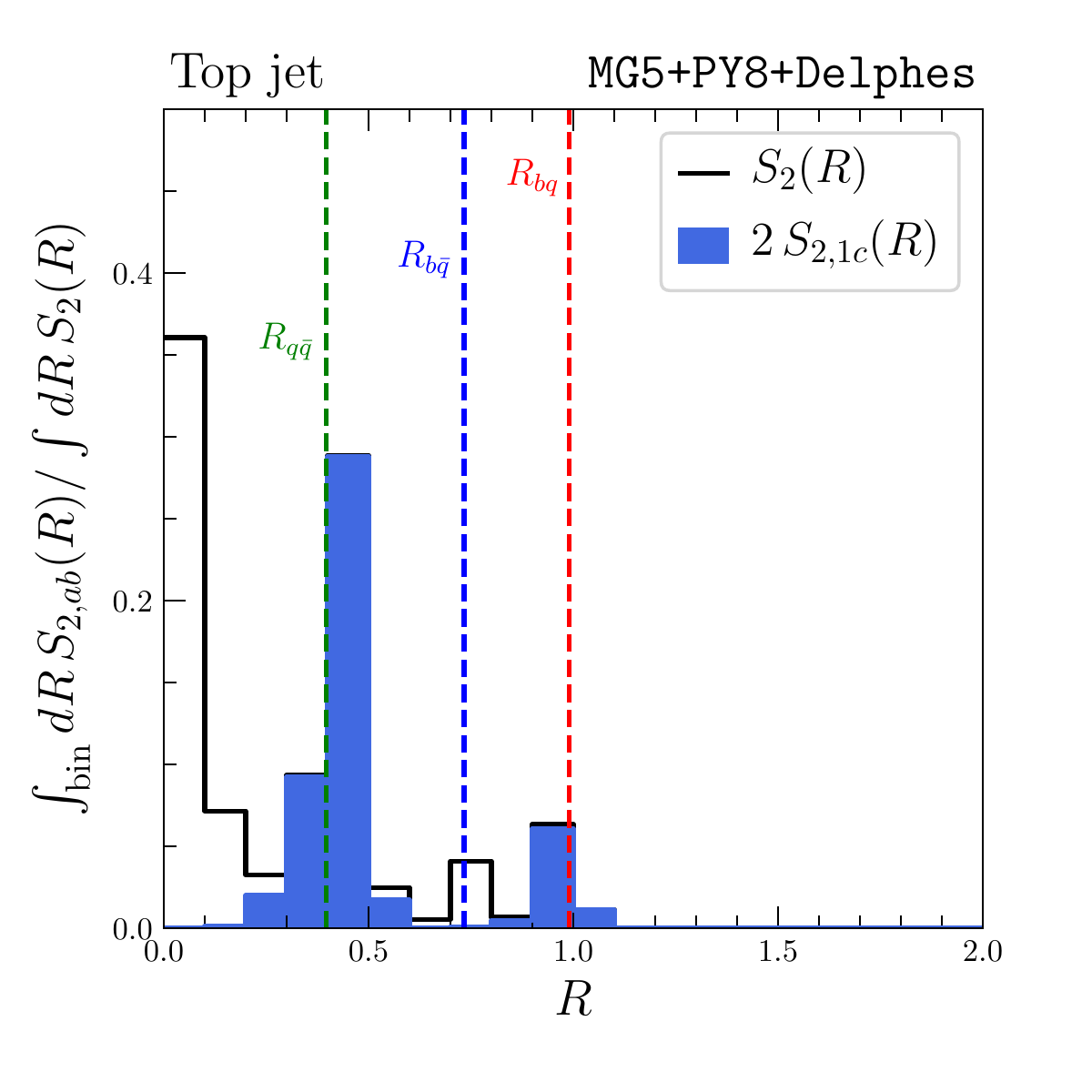}
\includegraphics[scale=0.5]{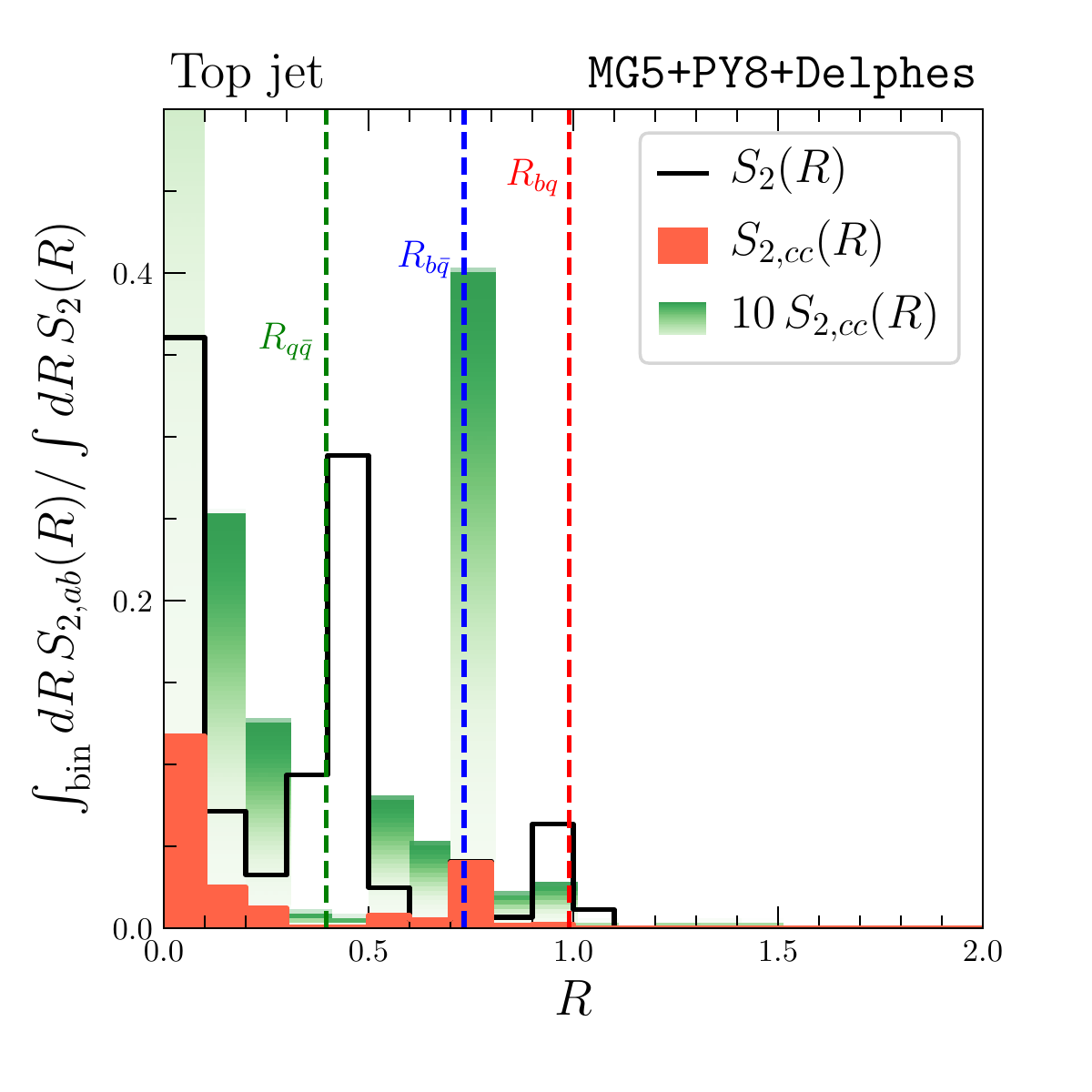}
\end{center}
\caption{
\label{fig:jet_spectra_leading_subjet_top1}
$S_{2,1c}$ and $S_{2,cc}$ distributions of the top jet in \figref{fig:jet_image_graph_on_jet_image_trimsoft_top1}.
The intensity of $S_{2,cc}$ is much smaller than $S_2$ because the subleading $p_T$ 
jets have small transverse momenta. 
The magnified distribution of $S_{2,cc}$ is shown in the green histogram.
The dashed lines are the characteristic angular scales at the parton level.}
\end{figure}

\begin{figure}
\begin{center}
\includegraphics[scale=0.5]{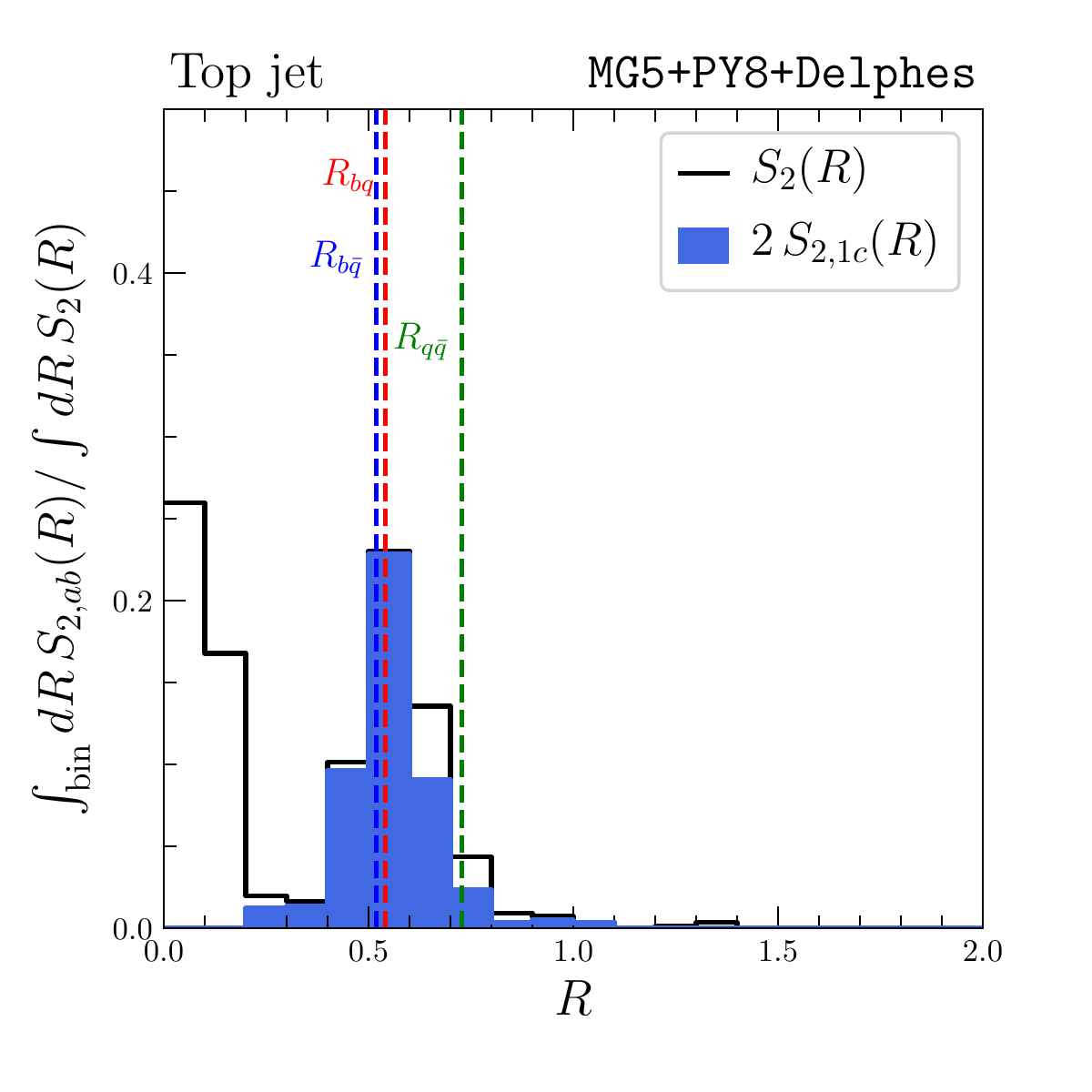}
\includegraphics[scale=0.5]{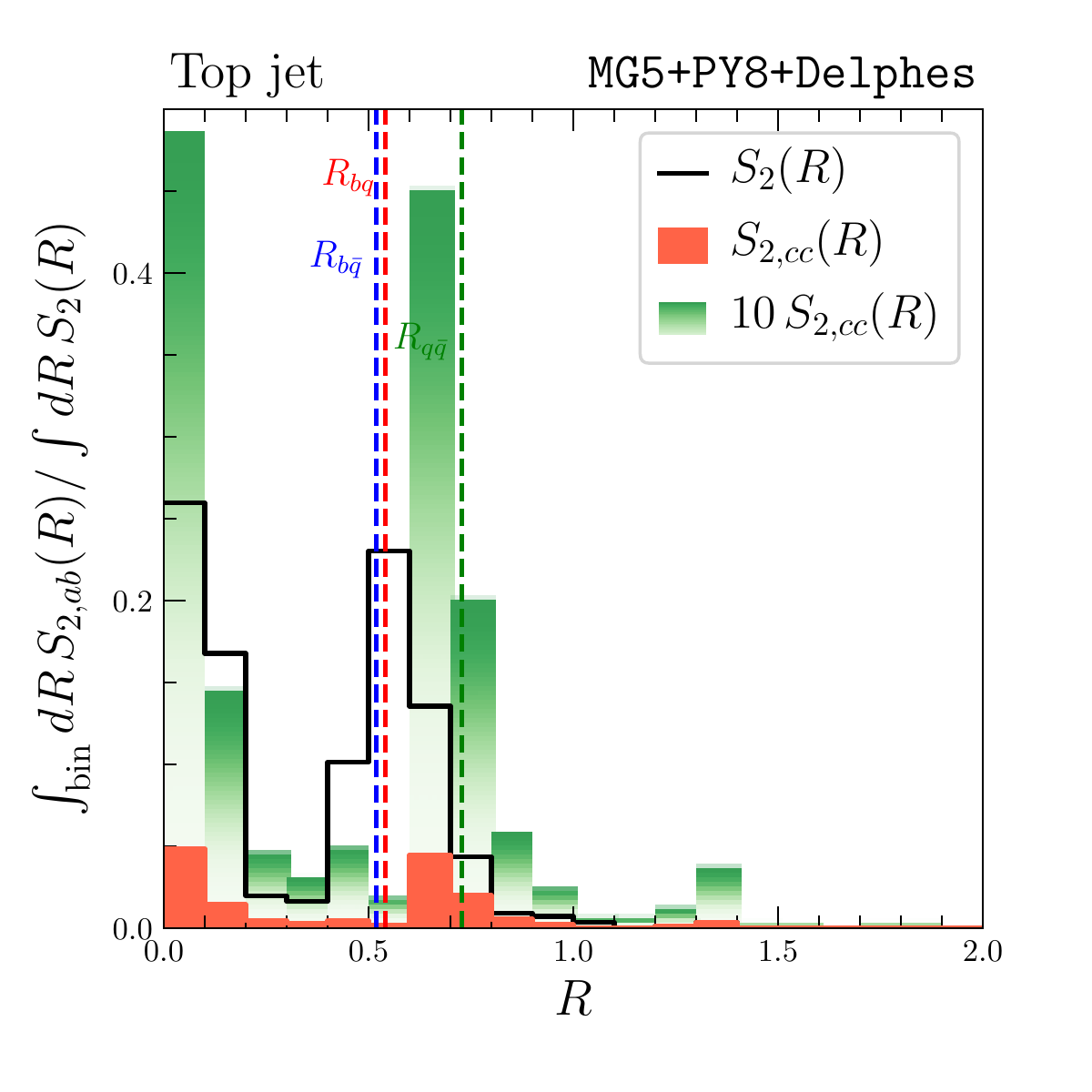}
\end{center}
\caption{
\label{fig:jet_spectra_leading_subjet_top2}
Same as \figref{fig:jet_spectra_leading_subjet_top1} but the top jet in  \figref{fig:jet_image_graph_on_jet_image_trimsoft_top2}.
}
\end{figure}

The information of the three-prong substructure is more clearly encoded in $S_{2,11}$, $S_{2, 1c}$, and $S_{2,cc}$. 
The two-point correlations of the top jets corresponding to  \figref{fig:jet_image_graph_on_jet_image_trimsoft_top1} and \figref{fig:jet_image_graph_on_jet_image_trimsoft_top2} are shown in \figref{fig:jet_spectra_leading_subjet_top1} and \figref{fig:jet_spectra_leading_subjet_top2}, respectively.
This decomposition of a given jet into $\jet_1$ and $\jet \setminus \jet_1$ factorizes the identification of a three-prong structure into that of two-prong substructures and its relative position from the $\jet_1$.
Those $S_{2,ab}$ in parton level are as follows, 
\begin{eqnarray}
S_{2,11}(R)
& = &
p_{T,i_1}^2 \delta(R),
\\
2\, S_{2,1c}(R)
& = &
2 p_{T,i_1} p_{T,i_2} \delta(R - R_{i_1 i_2})
+
2 p_{T,i_1} p_{T,i_3} \delta(R - R_{i_1 i_3}),
\\
S_{2,cc}(R)
& = &
(p_{T,i_2}^2 + p_{T,i_3}^2) \delta(R)
+
2 p_{T,i_2} p_{T,i_3} \delta(R - R_{i_2 i_3}),
\end{eqnarray}
where $i_k$ is the $k$-th leading $p_T$ parton.
\Figref{fig:jet_spectra_leading_subjet_top1} shows that the two peaks are in $S_{2,1c}$ and the other two peaks are in $S_{2,cc}$.
\Figref{fig:jet_spectra_leading_subjet_top2} is the case where values of $R_{bq}$ and $R_{b\bar{q}}$ are similar.
The $S_{2,cc}$ distribution has a peak at $R \approx 0.6$, and the peak intensity is comparable to that of the peak at $R = 0$ because the $\jet \setminus \jet_1$ has a two-prong substructure. 
In addition, the $S_{2,1c}$ distribution suggests that the high $p_T$ constituents of $\jet \setminus \jet_1$ are away from $\jet_1$ by a distance of $0.5$.
Note that the analysis on $S_{2,1c}$ is essentially telescoping jets \cite{Chien:2013kca,Chien:2017xrb} with respect to $\jet_1$.

\section{Morphological Analysis of Soft Emissions}
\label{sec:minkowski_functionals}

The number of particles of top jets and QCD jets is significantly different. 
For the boosted top quark decaying hadronically, i.e., $t\rightarrow  b W\rightarrow b q \bar{q}'$, the significant fraction of energy goes to color singlet $W$ boson.
The number of particles in a top jet is less than that of a gluon jet with the same jet mass and momentum, and the particles are concentrated near the quark directions.
The number of active pixels in the jet image, $N_{\mathrm{pixel}}$, is correlated to the number of particles in the jet, and therefore, it should be a crucial quantity of the jet image in the classification.
This quantity is IRC unsafe as $E\rightarrow 0$ and depends on the physics at a low energy scale, and its accuracy of the theoretical prediction is limited.\punctfootnote{Note that $N_{\mathrm{pixel}}$ in this paper is not calculated in the exact limit, $E\rightarrow 0$. The electronic calorimeter and hadronic calorimeter simulations have energy thresholds of 0.5 GeV and 1 GeV, respectively.}
Indeed \py{} and \hw{} predict significantly different pixel distributions for gluon jets, even though they are tuned to the experimental data.

\begin{figure}
\begin{center}
\begin{tikzpicture}[scale=1.0,>=stealth, every node/.style={circle, draw, minimum size=0.75cm}]
% image points
\begin{scope}
\foreach \x in {0,...,9}
	\draw [gray] (-1.125+0.25*\x,-1.25) -- (-1.125+0.25*\x,1.25);
\foreach \y in {0,...,9}
	\draw [gray] (-1.25,-1.125+0.25*\y) -- (1.25,-1.125+0.25*\y);
\draw [fill=gray] (-0.625,-0.125) rectangle ++(0.25,0.25);
\draw [fill=gray] (-0.875,0.375) rectangle ++(0.25,0.25);
\draw [fill=gray] (0.625,-0.625) rectangle ++(0.25,0.25);
\end{scope}
% sum
\node [draw=none] at (1.5,0) {$+$};
% square
\begin{scope}[shift={(3.0,0)}]
\foreach \x in {0,...,9}
	\draw [gray] (-1.125+0.25*\x,-1.25) -- (-1.125+0.25*\x,1.25);
\foreach \y in {0,...,9}
	\draw [gray] (-1.25,-1.125+0.25*\y) -- (1.25,-1.125+0.25*\y);
\foreach \x in {-1,...,1}
	\foreach \y in {-1,...,1}
		\draw [fill=gray] (-0.125+0.25*\x,-0.125+0.25*\y) rectangle ++(0.25,0.25);
\end{scope}
% arrow
\draw [->] (4.5,0) -- (5,0);
% pixel
\begin{scope}[shift={(6.5,0)}]
\foreach \x in {0,...,9}
	\draw [gray] (-1.125+0.25*\x,-1.25) -- (-1.125+0.25*\x,1.25);
\foreach \y in {0,...,9}
	\draw [gray] (-1.25,-1.125+0.25*\y) -- (1.25,-1.125+0.25*\y);
\foreach \x in {-1,...,1}
	\foreach \y in {-1,...,1}
		\draw [fill=gray] (-0.625+0.25*\x,-0.125+0.25*\y) rectangle ++(0.25,0.25);
\foreach \x in {-1,...,1}
	\foreach \y in {-1,...,1}
		\draw [fill=gray] (-0.875+0.25*\x,0.375+0.25*\y) rectangle ++(0.25,0.25);
\foreach \x in {-1,...,1}
	\foreach \y in {-1,...,1}
		\draw [fill=gray] (0.625+0.25*\x,-0.625+0.25*\y) rectangle ++(0.25,0.25);
\end{scope}
\end{tikzpicture}
\end{center}
\caption{
\label{fig:Minkowski_sum}
An illustration of the Minkowski sum in \eqref{eqn:Minkowski_sum}. 
The most left figure shows the active pixels $\mathcal{P}^{(0)}$, the figure at the center shows the pixels whose centers are $B^{(1)}$, and $\mathcal{P}^{(1)}$ is shown in the right figure. 
}
\end{figure}
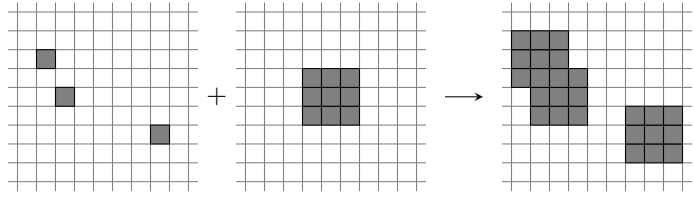

To generalize the idea of $N_{\mathrm{pixel}}$, we introduce a morphological analysis of soft emissions on jet images.
We consider two morphological operations: dilation, and filtering.
Let $N^{(i)}$ be a number of pixels in a dilated image,
\begin{equation}
N^{(i)} = \#(\mathcal{V}^{(i)}), \quad i \in \{ 0, 1, \cdots\},
\end{equation}
where $\mathcal{V}^{(i)}$ is the Minkowski sum of the set of the $(\eta, \phi)$ coordinate vectors $\vec{R}_i$ of the active pixels, $\mathcal{V}^{(0)}$, and a set of discrete coordinate vectors on a square for dilation, i.e.,
\begin{eqnarray}
\label{eqn:Minkowski_sum}
\mathcal{V}^{(i)} 
& = &
\mathcal{V}^{(0)} + \Delta R \times B^{(i)} 
= 
\{ a+\Delta R \, b \ \vert  \ a\in \mathcal{V}^{(0)}, b\in B^{(i)}  \},
\\
B^{(i)}
& = & \{ (k, l) | k, l \in \{-i, -i+1, \cdots, i-1, i \} \}. 
\end{eqnarray}
We denote a set of pixels whose centers belong to $\mathcal{V}^{(i)}$ as $\mathcal{P}^{(i)}$.
Note that $\mathcal{P}^{(0)}$ is identical to the set of active pixels, and $N^{(0)}$ is $N_{\pixel}$.
The set $\mathcal{P}^{(i)}$ is then a cover of the jet image, i.e., it is a union of the squares that attached to each active pixel.
The covers obey a recurrence relation that $\mathcal{P}^{(i)}$ includes pixels in $\mathcal{P}^{(i-1)}$ and those touching one of the edges or corners of $\mathcal{P}^{(i-1)}$.
This morphological mapping is illustrated in \figref{fig:Minkowski_sum}.
Note that $N^{(i)}$ is proportional to the area $A^{(i)}$ of pixels in the cover $\mathcal{P}^{(i)}$ because each pixel has the same area $(\Delta R)^2$, i.e.,
\begin{equation}
A^{(i)}=(\Delta R)^2 \times N^{(i)}.
\end{equation}
For the analysis of soft activity, we consider a filtered image whose active pixels have $p_T$ larger than $E$.
Let $N^{(i)}(E)$ be the number of active pixels in the filtered image.
If we choose sufficiently large threshold $E$, the number $N^{(i)}(E)$ is relatively stable against the choice of the event simulators.
The difference between the values of $N^{(i)}$ and $N^{(i)}(E)$ will provide us geometric information about the soft activity.

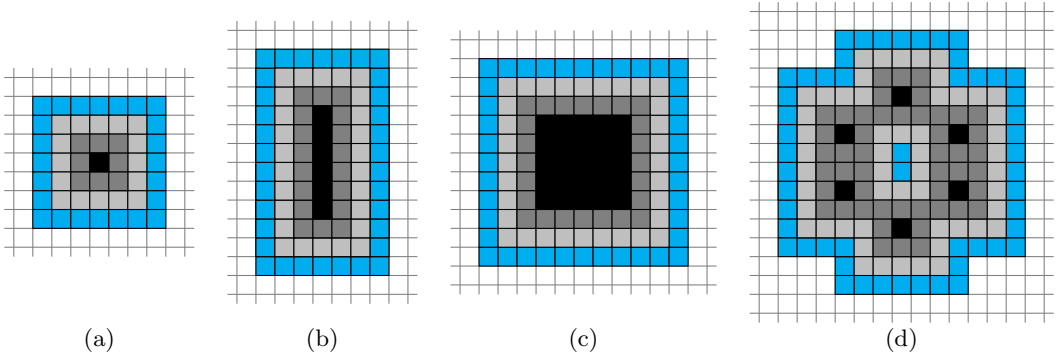
\begin{figure}
\begin{center}
\begin{tabular}{cccc}
\begin{tikzpicture}[scale=1.0,>=stealth, every node/.style={circle, draw, minimum size=0.75cm},
baseline={([yshift=-.5ex]current bounding box.center)}]
% grid
\foreach \x in {-4,...,5}
	\draw [gray] (-0.125+0.25*\x,-1.25) -- (-0.125+0.25*\x,1.25);
\foreach \y in {-4,...,5}
	\draw [gray] (-1.25,-0.125+0.25*\y) -- (1.25,-0.125+0.25*\y);
% n3
\foreach \x in {-3,...,3}
	\foreach \y in {-3,...,3}
		\draw [fill=cyan] (-0.125+0.25*\x,-0.125+0.25*\y) rectangle ++(0.25,0.25);
% n2
\foreach \x in {-2,...,2}
	\foreach \y in {-2,...,2}
		\draw [fill=lightgray] (-0.125+0.25*\x,-0.125+0.25*\y) rectangle ++(0.25,0.25);
% n1
\foreach \x in {-1,...,1}
	\foreach \y in {-1,...,1}
		\draw [fill=gray] (-0.125+0.25*\x,-0.125+0.25*\y) rectangle ++(0.25,0.25);
% n0
\draw [fill=black] (-0.125,-0.125) rectangle ++(0.25,0.25);
\end{tikzpicture}
&
\begin{tikzpicture}[scale=1.0,>=stealth, every node/.style={circle, draw, minimum size=0.75cm},
baseline={([yshift=-.5ex]current bounding box.center)}]
% grid
\foreach \x in {-5,...,9}
	\draw [gray] (-1.25,-0.125+0.25*\x) -- (1.25,-0.125+0.25*\x);
\foreach \y in {-4,...,5}
	\draw [gray] (-0.125+0.25*\y,-1.5) -- (-0.125+0.25*\y,2.25);
% n2
\foreach \x in {-3,...,3}
	\foreach \y in {-3,...,3}
		\foreach \pt in {0,...,5}
			\draw [fill=cyan] (-0.125+0.25*\y,-0.375+0.25*\pt+0.25*\x) rectangle ++(0.25,0.25);
% n2
\foreach \x in {-2,...,2}
	\foreach \y in {-2,...,2}
		\foreach \pt in {0,...,5}
			\draw [fill=lightgray] (-0.125+0.25*\y,-0.375+0.25*\pt+0.25*\x) rectangle ++(0.25,0.25);
% n1
\foreach \x in {-1,...,1}
	\foreach \y in {-1,...,1}
		\foreach \pt in {0,...,5}
			\draw [fill=gray] (-0.125+0.25*\y,-0.375+0.25*\pt+0.25*\x) rectangle ++(0.25,0.25);
% n0
\foreach \pt in {0,...,5}
	\draw [fill=black] (-0.125,-0.375+0.25*\pt) rectangle ++(0.25,0.25);
\end{tikzpicture}
&
\begin{tikzpicture}[scale=1.0,>=stealth, every node/.style={circle, draw, minimum size=0.75cm},
baseline={([yshift=-.5ex]current bounding box.center)}]
% grid
\foreach \x in {-6,...,7}
	\draw [gray] (-0.125+0.25*\x,-1.75) -- (-0.125+0.25*\x,1.75);
\foreach \y in {-6,...,7}
	\draw [gray] (-1.75,-0.125+0.25*\y) -- (1.75,-0.125+0.25*\y);
% n3
\foreach \x in {-5,...,5}
	\foreach \y in {-5,...,5}
		\draw [fill=cyan] (-0.125+0.25*\x,-0.125+0.25*\y) rectangle ++(0.25,0.25);
% n2
\foreach \x in {-4,...,4}
	\foreach \y in {-4,...,4}
		\draw [fill=lightgray] (-0.125+0.25*\x,-0.125+0.25*\y) rectangle ++(0.25,0.25);
% n1
\foreach \x in {-3,...,3}
	\foreach \y in {-3,...,3}
		\draw [fill=gray] (-0.125+0.25*\x,-0.125+0.25*\y) rectangle ++(0.25,0.25);
% n0
\foreach \ptx in {-2,...,2}
	\foreach \pty in {-2,...,2}
	\draw [fill=black] (-0.125+0.25*\ptx,-0.125+0.25*\pty) rectangle ++(0.25,0.25);
\end{tikzpicture}
&
\begin{tikzpicture}[scale=1.0,>=stealth, every node/.style={circle, draw, minimum size=0.75cm},
baseline={([yshift=-.5ex]current bounding box.center)}]
% grid
\foreach \x in {-7,...,8}
	\draw [gray] (-0.125+0.25*\x,-2.0) -- (-0.125+0.25*\x,2.25);
\foreach \y in {-7,...,9}
	\draw [gray] (-2.0,-0.125+0.25*\y) -- (2.0,-0.125+0.25*\y);
% n3
\foreach \x in {-3,...,3}
	\foreach \y in {-3,...,3}
	{
		\draw [fill=cyan] (-0.875+0.25*\x,-0.375+0.25*\y) rectangle ++(0.25,0.25);
		\draw [fill=cyan] (0.625+0.25*\x,-0.375+0.25*\y) rectangle ++(0.25,0.25);
		\draw [fill=cyan] (-0.875+0.25*\x,0.375+0.25*\y) rectangle ++(0.25,0.25);
		\draw [fill=cyan] (0.625+0.25*\x,0.375+0.25*\y) rectangle ++(0.25,0.25);
		\draw [fill=cyan] (-0.125+0.25*\x,-0.875+0.25*\y) rectangle ++(0.25,0.25);
		\draw [fill=cyan] (-0.125+0.25*\x,0.875+0.25*\y) rectangle ++(0.25,0.25);
	}
% n2
\foreach \x in {-2,...,2}
	\foreach \y in {-2,...,2}
	{
		\draw [fill=lightgray] (-0.875+0.25*\x,-0.375+0.25*\y) rectangle ++(0.25,0.25);
		\draw [fill=lightgray] (0.625+0.25*\x,-0.375+0.25*\y) rectangle ++(0.25,0.25);
		\draw [fill=lightgray] (-0.875+0.25*\x,0.375+0.25*\y) rectangle ++(0.25,0.25);
		\draw [fill=lightgray] (0.625+0.25*\x,0.375+0.25*\y) rectangle ++(0.25,0.25);
		\draw [fill=lightgray] (-0.125+0.25*\x,-0.875+0.25*\y) rectangle ++(0.25,0.25);
		\draw [fill=lightgray] (-0.125+0.25*\x,0.875+0.25*\y) rectangle ++(0.25,0.25);
	}
% n1
\foreach \x in {-1,...,1}
	\foreach \y in {-1,...,1}
	{
		\draw [fill=gray] (-0.875+0.25*\x,-0.375+0.25*\y) rectangle ++(0.25,0.25);
		\draw [fill=gray] (0.625+0.25*\x,-0.375+0.25*\y) rectangle ++(0.25,0.25);
		\draw [fill=gray] (-0.875+0.25*\x,0.375+0.25*\y) rectangle ++(0.25,0.25);
		\draw [fill=gray] (0.625+0.25*\x,0.375+0.25*\y) rectangle ++(0.25,0.25);
		\draw [fill=gray] (-0.125+0.25*\x,-0.875+0.25*\y) rectangle ++(0.25,0.25);
		\draw [fill=gray] (-0.125+0.25*\x,0.875+0.25*\y) rectangle ++(0.25,0.25);
	}
% n0
\draw [fill=black] (-0.875,-0.375) rectangle ++(0.25,0.25);
\draw [fill=black] (0.625,-0.375) rectangle ++(0.25,0.25);
\draw [fill=black] (-0.875,0.375) rectangle ++(0.25,0.25);
\draw [fill=black] (0.625,0.375) rectangle ++(0.25,0.25);
\draw [fill=black] (-0.125,-0.875) rectangle ++(0.25,0.25);
\draw [fill=black] (-0.125,0.875) rectangle ++(0.25,0.25);
\end{tikzpicture}
\\
\begin{subfigure}{0.1\textwidth}
\caption{
\label{fig:examples:1}
}
\end{subfigure} & 
\begin{subfigure}{0.1\textwidth}
\caption{
\label{fig:examples:2}
}
\end{subfigure} & 
\begin{subfigure}{0.1\textwidth}
\caption{
\label{fig:examples:3}
}
\end{subfigure} & 
\begin{subfigure}{0.1\textwidth}
\caption{
\label{fig:examples:4}
}
\end{subfigure} 
\end{tabular}
\end{center}
\caption{
\label{fig:examples}
Illustrations of $\mathcal{P}^{(i)}$ for (a) an isolated pixel, (b) a line of pixels,  (c)  a $5\times5$ square of pixels, (d) a ring of six pixels. 
For each plot, black pixels belongs to $P^{(0)}$; dark gray, light gray, blue pixels are the difference 
 $P^{(i)} \setminus P^{(i-1)}$ for $i=1,2,3$, respectively.
}
\end{figure}

The sequence of $N^{(i)}$ gives a quantitative description of the spatial distribution of pixels in the jet.
Before going into some mathematical background, let us capture the idea using simple examples. 
Consider the relations between $N^{(0)}$ and $N^{(1)}$ of \figref{fig:examples:1}, \figref{fig:examples:2}, and \figref{fig:examples:3}:
\begin{enumerate}
 \item Active pixels are separated by two or more pixels. 
 \begin{equation}
	N^{(1)} = 9 N^{(0)}
 \end{equation}
This corresponds to the limit of sparse and scattered pixels. 
\item  Active pixels are aligned on a line. 
\begin{equation}
N^{(1)} = 3 N^{(0)} + 6
\end{equation}
This case is the limit when soft activities come from a very narrow color string between two quarks at each end.
\item  Active pixels are clustered in a square. 
 \begin{equation}
 N^{(1)} \sim ( \sqrt{N^{(0)}}+2 )^2
 \end{equation}
This is the limit of a one-prong jet such as quark jet. 
\end{enumerate}
The ratio $N^{(1)}/N^{(0)}$ in large $N^{(0)}$ limit is approximately 9, 3, 1, respectively.
If pixel clusters appear at small angular scale, $N^{(1)}/N^{(0)}$ is reduced.
Therefore, $N^{(1)}/N^{(0)}$ quantifies the level of isolation of the pixels.

\begin{figure}
\begin{center}
\includegraphics[scale=0.5]{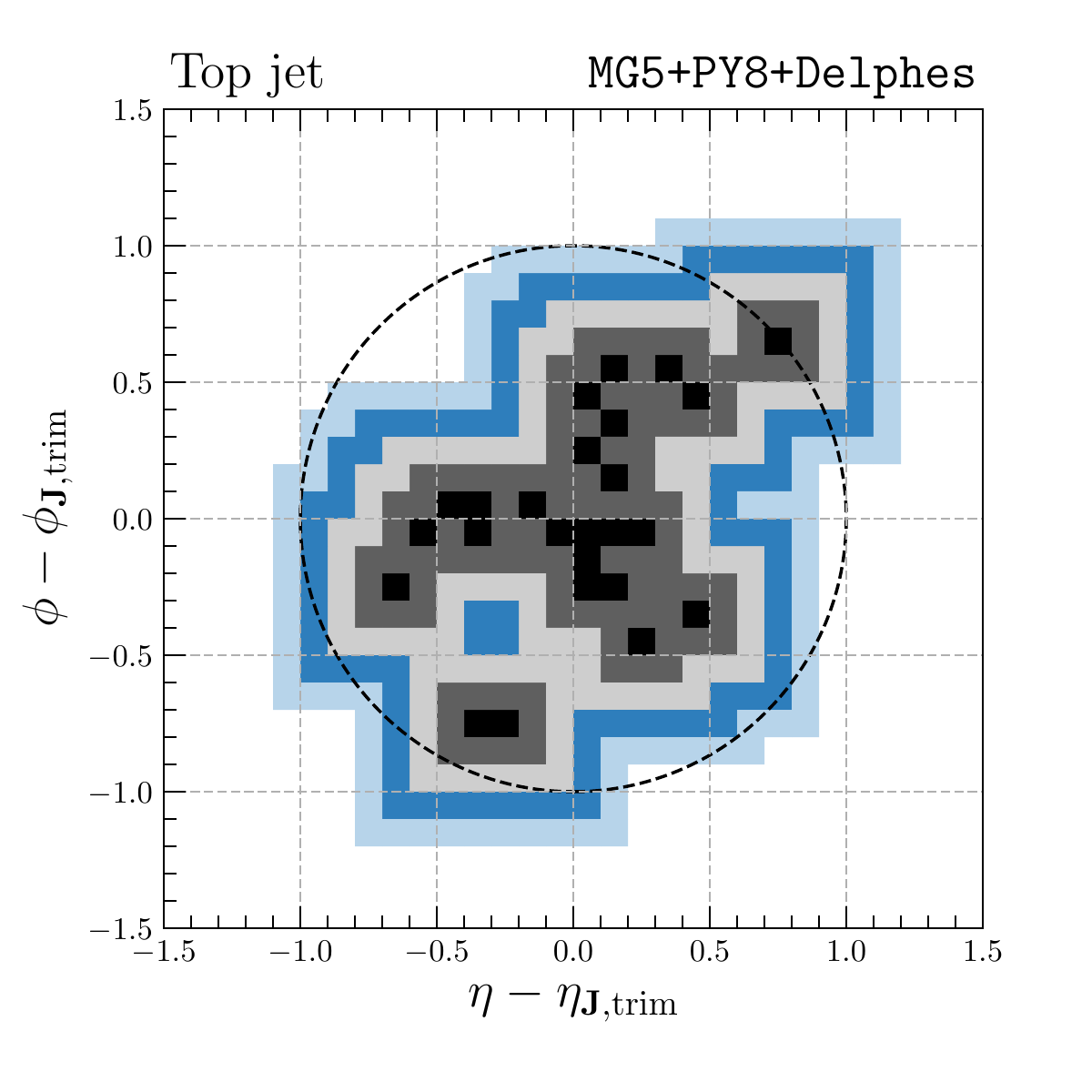}
\includegraphics[scale=0.5]{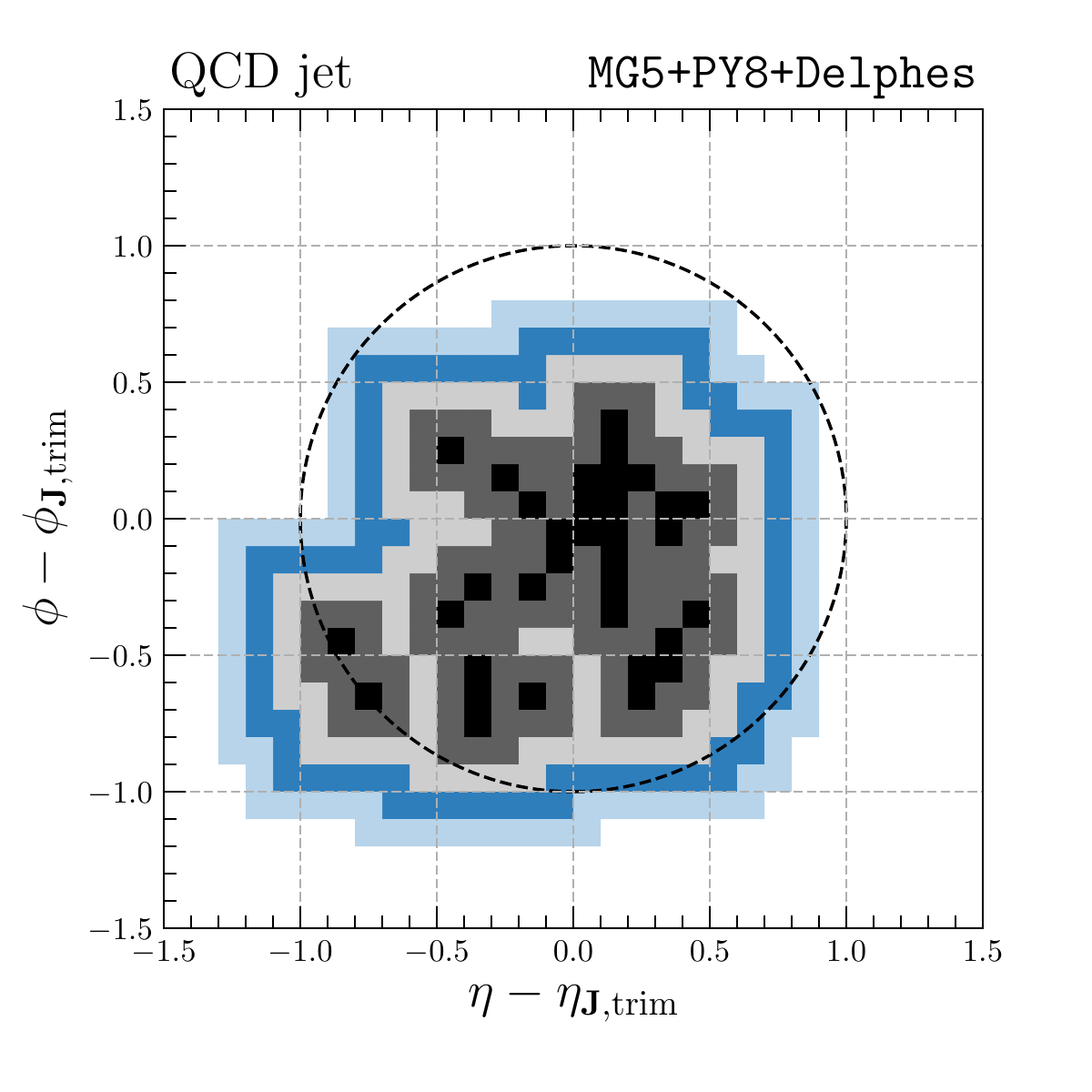}
\end{center}
\caption{
\label{fig:jet_image_graph_on_jet_image_leadingsubjet}
The Minkowski sum $\mathcal{P}^{(i)}$ of the top jet image and QCD jet image. 
For each plot, black pixels belongs to $P^{(0)}$  dark gray, light gray, blue pixels are the difference 
between $P^{(i)}-P^{(i-1)}$ for  $i=1,2,3$, respectively.
}
\end{figure}

\Figref{fig:jet_image_graph_on_jet_image_leadingsubjet} shows $\mathcal{P}^{(i)}$ of top and QCD jet images in \figref{fig:jet_image_graph_on_jet_image_trimsoft_top2} and \figref{fig:jet_image_graph_on_jet_image_trimsoft_qcd1}, respectively.
One quick observation is that $\mathcal{P}^{(i)}$ has some non-trivial structures for small $i$ ($i=0,1$),  but the pixels quickly merge into a single cluster as the index $i$ increases.
In the large angular scale ($i\geq 2$), the only relevant physics for the top jet vs.~QCD jet classification is the color charge of the parent parton, and the $N^{(i)}$ does not carry significant additional information. 
In the next section, we show that $N^{(0)}$ and $N^{(1)}$ are sufficient to describe the soft structure contributing to the top jet and QCD jet classifier modeled by CNN.

The analysis based on the pixels can be generalized to the particle level analysis with a continuous parameter $R$ as follows. 
Let $\mathcal{P}(R)$ be a cover of particles on $(\eta,\phi)$ plane.
\begin{equation}
\mathcal{P}(R)= \bigcup_{i \in \jet} B_i(R), 
\end{equation}
where $B_i(R)$ is a disk with radius $R$ and whose center is the direction vector $\vec{R}_i$ of a particle $i$.
The area $A(R)$ of the cover $\mathcal{P}(R)$ is a quantity related to $N^{(i)}$, i.e., $A^{(i)}$ can be considered as a discrete analog of $A(R)$.

The change of $A(R)$ with respect to $R$ also quantifies the spacial distribution of particles.
As far as all the disks are isolated, $A(R)/ (\pi R^2)$  is the number of particles.
The ratio $A(R)/ R^2$ decreases at the scale where the disks start overlapping.
Therefore, the profile of $A(R)/ R^2$ along $R$ encodes all the distances between particles.

A more general description of these morphological measures can be obtained from the integral geometry.
According to Hadwiger's theorem \cite{hadwigeb1956integralsatze}, any geometric measure that has a notion of the size of a polyconvex set in Euclidean space $\mathbb{R}^d$ can be described by a set of functions called ``Minkowski functionals."
The polyconvex set is a finite union of closed and bounded convex bodies.
More precisely, the geometric measure $v$ should satisfy the following properties,
\begin{itemize}
\item
\textbf{Valuation:} $v$ has a notion of size of a set. 
The value of $v$ of the empty set $\phi$ is zero, i.e, $v(\phi) = 0$, and $v$ satisfies the following inclusion-exclusion property,
\begin{equation}
v(B_1 \cup B_2) = v(B_1) + v(B_2) - v(B_1 \cap B_2)
\end{equation}
where $B_1$ and $B_2$ are polyconvex sets.
\item
\textbf{Invariance:}
$v$ is invariant under any rotation and translation.
\item
\textbf{Continuity:} For any sequence of polyconvex sets $B_n$ that converges\footnote{The convergence is defined in terms of the Hausdorff metric.} to $B$, its valuation $v(B_n)$ also converges to $v(B)$.
\end{itemize}
Such geometric measures can be represented as a linear combination of $d+1$ Minkowski functionals $M_i$,
\begin{equation}
v(B) = \sum_{i=0}^d c_i M_i(B).
\end{equation}
In $d=2$, we have three Minkowski functionals: area, perimeter, and Euler characteristic. 
Since $\mathcal{P}(R)$ on $(\eta,\phi)$ plane is a finite union of closed and bounded convex bodies $B_i(R)$, its geometry can be described by the Minkowski functionals.
We already discussed the area $A(R)$ of $\mathcal{P}(R)$, and its perimeter $L(R)$ and Euler characteristic $\chi(R)$ are also useful quantities.
The discrete analog of $L(R)$ and $\chi(R)$ can be used for analyzing jet images.

The Minkowski functionals show that $N^{(1)}$ carries independent information to $N^{(0)}$.
We denote the perimeter and Euler characteristic of $\mathcal{P}^{(i)}$ as $L^{(i)}$ and $\chi^{(i)}$.
If all the active pixels of a jet image are isolated enough, we may represent $A^{(1)}$ as $A^{(0)}+  L^{(0)} \cdot (\Delta R) + 4 \chi^{(0)} \cdot (\Delta R)^2 $ since $A^{(1)}$ is a valuation of $\mathcal{P}^{(0)}$.\punctfootnote{
For the rectangle shape pixels, the term proportional to $4\chi^{(0)}$ corresponds to the number of pixels that touch only the corner of the pixels in $\mathcal{P}^{(0)}$.}
The relation between $N^{(0)}$ and $N^{(1)}$ is then as follows,
\begin{equation}
 N^{(1)} = N^{(0)}+ L^{(0)} / (\Delta R) + 4 \chi^{(0)}. 
\end{equation}
Note that this relation only holds when the squares attached to active pixels do not overlap each other.
Once some squares start to overlap, the relation begins to deviate, and the persistence of this relation can be used as a morphological indicator for the topological change of $\mathcal{P}^{(i)}$.
Therefore, $N^{(0)}$ and $N^{(1)}$ are effective variables for analyzing the geometry of soft particles of the jet image.

\Figref{fig:examples:4} is an example that the sequence of $\mathcal{P}^{(i)}$ shows a non-trivial topological change.
The sequence starts with six isolated pixels, $\mathcal{P}^{(1)}$ and $\mathcal{P}^{(2)}$ 
are a ring, and $\mathcal{P}^{(3)}$ is a single large cluster. The Euler characteristic $\chi^{(i)}$ and the perimeter $L^{(i)}$ of $\mathcal{P}^{(i)}$ are as follows.
\begin{eqnarray}
\chi^{(i)}
& = &
(6, \ 0, \ 0, \ 1, \cdots ) 
\\
\frac{L^{(i)}}{\Delta R}
& = &
(24,\ 52,\ 52,\ 54, \cdots). 
\end{eqnarray}
The non-monotonic behavior of the sequence of Minkowski functionals for analyzing the topology of point distributions is often discussed in other literature \cite{Mecke:1994ax,Schmalzing:1995qn}. 
Utilizing this topological information for jet classification problems or global event topology analysis might be interesting, but the full analysis of the sequence of the Minkowski functionals is outside the scope of this paper.

Morphological analysis have been applied in physics to quantify the distribution of the objects.   
In \cite{Mecke:1994ax,Schmalzing:1995qn}, Minkowski functionals are used to identify the structure of the astrophysical objects.
In more recent papers, persistence topology turns out to be useful tool for charcterizing seemingly random distribution of the points and applied in analysis of cosmic microwave background \cite{Cole:2017kve} and string landscape \cite{Cole:2018emh}.
It is tempting to consider other roles of morphological analysis with Minkowski functionals in jet classifications.

\section{Top Tagger based on Relation Network and Jet Morphology}
\label{sec:s2_top_tagger}

In this section, we describe our setup of classifiers trained on the inputs discussed in the previous sections, $S_{2,ab}$ and $N^{(i)}$. 
These inputs are derivable from jet images, so the CNN performs better than those RNs in principle.
We show that the deep learning on the small number of derived inputs reproduces the performance of the CNN. 
Therefore, those inputs are associated with the relevant physics for solving the classification problem. 
Moreover, the network using the derived inputs typically has less overfitting than that using the raw inputs.

\subsection{Training Data and Model Implementation}
\label{sec:network_setup}

We simulate top jet and QCD jet samples by \texttt{Madgraph5} \cite{Alwall:2014hca}, followed by \texttt{Pythia 8} (\py{}) \cite{Sjostrand:2014zea} or \texttt{Herwig 7} (\hw{}) \cite{Bellm:2015jjp,Bahr:2008pv}.
The detector response of generated events is simulated by \texttt{Delphes} \cite{deFavereau:2013fsa}.
Jets are reconstructed by the anti-$k_T$ algorithm with radius parameter $R_\jet=1.0$.
The jet constituents are calorimeter towers with angular resolution approximately $\Delta R = 0.1$.
The details of the sample preparation are explained in \appendixref{app:mc}.

We categorize the inputs to the RNs into the four sets: $\bm{x}_{\trim}$, $\bm{x}_{\jet_1}$, $\bm{x}_{\mathrm{kin}}$, and $\bm{x}_{\mathrm{geometry}}$.
\begin{itemize}
\item $\bm{x}_{\trim}$ is a set of discretized $S_{2,\trim}$ and $S_{2,\soft}$ up to angular scale $R=1.5$,
\begin{eqnarray}
\bm{x}_{\trim}
& = &
( S_{2,\trim}^{(i)} | i = 0, \cdots, 14 ) \oplus (S_{2,\soft}^{(i)} | i = 0, \cdots, 14 ),
\end{eqnarray}
where $S_{2,ab}^{(i)}$ is the binned spectrum of $S_{2,ab}$, with bin size $\Delta R$ in order to keep the same angular resolution to the jet image,
\begin{equation}
S_{2,ab}^{(i)} = \frac{1}{\Delta R} \int_{i \Delta R}^{(i+1) \Delta R} dR\, S_{2,ab}(R)
= \frac{1}{\Delta R} \sum_{\substack{j \in \jet_a, k \in \jet_b \\ R_{jk} \in [i \Delta R, (i+1) \Delta R)}} p_{T,j} p_{T,k}. 
\end{equation}
We may consider the angular scale up to the diameter $2R_{\jet} = 2.0$, but $S_{2,\trim}^{(i)}$ and $S_{2,\soft}^{(i)}$ at such large $R$ are less useful \cite{Chakraborty:2019imr}.

\item 
 $\bm{x}_{\jet_1}$  is a set of discretized $S_{2,11}$, $S_{2,1c}$ and $S_{2,cc}$ as follows,
\begin{eqnarray}
\bm{x}_{\jet_1}
&=&
(S_{2,11}^{(i)} | i = 0, \cdots, 3 ) \oplus ( S_{2,1c}^{(i)} | i = 0, \cdots, 9 ) \oplus ( S_{2,cc}^i | i = 0, \cdots, 14 ).
\end{eqnarray}
Again, we consider spectra only up to the relevant angular scales.
For $S_{2,11}^{(i)}$ and $S_{2,cc}^{(i)}$, the scale is the diameter of the corresponding subjet but too large angular scale is ignored.
For $S_{2,1c}^{(i)}$,  the scale is the jet radius because it is the correlation between the core part $\jet_1$ and its surroundings.

\item 
$\bm{x}_{\mathrm{kin}}$ is a set of global inputs,
\begin{equation}
\bm{x}_{\mathrm{kin}} = (
p_{T,\jet},
m_{\jet},
p_{T,\jet_{\trim}},
m_{\jet_{\trim}},
p_{T,\jet \setminus \jet_1},
m_{\jet \setminus \jet_1}
).
\end{equation}
In addition to the transverse momenta, we include the masses as the inputs because $2 m_{\jet_a}/ p_{T, \jet_a}$ is a characteristic angular scale of $\jet_a$.

\item $\bm{x}_{\mathrm{geometry}} $ is a set of the numbers of pixels of the jet images $\mathcal{P}^{(0)}$ and $\mathcal{P}^{(1)}$,
 \begin{eqnarray}
\bm{x}_{\mathrm{geometry}} 
& = & (
N^{(0)}, 
N^{(1)},
N^{(0)}(4\,\mathrm{GeV}),
N^{(1)}(4\,\mathrm{GeV})
).
\end{eqnarray}
\end{itemize}

We modularize the implementations of the model outputs $\bm{u}'=\phi^u(\bm{x})$ to avoid the curse of dimensionality.
When inputs are too many, there is a potential danger of overfitting due to sparsely distributed samples.
In our previous work, we use $\sim 40$ inputs for the classification of Higgs jets and QCD jets \cite{Lim:2018toa,Chakraborty:2019imr}. 
The inputs for the classification of top jets and QCD jets are increased to $\sim 70$, and training of a simple MLP classifier on these inputs may have difficulties.
Therefore, we compress $\bm{x}_{\trim}$ and $\bm{x}_{\jet_1}$ to a smaller number of hidden variables $\bm{h}_{\trim}$ and $\bm{h}_{\jet_1}$ by a neural network and get $\bm{u}'$ from them.
The following is the closed-form expression of $\RN_{S_2}$ that uses only the IRC safe inputs: $\bm{x}_{\trim}$, $\bm{x}_{\jet_1}$, and $\bm{x}_{\mathrm{kin}}$.
\begin{eqnarray}
\bm{h}_{\trim} 
& = & 
\MLP_{\trim}( \bm{x}_{\trim}, \bm{x}_{\mathrm{kin}} ; \bm{\theta}_{\trim}),
\\
\bm{h}_{\jet_1} 
& = & 
\MLP_{\jet_1}( \bm{x}_{\jet_1}, \bm{x}_{\mathrm{kin}} ; \bm{\theta}_{\jet_1}),
\\
\bm{u}' 
& = &
\MLP_{\mathrm{logit}}( \bm{h}_{\trim}, \bm{h}_{\jet_1}, \bm{x}_{\mathrm{kin}} ; \bm{\theta}_{\mathrm{logit}}),
\end{eqnarray}
where $\MLP_a$ is a multilayer perceptron (MLP) and $\bm{\theta}_a$ are its trainable parameters.
We provide $\bm{x}_{\kin}$ to each network to tell the characteristic angular scales directly.
We use the exponential linear unit (ELU) \cite{DBLP:journals/corr/ClevertUH15} as the activation function of each MLP.
The dimensions of $\bm{h}_{\trim}$ and $\bm{h}_{\jet_1}$ are 5. 
The output $\bm{u}'$ is a dimension two vector and will be transformed into the softmax outputs for the binary classification purpose.
\begin{equation}
\hat{y}^i = \frac{\mathrm{exp}(u'_i)}{\mathrm{exp}(u'_0) + \mathrm{exp}(u'_1)}, \quad i = 0,1
\end{equation}

When the geometric information $\bm{x}_{\mathrm{geometry}}$ is included in the inputs, we use them as arguments of $\MLP_{\mathrm{logit}}$,
\begin{eqnarray}
\bm{u}' 
& = &
\MLP_{\mathrm{logit}}( \bm{h}_{\trim}, \bm{h}_{\jet_1}, \bm{x}_{\mathrm{kin}}, \bm{x}_{\mathrm{geometry}}  ; \bm{\theta}_{\mathrm{logit}})
\end{eqnarray}
We consider three additional relation networks that uses the geometric information: $\RN_{S_2,N^{(0)}}$, $\RN_{S_2,N^{(0)},N^{(0)}(4\,\mathrm{GeV})}$, and $\RN_{S_2,N^{(0)},N^{(1)}}$.
Their inputs are listed in \tableref{table:RN_inputs}.
The detailed implementations of these RNs are in \appendixref{app:implementation:rn}.

\begin{table}
\begin{center}
\begin{tabular}{lccccccc}
\toprule
\multirow{2}{*}{model} & 
\multirow{2}{*}{$\bm{x}_{\mathrm{kin}}$} &
\multirow{2}{*}{$\bm{x}_{\trim}$} & 
\multirow{2}{*}{$\bm{x}_{\jet_1}$} & 
\multicolumn{4}{c}{$\bm{x}_{\mathrm{geometry}}$} 
\\
\cmidrule(lr){5-8}
& & & & $N^{(0)}$ & $N^{(0)}(4\,\mathrm{GeV})$ & $N^{(1)}$ & $N^{(1)}(4\,\mathrm{GeV})$ \\
\midrule
$\RN_{S_2}$ & $\bigcirc$ & $\bigcirc$ & $\bigcirc$ & & & &
\\ 
$\RN_{S_2,N^{(0)}}$ & $\bigcirc$ & $\bigcirc$ & $\bigcirc$ & $\bigcirc$ & & & 
\\
$\RN_{S_2,N^{(0)},N^{(0)}(4\,\mathrm{GeV})}$ & $\bigcirc$ & $\bigcirc$ & $\bigcirc$ & $\bigcirc$ & $\bigcirc$ & & 
\\
$\RN_{S_2,N^{(0)},N^{(1)}}$ & $\bigcirc$ & $\bigcirc$ & $\bigcirc$ & $\bigcirc$ & $\bigcirc$ & $\bigcirc$ & $\bigcirc$
\\
\bottomrule
\end{tabular}
\end{center}
\caption{
\label{table:RN_inputs}
The list of inputs used in each RN.
The circle represents that the given input is used.
The CNN trained on jet images can utilize all this  information.
}
\end{table}

The softmax output is trained by minimizing the cross-entropy loss function.
In addition, we marginalize the $p_{T,\jet}$ distribution in the classification because the top jet samples and QCD jet samples have different $p_{T,\jet}$ distributions.
To do this, we train networks in a way that interpolates binary classifiers for the jets at given $p_{T,\jet}$.
The corresponding cross-entropy loss function $\mathcal{L}_{\mathrm{CE}}$ is as follows.
\begin{eqnarray}
\mathrm{CE}(p_{T,\jet};\bm{\theta})
& = &
- \frac{1}{2} \sum_{Y=\mathrm{top, QCD}}\int d \tilde{\bm{x}} f_{\tilde{\bm{x}}|p_{T,\jet}}(\tilde{\bm{x}};Y) \,  \sum_{i=0,1} y_Y^i \log \hat{y}^i (\bm{x};\bm{\theta})
\\
\mathcal{L}_{\mathrm{CE}}(\bm{\theta})
& = &
\frac{1}{p_{T,\jet}^{\max} - p_{T,\jet}^{\min}} \int_{p_{T,\jet}^{\min}}^{p_{T,\jet}^{\max}} d p_{T,\jet} \,
\mathrm{CE}(p_{T,\jet};\bm{\theta})
\end{eqnarray}
where $Y$ is a category label, $y_{\mathrm{top}} = (1,0)$, and $y_{\mathrm{QCD}} = (0,1)$.
The function $f_{\tilde{\bm{x}}|p_{T,\jet}}(\bm{\tilde{x}}; Y)$ is the conditional probability density of $\tilde{\bm{x}}$ given $p_{T,\jet}$, and $\tilde{\bm{x}}$ is $\bm{x}$ without $p_{T,\jet}$.

The integral can be approximated by a Monte-Carlo integration,
\begin{eqnarray}
\label{eqn:loss_cross_entropy}
\mathcal{L}_{\mathrm{CE}}(\bm{\theta})
& \approx &
- \frac{1}{2} \sum_{Y=\mathrm{top,QCD}} \sum_{ i_Y = 1}^{N_Y} 
\frac{1}{f_{p_{T,\jet}} ( p_{T,\jet}^{[i_Y]};Y )}  \,  \sum_{i=0,1} y_Y^i \log \hat{y}^i (\bm{x}^{[i_Y]};\bm{\theta})
\end{eqnarray}
where $f_{p_{T,\jet}}(p_{T,\jet}; Y)$ is the probability density function of $p_{T,\jet}$ given $Y$, and the variables with superscript $[i_Y]$ is the value at the $i_Y$-th sample in the training dataset of $Y$.
The probability density function $f_{p_{T,\jet}}(p_{T,\jet}; Y)$ is modeled by kernel density estimation (KDE) described in \appendixref{app:kde}.
The resulting loss function is essentially a cross-entropy with samples whose $p_{T,\jet}$ distribution is reweighted to be uniform.
In addition to this cross-entropy loss, $L_2$ regularizer $\mathcal{L}_\mathrm{reg}$ \cite{NIPS1988_156,NIPS1990_323,NIPS1991_563} with the weight decay constant $\lambda = 0.001$ is added to regularize $\MLP_a$.
\begin{equation}
\mathcal{L}_\mathrm{reg} = \frac{\lambda}{2} \sum_a |\bm{W}_a|^2 
\end{equation}
where $\bm{W}_a$ are the weights of hidden layers in $\MLP_a$.

The training setup is as follows.
We minimize the loss function $\mathcal{L}(\bm{\theta})=\mathcal{L}_{\mathrm{CE}}(\bm{\theta})+\mathcal{L}_{\mathrm{reg}}(\bm{\theta})$ by ADAM optimizer \cite{ADAM} with learning rate 0.001, the first moment exponential moving average coefficient $\beta_1 =0.9$, the second moment exponential moving average coefficient $\beta_2=0.999$, and stabilization constant $\epsilon=10^{-7}$.
Batched samples are used in order to reduce overfitting.
The weights of the MLP are initialized by the He initializer \cite{He_2015_ICCV}, and the biases are initialized to be zero.
We will use early stopping for the termination criterion, but there is a chance that the network is mildly overfitted to the validation dataset during learning the features of the rare events.
If the gradients from the rare events distort the trained results for the dominant events, the network parameters have to be corrected again, and the training becomes noisy.
The random overfitting to the validation sample occurs during this noisy learning on the rare events.
To avoid this artifact, we use the exponential moving averages $\hat{\bm{\theta}}^{(t)}$ of the trainable parameters $\bm{\theta}^{(t)}$ at the epoch $t$ for the validation and testing.
The details of the moving average can be found in \appendixref{app:moving_avg}.
We monitor the loss $\mathcal{L}_\mathrm{tot}(\hat{\bm{\theta}})$ of the validation samples during the training and terminate the training if the loss function does not improve during 50 latter epochs. 
The networks and training setup is implemented in Keras \cite{chollet2015keras} with TensorFlow \cite{tensorflow2015-whitepaper} backend.
Optimization on the batch number is performed by the grid search.
We iterate the training for batch numbers, 20, 50, and 100 and two different random number seeds.

The results of the RN-based classifier will be compared to that of a CNN-based classifier.
The CNN model is similar to that of the previous paper \cite{Chakraborty:2019imr} but with more nodes and layers.
The closed-form expression of the CNN is as follows,
\begin{eqnarray}
\bm{h}_{\mathrm{image}}
& = &
\mathrm{CNN}_{\mathrm{image}}( \bm{x}_{\mathrm{image}}; \bm{\theta}_{\mathrm{CNN}})
\\
\bm{u}' & = &
\mathrm{MLP}_{\mathrm{logit}}( \bm{h}_{\mathrm{image}}, \bm{x}_{\mathrm{kin}}; \bm{\theta}_{\mathrm{MLP}} ),
\end{eqnarray}
where $\bm{x}_{\mathrm{image}}$ is energy deposits of the preprocessed jet image described in \appendixref{app:implementation:cnn}.
The module $\mathrm{CNN}_{\mathrm{image}}$ consists of 6 two-dimensional convolutional layers with $3\times3$ filters and ELU activations.
We insert two $2\times2$ max-pooling layers after the third and sixth convolutional layers.
The $\bm{h}_{\mathrm{image}}$ are the flattened outputs of the $\mathrm{CNN}_{\mathrm{image}}$.
The model outputs $\bm{u}'$ are from an MLP analyzing $\bm{h}_{\mathrm{image}}$ together with the kinematic information $\bm{x}_{\mathrm{kin}}$.
The detailed implementation of this CNN is in \appendixref{app:implementation:cnn}.
The training setup is the same as that of the RNs, but we check batch numbers 100, 200, and 500 instead because of the limitation of GPU memory.

\subsection{Classification results}

\begin{figure}
\begin{center}
\includegraphics[width=0.6\textwidth]{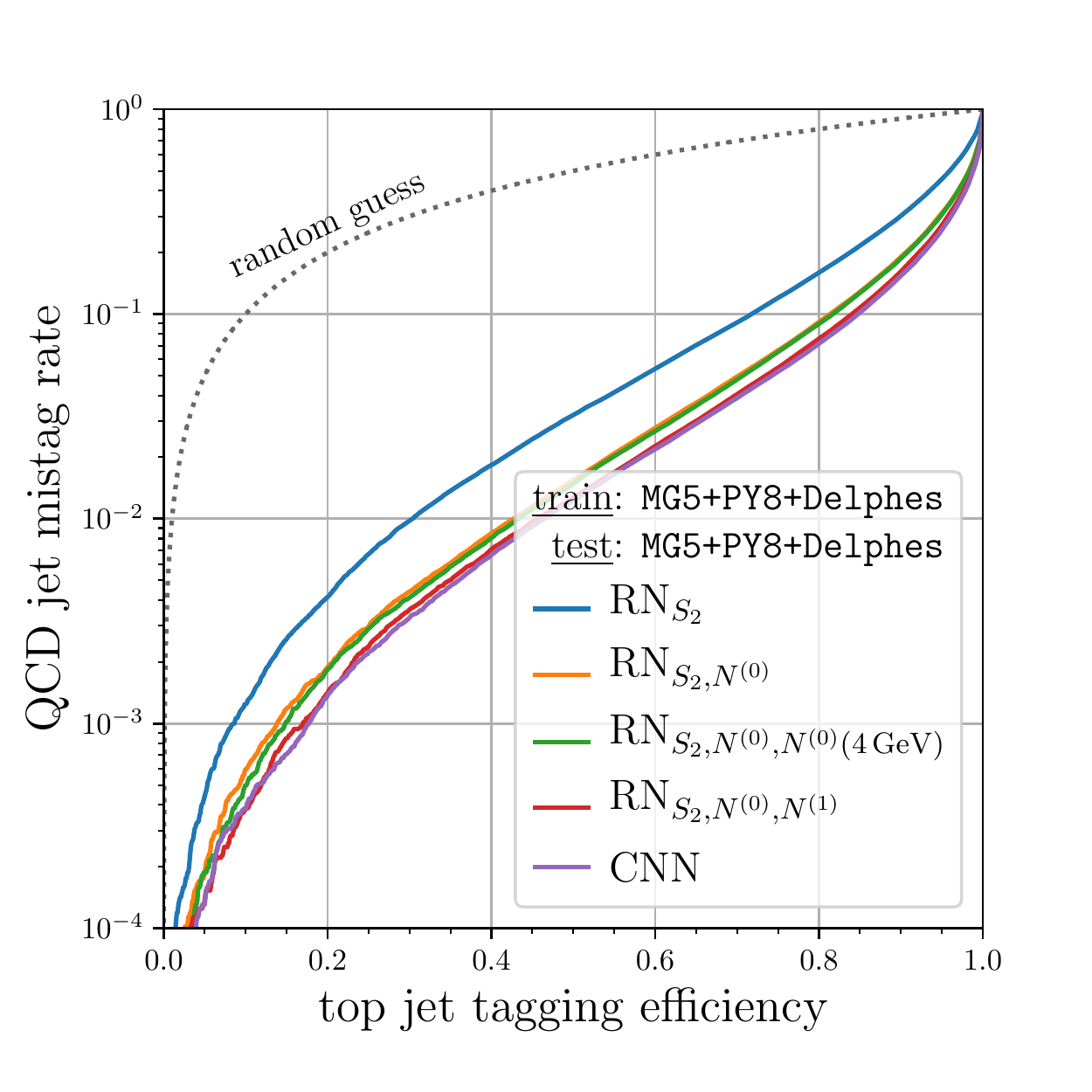}
\end{center}
\caption{
\label{fig:ROC:all}
The ROC curves of the networks trained on \py{} samples. 
}
\end{figure}

\Figref{fig:ROC:all} shows the ROC curves of the networks trained on \py{} samples.
The AUC, which is the upper area of each curve, of $\RN_{S_2}$, $\RN_{S_2,N^{(0)}}$, $\RN_{S_2,N^{(0)},N^{(1)}}$, and $\CNN$ are 0.8990, 0.9352, 0.9442,  and 0.9465, respectively.
There is a large gap between the ROC curves of $\RN_{S_2}$ and $\CNN$. 
This gap is partially filled by including an additional input $N^{(0)}$, as shown in the ROC curve of $\RN_{S_2,N^{(0)}}$.
Surprisingly, when we consider all the geometric inputs $\bm{x}_{\mathrm{geometry}}$, the ROC curve of $\RN_{S_2,N^{(0)},N^{(1)}}$ is almost equal to that of the CNN.
Therefore, the inputs $\bm{x}_{\trim}$, $\bm{x}_{\jet_1}$, $\bm{x}_{\kin}$, and $\bm{x}_{\mathrm{geometry}}$ can be considered as useful middle-level variables for modeling the top jet classifier.

\begin{figure}
\begin{center}
\includegraphics[width=0.49\textwidth]{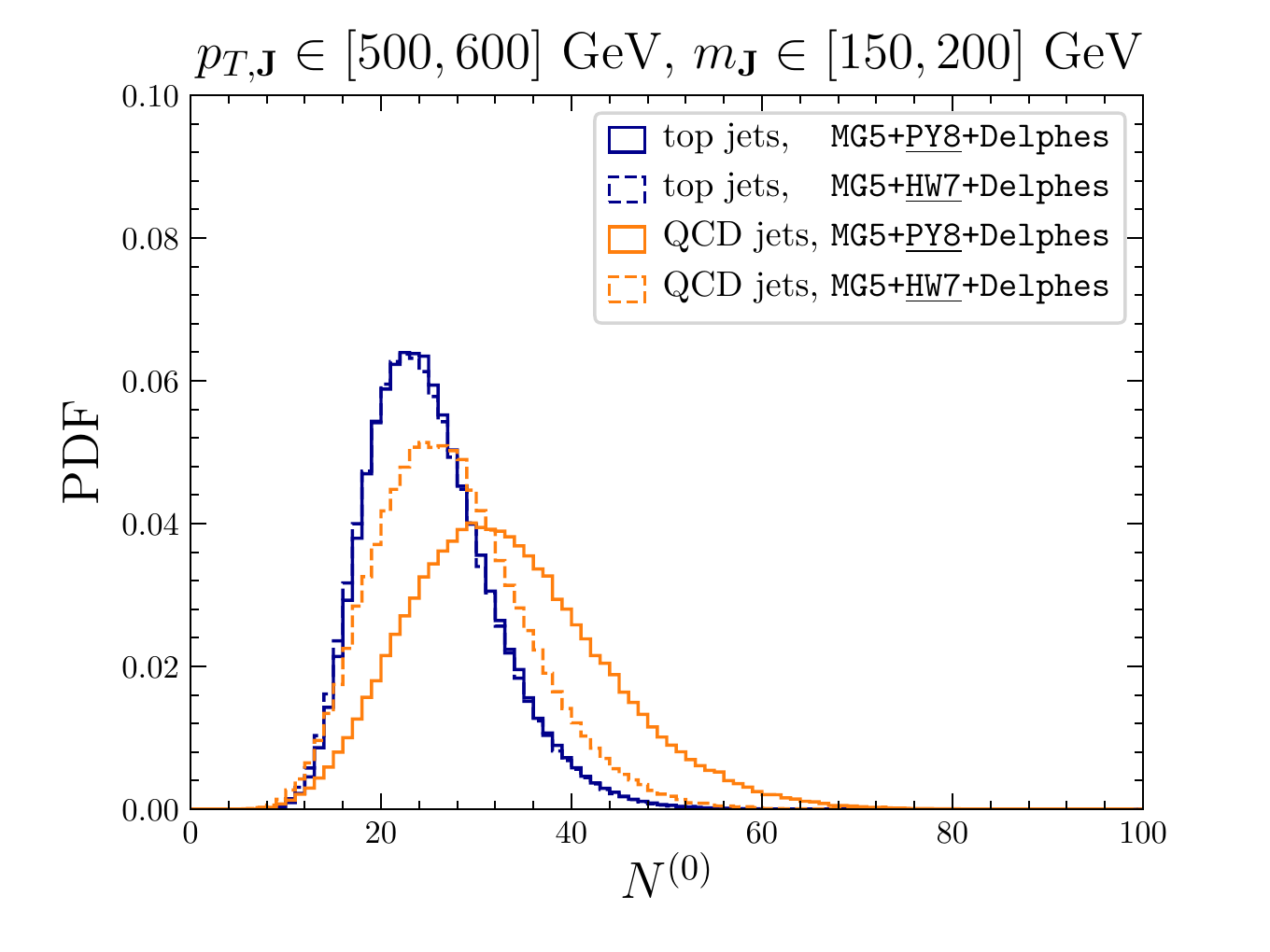}
\includegraphics[width=0.49\textwidth]{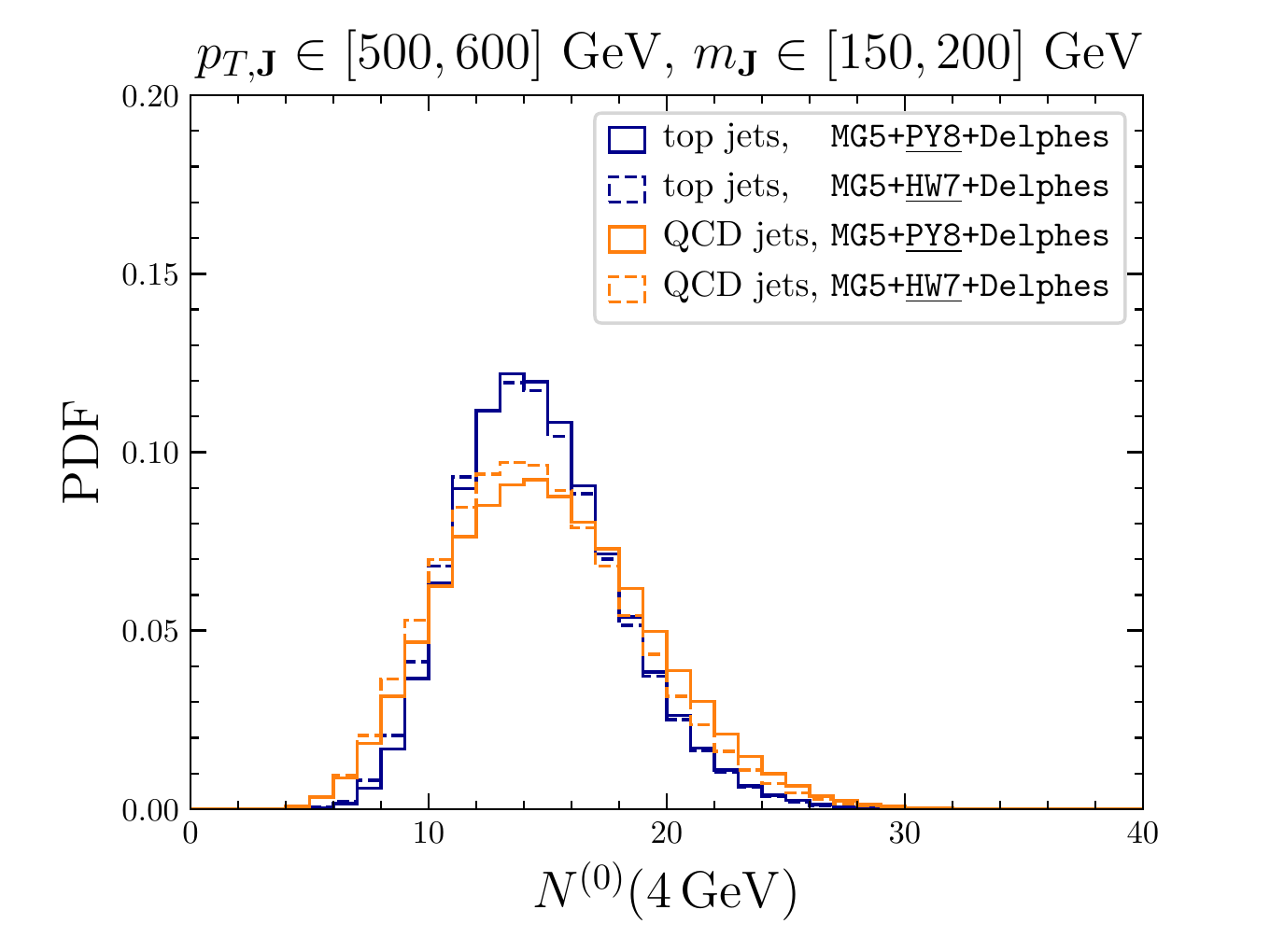}
\end{center}
\caption{
\label{fig:n_pixel}
$N^{(0)}$ and $N^{(0)}(4\,\mathrm{GeV})$ distributions of \py{} and \hw{} data sets.
The blue histograms are for the top jets, and the orange histograms are for QCD jets.
The solid lines are for \py{} generated samples, and the dashed lines are for \hw{} generated samples.
}
\end{figure}

The reason for a big gap between the ROC curves of $\RN_{S_2}$ and $\CNN$ is the difference in $N^{(0)}$ distributions between top jet samples and QCD jet samples.
The QCD jets in this paper are leading $\pt$ jets of $pp \rightarrow jj$ so that they are mostly gluon jets, which have a large $N^{(0)}$ than a jet from a color triplet parton.
In addition, \py{} predicts significantly higher $N^{(0)}$ of gluon jets than \hw{}, as in \figref{fig:n_pixel}.
Similar situations have been pointed out for the counting variables such as the charged track multiplicity \cite{Gallicchio:2012ez}, and the soft drop multiplicity \cite{Frye:2017yrw}.

The situation may be compared with the classification of Higgs jets and QCD jets, studied in \cite{Lim:2018toa}.
In this case, the difference between the ROC curves of $\RN_{S_2}$ and $\CNN$ is tiny. 
QCD jet samples in the study are leading $\pt$ jets of $pp \rightarrow Zj$ with invisibly decaying $Z$ boson, and most of the samples are the quark jets.  
The difference in the $N^{(0)}$ distribution of the Higgs jets and QCD jets is small, and therefore, $N^{(0)}$ does not play an important role there.

The remaining gap between the ROC curves of $\RN_{S_2,N^{(0)}}$ and $\CNN$ is almost filled by including $N^{(1)}$ in the analysis.
As discussed in the previous section, the ratio $N^{(1)}/N^{(0)}$ is a morphological measure that quantifies the level of clustering of the pixels.
Therefore, $N^{(1)}$ is useful for distinguishing compact top jets from QCD jets whose number of pixels is the same.
The similarity of two ROC curves indicates that the information summarized in the Minkowski functionals is used in the jet image analysis.

\begin{figure}
\begin{center}
\includegraphics[scale=0.6]{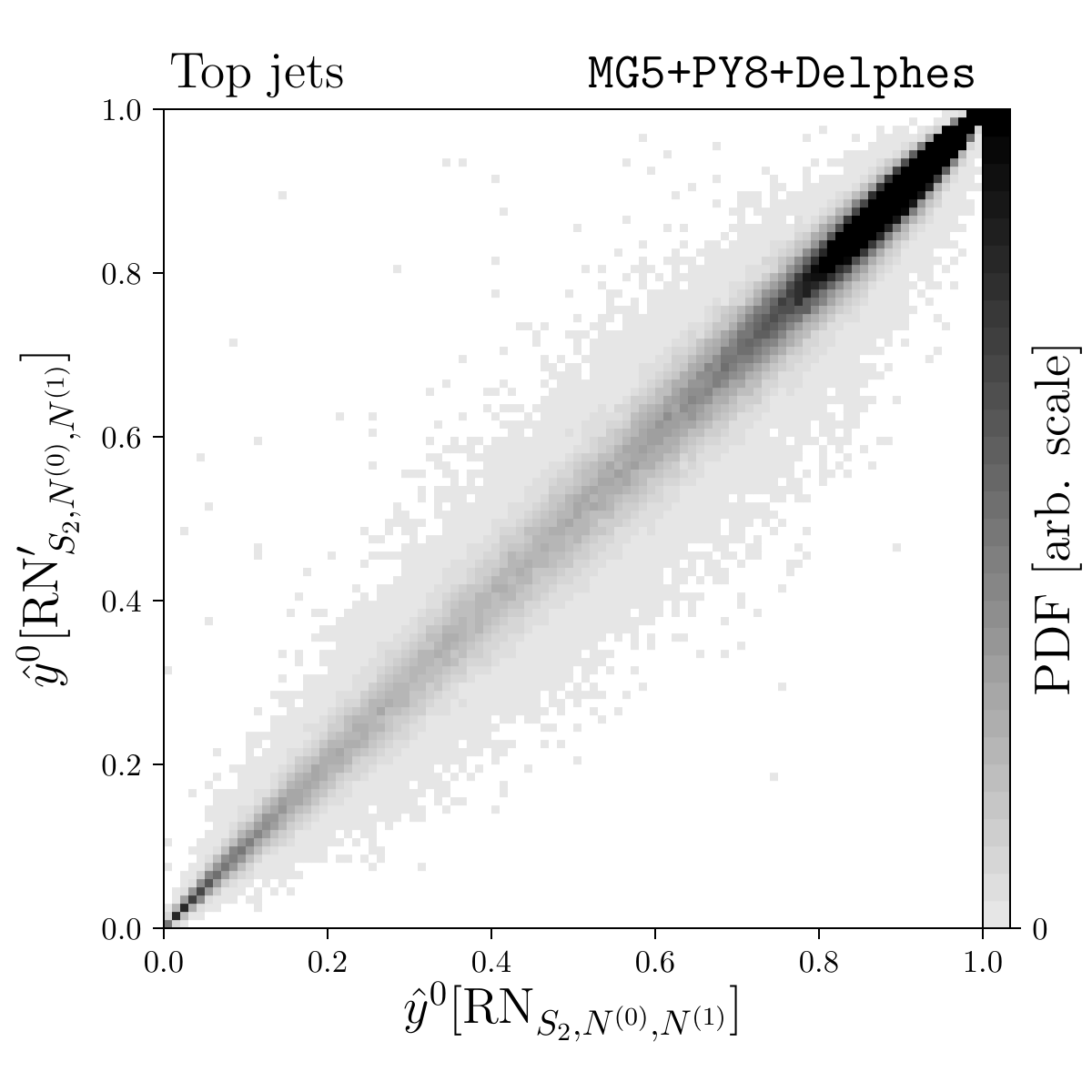} \includegraphics[scale=0.6]{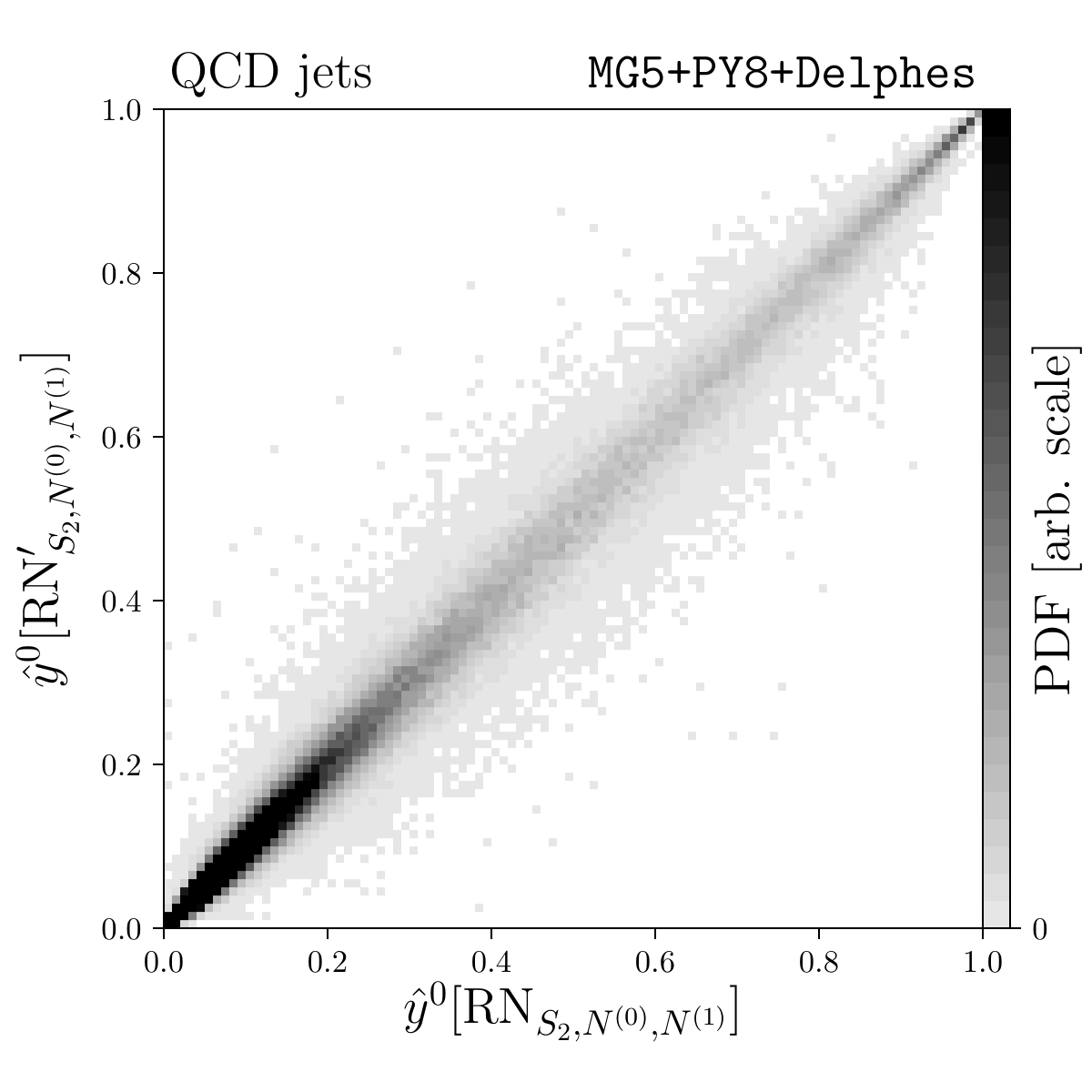}
\end{center}
\begin{center}
\includegraphics[scale=0.6]{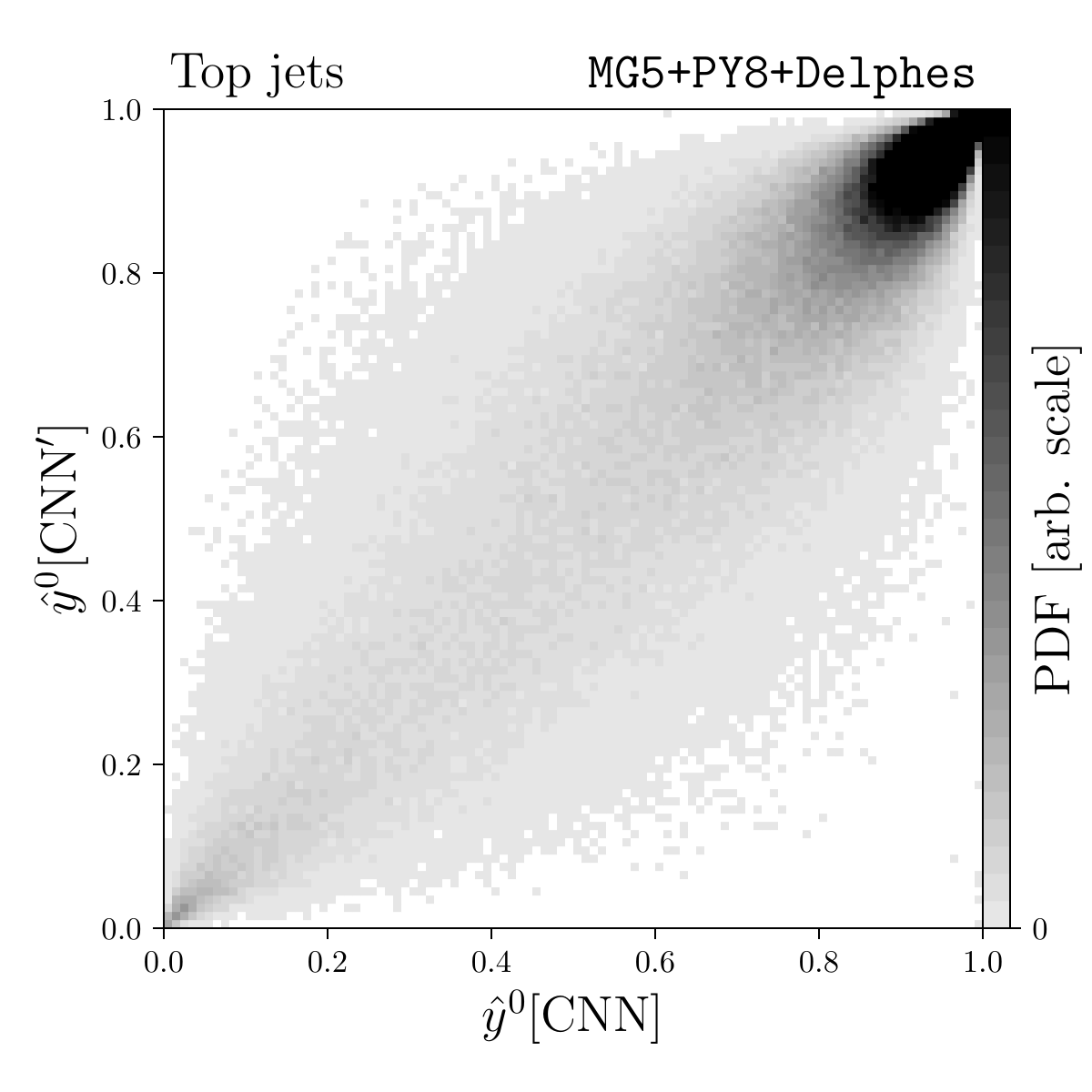} \includegraphics[scale=0.6]{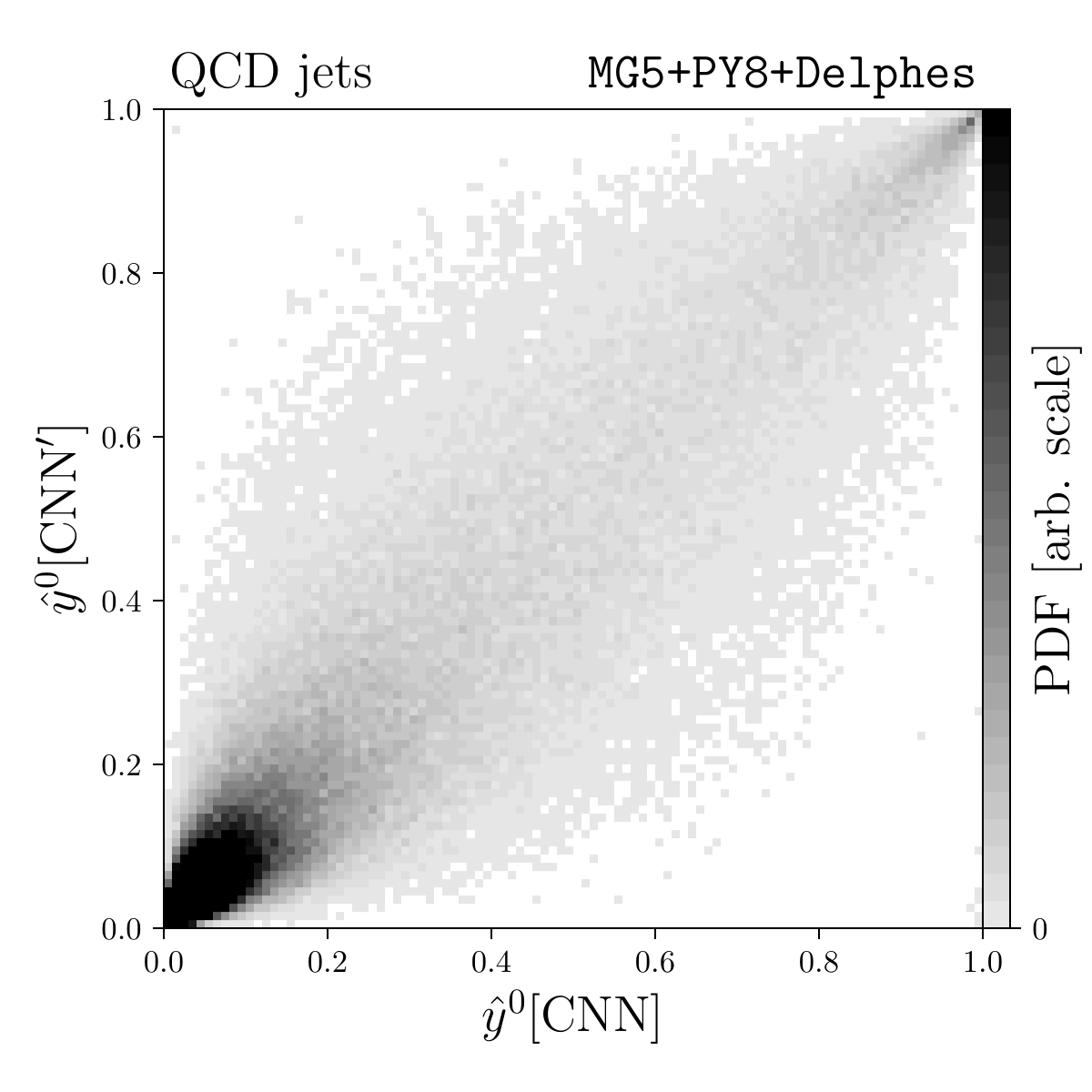}
\end{center}
\caption{
\label{fig:pred_scatter}
The distribution of softmax output of the classifiers for different random seed. upper two figures for 
$\hat{y}^0[\RN_{S_2,N^{(0)},N^{(1)}}]$, and the bottom two figures for  $\hat{y}^0[\RN_{S_2,N^{(0)},N^{(1)}}]$.  
The left figures are top jets and the right figures are QCD jets. 
}
\end{figure}

Not only $\RN_{S_2,N^{(0)},N^{(1)}}$ gives a comparable result to $\CNN$, but it is also significantly more stable. 
We compare the softmax output $\hat{y}^0$ of a network $\NN$ and the output of the same network trained with a different random seed, and we call the alternative output $\hat{y}'^{\,0}$.
The change of the seed affects the shuffling of the events between batches and alters the initialization of the network.
Since the training of the neural network is not a convex optimization in general, the network output difference $\Delta\hat{y}^0[\NN] = \hat{y}'^{\,0}[\NN] - \hat{y}^0[\NN] \neq 0$.
In \figref{fig:pred_scatter}, we show the histogram of two outputs $(\hat{y}^0, \hat{y}'^{\,0})$ for $\RN_{S_2,N^{(0)},N^{(1)}}$ and $\CNN$.
The distribution for $\RN_{S_2,N^{(0)},N^{(1)}}$ is narrower than that for $\CNN$.
This shows that training of $\RN_{S_2,N^{(0)},N^{(1)}}$ is more stable.

The better training stability of $\RN$ is due to the difference in the inputs of the functional model.
%$\CNN$ uses the jet image itself, while $\RN$ uses the derived inputs from the jet image. 
The pair of preprocessed jet image and $\bm{x}_{\kin}$ contain more information than the two-point energy correlations and Minkowski functionals.
Hence, $\CNN$ could approximate a wider variety of functions of jet constituents than $\RN$.
In other words, the training of CNN requires more effort in order to scan over larger space of functions.
The training of a simpler model is much stable than that of a complex model because of less number of inputs and trainable parameters.
A simpler model has a potential danger of underfitting, but it is less severe in $\RN$ because $S_{2,ab}$ and $N^{(i)}$ are reasonable set for describing functional space of energy correlation and geometry of the jet constituents, respectively.

\begin{figure}
\begin{center}
\includegraphics[scale=0.45]{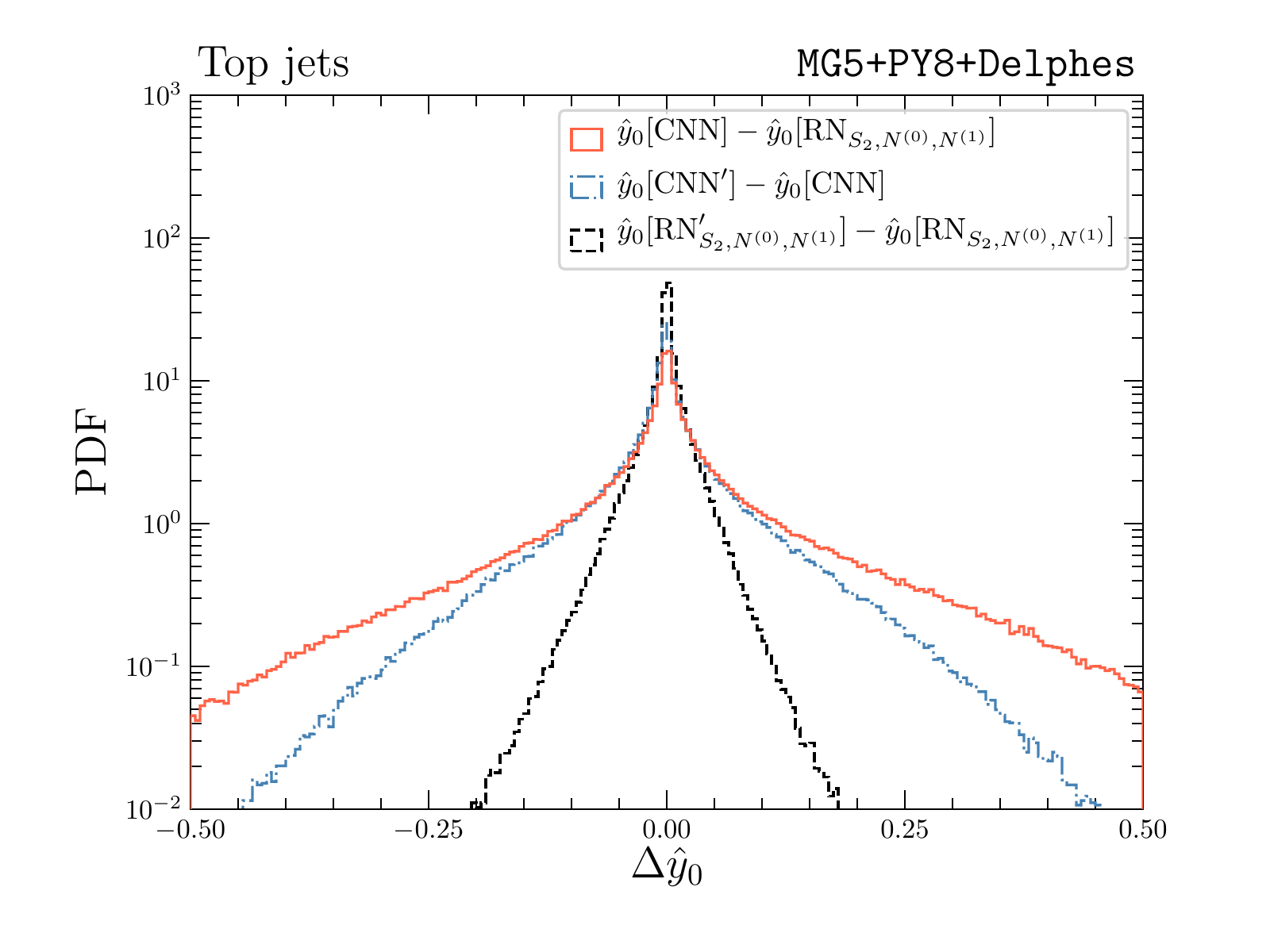}
\includegraphics[scale=0.45]{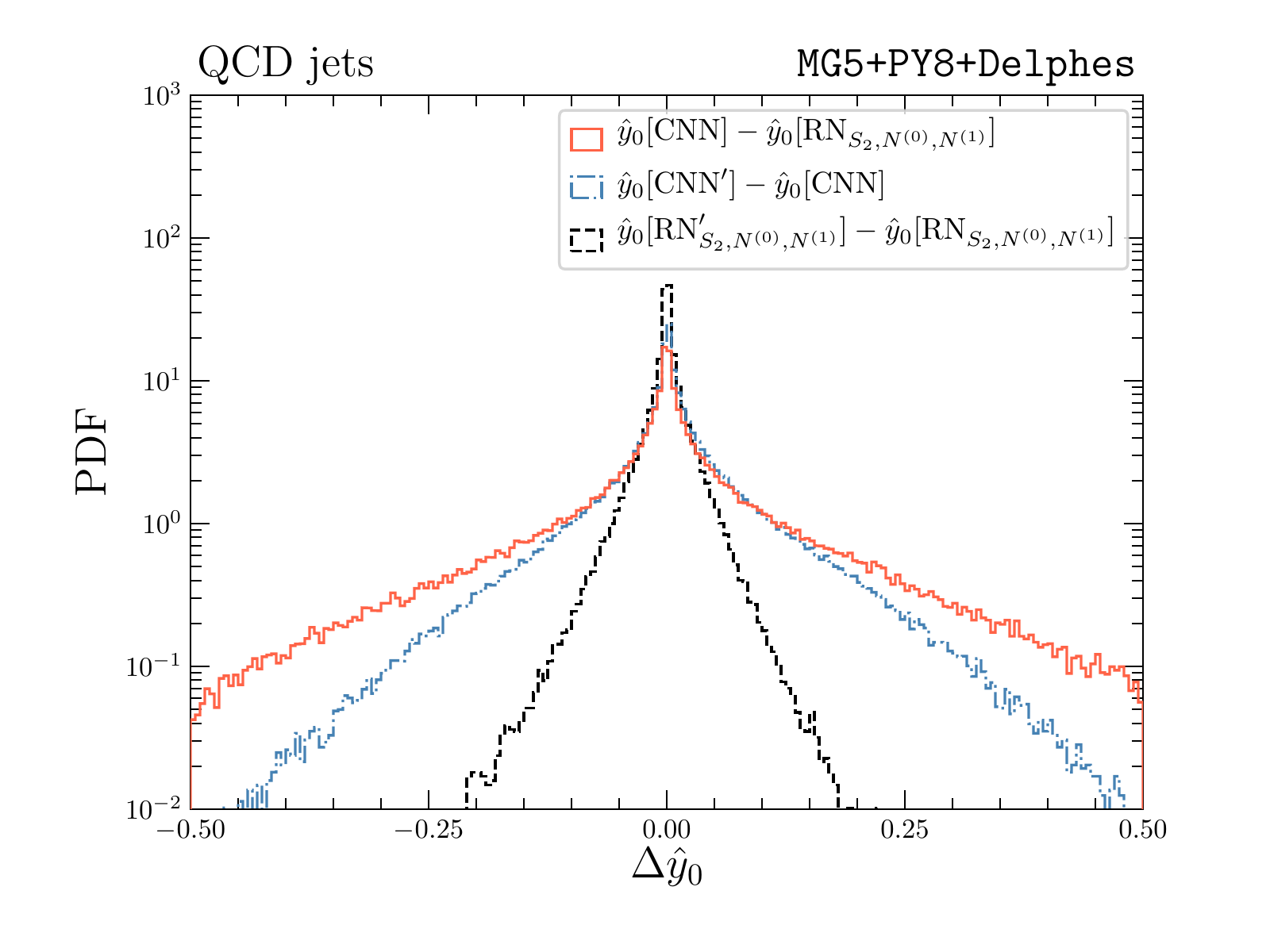}
\end{center}
\caption{
\label{fig:rn_cnn}
The difference of softmax output for various models, $\Delta \hat{y}^0[\RN]$, $\Delta \hat{y}^0[\CNN]$, and 
$\Delta \hat{y}^0[\CNN,\RN]$ for top jets and QCD jets. 
}
\end{figure}

We now compare the outputs $\hat{y}^0[\CNN]$ and $\hat{y}^0[\RN] = \hat{y}^0[\RN_{S_2,N^{(0)},N^{(1)}}]$.\footnote{ 
From here, we denote $\RN_{S_2,N^{(0)},N^{(1)}}$ as $\RN$.
}
\Figref{fig:rn_cnn} shows the distributions of 
$\Delta \hat{y}^0[\RN]$, 
$\Delta \hat{y}^0[\CNN]$, and
$\Delta \hat{y}^0[\CNN,\RN] =\hat{y}^0[\CNN] - \hat{y}^0[\RN]$.
The mean and standard deviation of these differences are summarized in \tableref{table1}. 
All the $\Delta \hat{y}^0$ distributions are sharply peaked approximately at $\Delta \hat{y}^0 = 0$, which indicates that the classifiers make the same decision for the majority of the events. 
Since the training of the $\CNN$ is less stable than that of $\RN$, the standard deviation $\sigma(\Delta \hat{y}^0[\CNN])$ of $\Delta \hat{y}^0[\CNN]$ is much larger than $\sigma(\Delta \hat{y}^0[\RN])$ of $\Delta \hat{y}^0[\RN]$.
The standard deviation of $\Delta \hat{y}^0[\CNN,\RN]$ is larger than the error $\sqrt{\sigma(\Delta \hat{y}^0(\CNN))^2 +\sigma(\Delta \hat{y}(\RN))^2}$, which is 0.091 for the top jet samples and 0.095 for the QCD jet samples.
This indicates that the outputs of $\RN$ and $\CNN$ are highly correlated, but there are still some differences. 
We repeat the same analysis on non-typical events, which satisfies $0.15<\hat{y}^0<0.85$ for one of $\RN$ or $\CNN$.
The results are similar, but the standard deviations are larger by a factor 1.5 because we removed samples easy to classify.

\begin{table}
\begin{center}
\begin{tabular}{lllll}
\toprule
\multirow{2}{*}{output difference}& \multicolumn{2}{c}{top jet samples} & \multicolumn{2}{c}{QCD jet samples} \\
\cmidrule(lr){2-3}
\cmidrule(lr){4-5}
 & average & deviation & average & deviation\cr
\midrule
\phantombox{$\Delta \hat{y}^0[\RN]$}{$\Delta \hat{y}^0[\CNN, \RN]$} $= \hat{y}'^0[\RN] - \hat{y}^0[\RN] $ & $-9.56\times 10^{-4}$ & 0.0271 & $-1.65\times 10^{-4}$ & 0.0279 \cr
\phantombox{$\Delta \hat{y}^0[\CNN]$}{$\Delta \hat{y}^0[\CNN, \RN]$} $= \hat{y}'^0[\CNN] - \hat{y}^0[\CNN] $ & $-1.46\times 10^{-3}$ & 0.0867 & $-6.14\times 10^{-3}$ & 0.0911\cr
$\Delta \hat{y}^0[\CNN, \RN]$ $= \hat{y}^0[\CNN] - \hat{y}^0[\RN] $  & $\phantom{-}6.98\times 10^{-3}$ & 0.141\phantom{0} & $\phantom{-}3.10\times 10^{-3}$ & 0.144\phantom{0}\cr
\phantombox{}{$\Delta \hat{y}^0[\CNN, \RN]=$} $\hat{y}'^0[\CNN] - \hat{y}^0[\RN] $ & $\phantom{-}5.51\times 10^{-3}$ & 0.137\phantom{0} & $\phantom{-}9.26\times 10^{-3}$ & 0.142\phantom{0}\cr
\midrule
\multicolumn{5}{l}{after selection: $0.15< \hat{y}^0[\RN_{S_2,N^{(0)},N^{(1)}}] <0.85$ or $0.15< \hat{y}^0[\CNN] <0.85$} \\
\midrule
\phantombox{$\Delta \hat{y}^0[\RN]$}{$\Delta \hat{y}^0[\CNN, \RN]$} $= \hat{y}'^0[\RN] - \hat{y}^0[\RN] $  & $-1.60\times 10^{-3}$ & 0.0403 & $-1.06\times 10^{-3}$ & 0.0409 \cr
\phantombox{$\Delta \hat{y}^0[\CNN]$}{$\Delta \hat{y}^0[\CNN, \RN]$} $= \hat{y}'^0[\CNN] - \hat{y}^0[\CNN] $ & $\phantom{-}3.64\times 10^{-3}$ & 0.129 & $\phantom{-}3.51\times 10^{-3}$ & 0.131\cr
$\Delta \hat{y}^0[\CNN, \RN] = \hat{y}^0[\CNN] - \hat{y}^0[\RN] $  & $\phantom{-}1.61\times 10^{-2}$ & 0.215 & $\phantom{-}3.50\times 10^{-3}$ & 0.217\cr
\phantombox{}{$\Delta \hat{y}^0[\CNN, \RN]=$} $\hat{y}'^0[\CNN] - \hat{y}^0[\RN] $  & $\phantom{-}1.97\times 10^{-2}$ & 0.206 & $\phantom{-}9.39\times 10^{-3}$ & 0.210\cr
\bottomrule
\end{tabular}
\end{center}
\caption{
\label{table1}
Average and standard deviation of the output difference $\Delta y^0$.
}
\end{table}

\begin{figure}
\begin{center}
\includegraphics[scale=0.5]{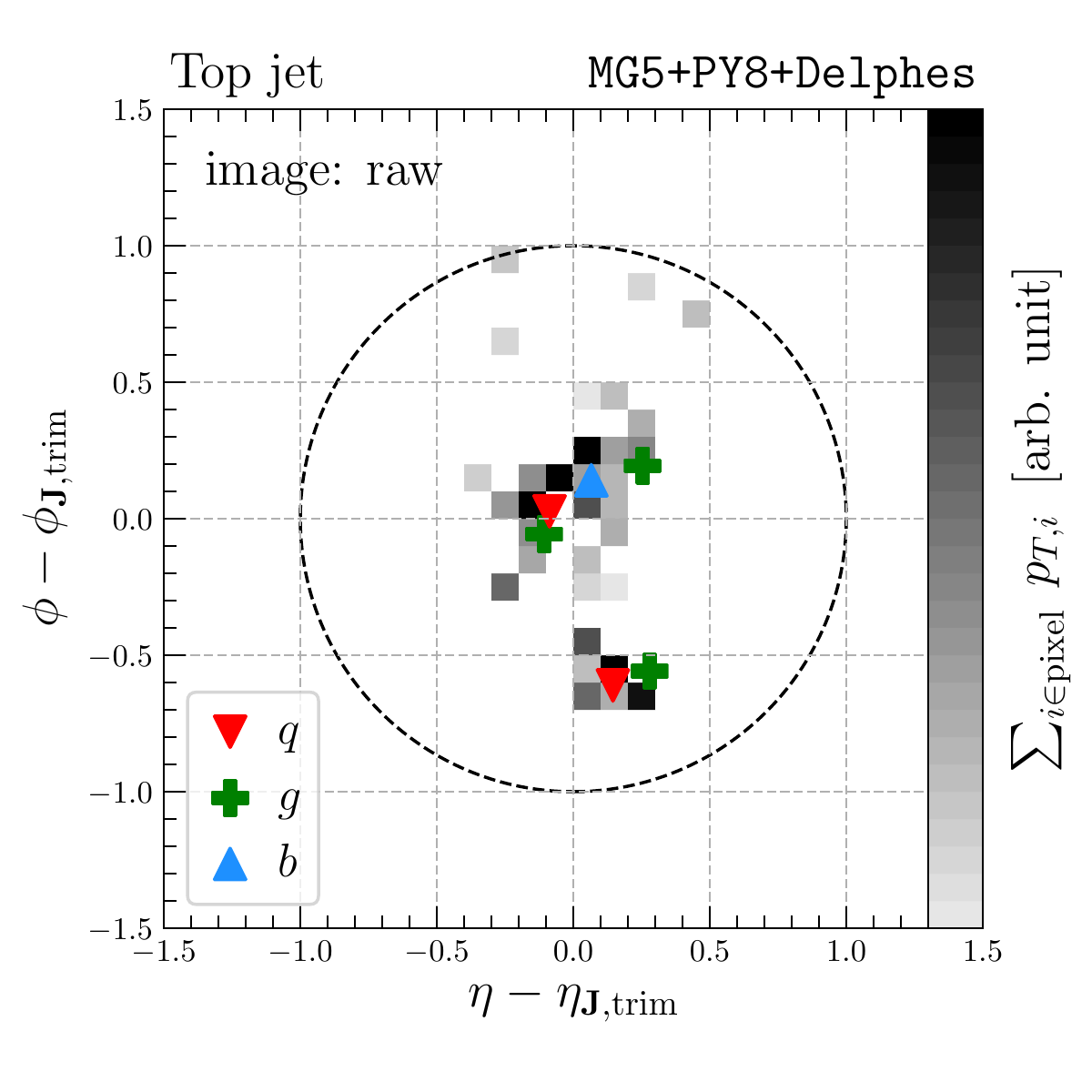}
\includegraphics[scale=0.5]{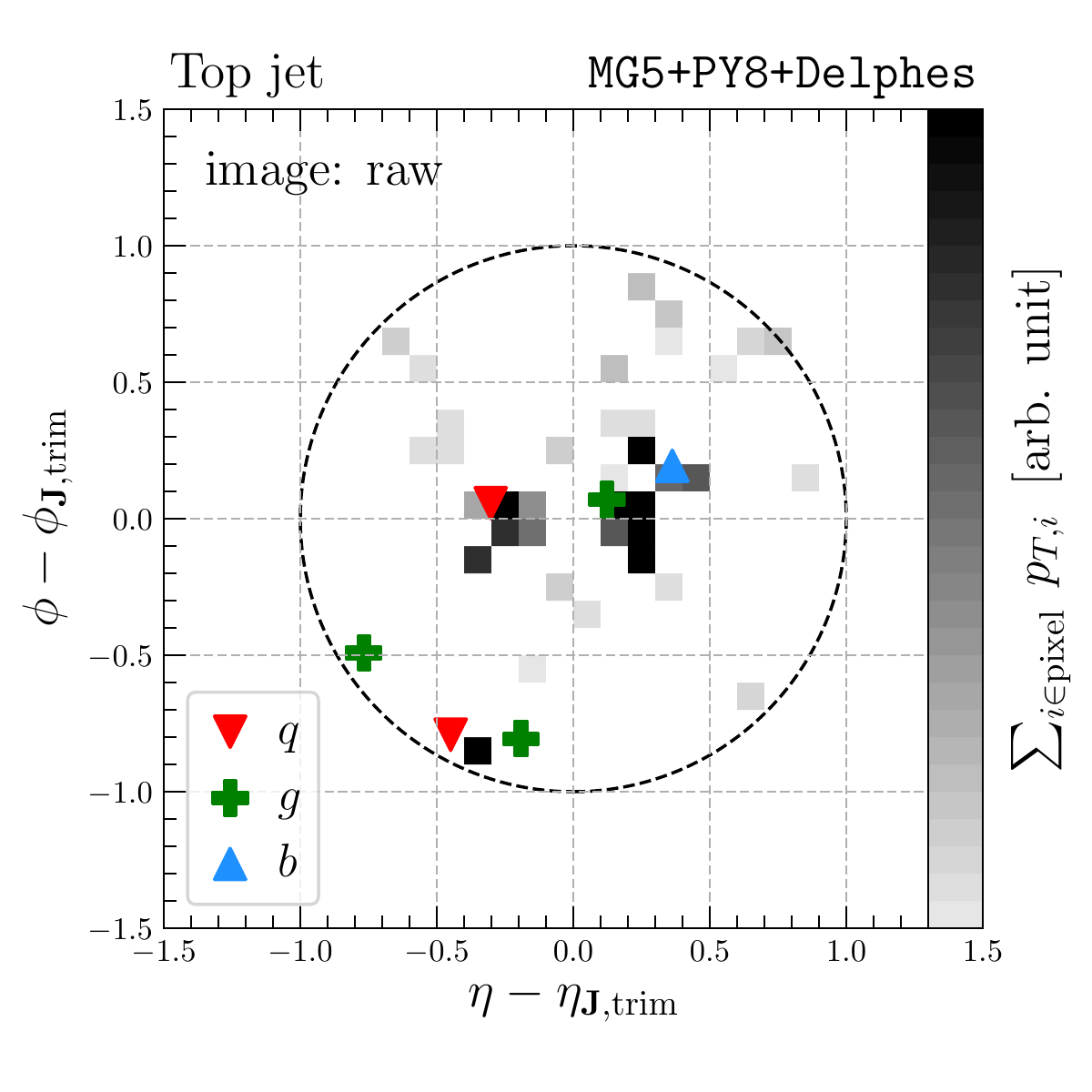}
\end{center}
\caption{
\label{fig:jet_image_confusing}
Jets images of top jets that $\RN_{S_2,N^{(0)},N^{(1)}}$ and $\CNN$ give different answers. 
The network outputs of each jet image are as follows:
(left) $\hat{y}^0[\RN_{S_2,N^{(0)},N^{(1)}}] = 0.0795$, $\hat{y}^0[\CNN] = 0.908$, $\hat{y}'^0[\CNN] = 0.883$,
(right) $\hat{y}^0[\RN_{S_2,N^{(0)},N^{(1)}}] = 0.836$, $\hat{y}^0[\CNN] = 0.0905$, $\hat{y}'^0[\CNN] = 0.101$,
}
\end{figure}

In order to understand the cases on which $\RN_{S_2,N^{(0)},N^{(1)}}$ and $\CNN$ gives us extremely different answers, we show two examples in \figref{fig:jet_image_confusing}.
To choose jets with stable $\CNN$ predictions, the selected jets have similar $\hat{y}^0[\CNN]$ and $\hat{y}'^0[\CNN]$.
For the left jet image, $\CNN$ judges the jet as a top jet while $\RN$ does not.\
The $b$ quark and one of the light quarks accidentally overlap in this event. 
This type of event is certainly not typical.
The probability that the angle $R_{bq}$ or $R_{b\bar{q}}$ is less than 0.2 is 5.6\% without considering spin-correlation.
For the right jet image, $\CNN$ judges that the jet is a QCD jet, but $\RN$ does not.
It is a two-prong jet with many soft radiations and a small $p_T$ subjet from a quark due to longitudinal decay of W boson.
In the longitudinal decay, one of the quark goes backward to the boost direction, but these jets suppressed in the phase space.
Because both of the top jets are not typical, it is not surprising the two different models give very different results for those events.

We have checked if more aggressive training on these rare events improves the performance, for example,  relaxing the regularizer setup.
The AUCs of $\RN$ and $\CNN$ with the weight decay constant $\lambda = 10^{-4}$ are 0.9461 and 0.9465, respectively.
There are tiny improvements in the classification performance, but it comes together with overfitting. 
The validation loss $\mathcal{L}(\hat{\bm\theta})$ and training loss $\mathcal{L}({\bm\theta})$ are 0.3044, and 0.2969 for $\RN$; 0.3201, and 0.2971 for $\CNN$, respectively.
The $\mathcal{L}(\hat{\bm\theta})$ and $\mathcal{L}({\bm\theta})$ in the original setup are 0.3076, and 0.3049 for $\RN_{S_2,N^{(0)},N^{(1)}}$; 0.3338, and 0.3371 for the $\CNN$.
The difference between the training and validation loss is much bigger in $\lambda = 10^{-4}$ setup, which is a sign of overfitting.

\subsection{Alternative vertex label choice}
\begin{figure}
\begin{center}
\begin{subfigure}{0.32\textwidth}
\includegraphics[scale=0.4]{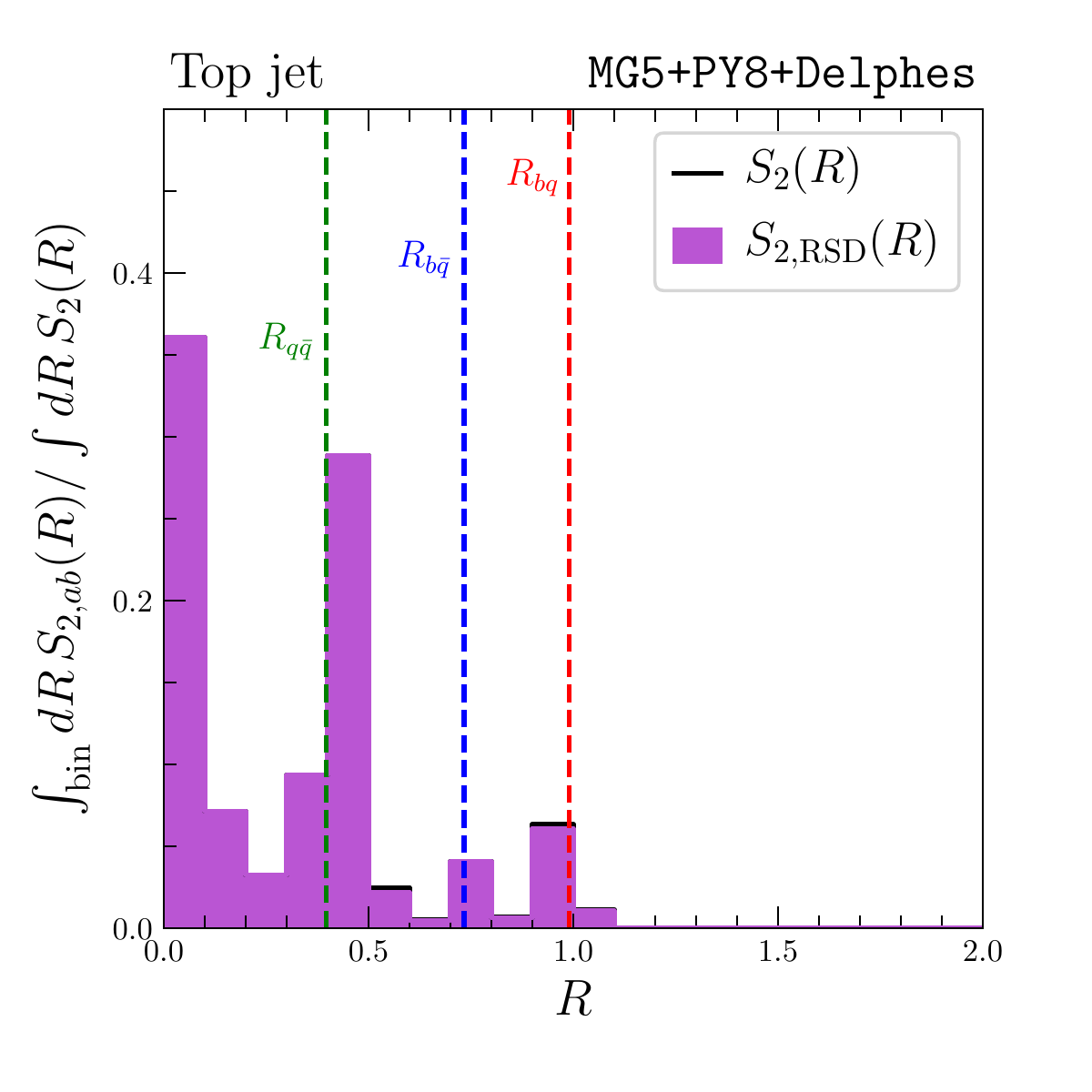}
\caption{\label{fig:jet_spectra_rsd_top1}}
\end{subfigure}
\begin{subfigure}{0.32\textwidth}
\includegraphics[scale=0.4]{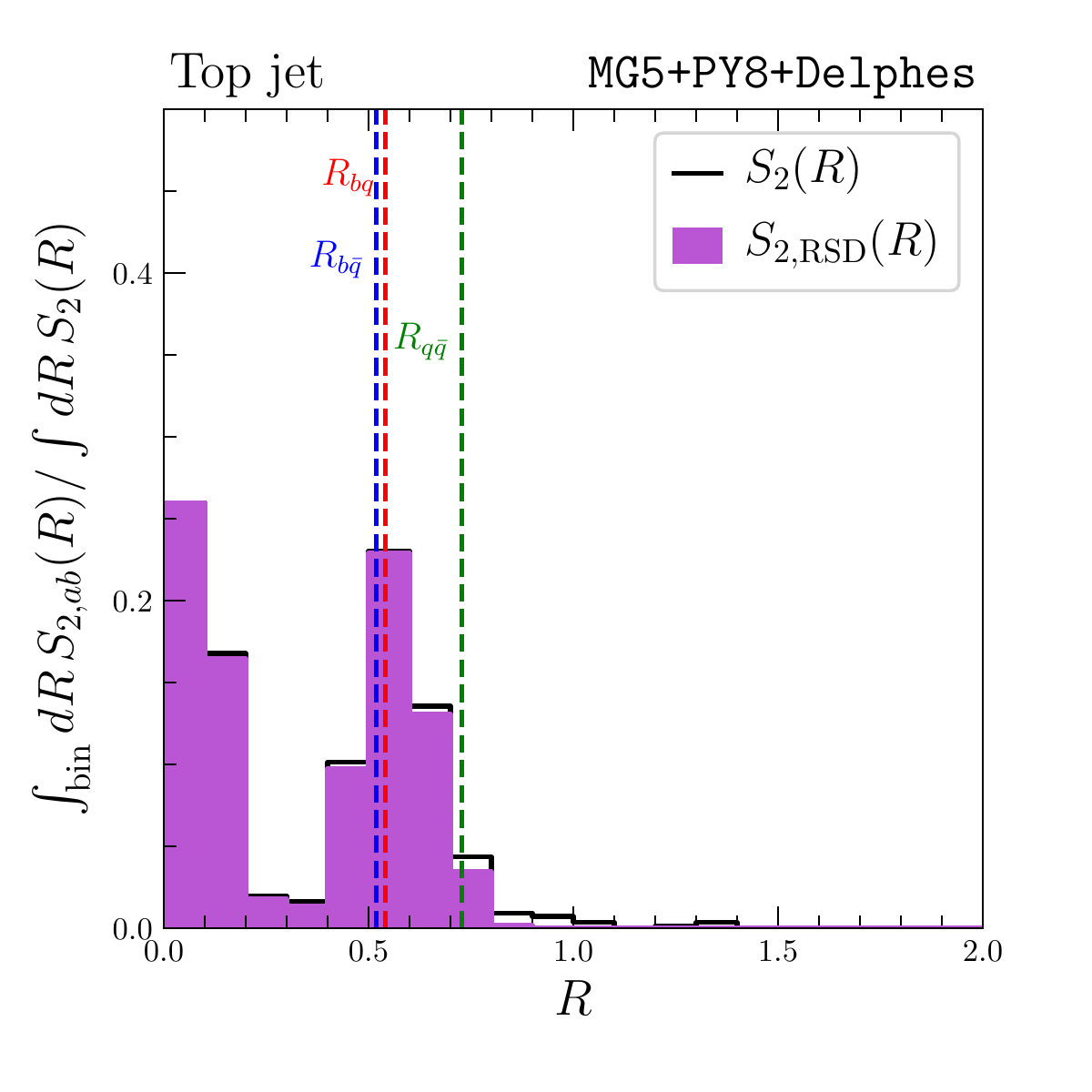}
\caption{\label{fig:jet_spectra_rsd_top2}}
\end{subfigure}
\begin{subfigure}{0.32\textwidth}
\includegraphics[scale=0.4]{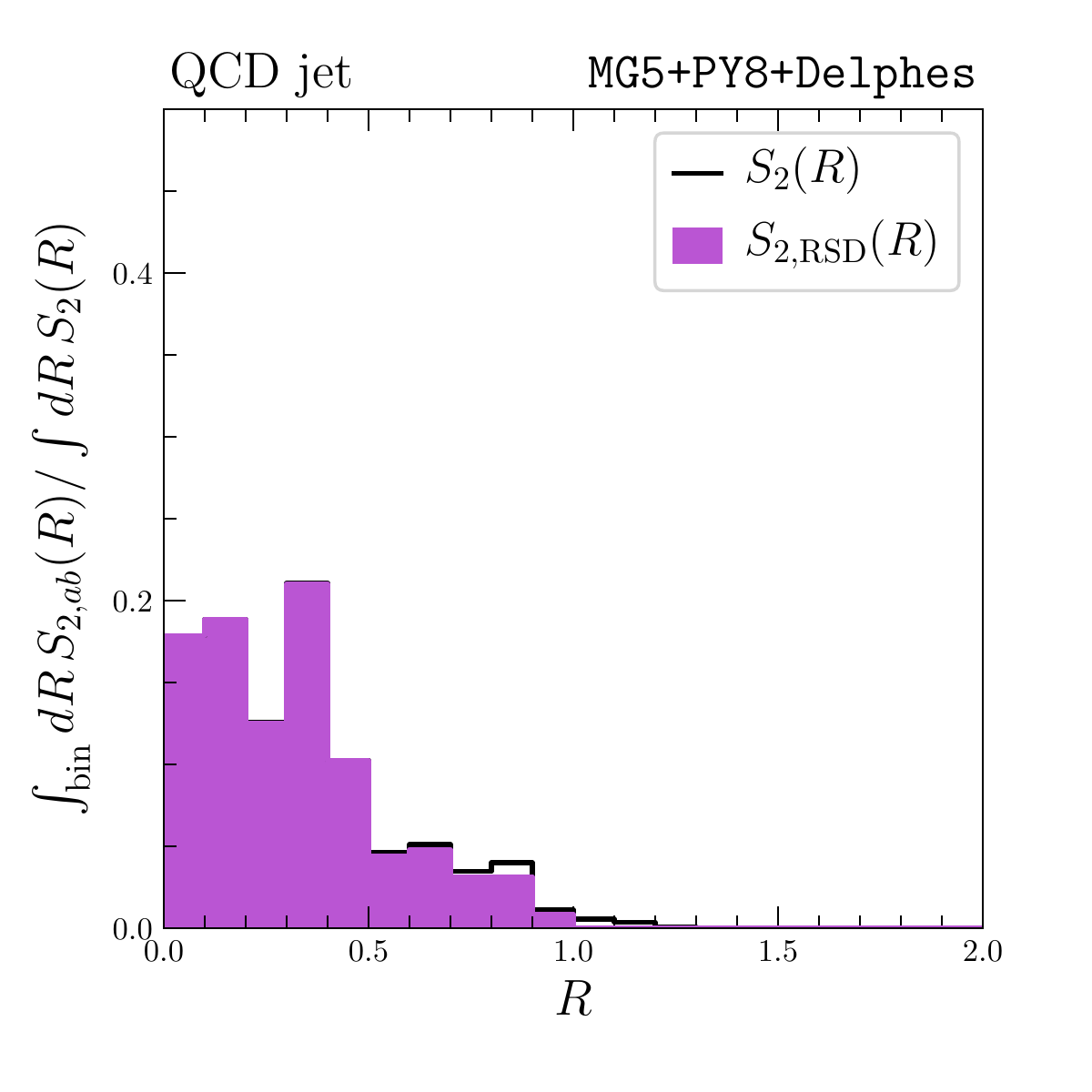}
\caption{\label{fig:jet_spectra_rsd_qcd1}}
\end{subfigure}
\end{center}
\caption{
\label{fig:jet_spectra_rsd}
The $S_2$ and $S_{2,\mathrm{RSD}}$ distributions of the top jets and the QCD jet in \figref{fig:jet_image_graph_on_jet_image_trimsoft}.
The dashed lines are the characteristic angular scales of the top jets in the parton level.
}
\end{figure}

Vertex label is a hyperparameter of the RNs, and we use labels based on the trimmed jet and leading $p_T$ subjet in order to explicitly identify hard substructures and subleading $p_T$ substructures.
Other labels may be used depending on the purpose.
For example, trimming may be replaced with recursive soft drop (RSD) \cite{Larkoski:2014wba,Dreyer:2018tjj} for better analytic tractability of the two-point energy correlations.
Let $\jet_r$ be the groomed jet by the RSD\footnote{We use soft drop parameters $z_{\mathrm{cut}} = 0.5$ and $\beta=1$, and fully inspect whole clustering history.}, and $r$ be its vertex label. 
We define $S_{2,\mathrm{RSD}}$ and $S_{2,\mathrm{RSD}^c}$, 
\begin{eqnarray}
S_{2,\mathrm{RSD}}(R) 
& = &
S_{2,rr}(R), 
\\
S_{2,\mathrm{RSD}^c}(R) 
& = &
S_{2}(R) - S_{2,rr}(R),
\end{eqnarray}
corresponding to $S_{2,\trim}$ and $S_{2,\soft}$, respectively. 
\Figref{fig:jet_spectra_rsd} shows the $S_{2,\mathrm{RSD}}$ distribution of the top jets and QCD jet in \figref{fig:jet_image_graph_on_jet_image_trimsoft}, but the difference is small.
The two-point energy correlation related to the soft activity that satisfies the soft drop condition may be included in $S_{2,\mathrm{RSD}}$.
%One difference between $S_{2,\trim}$ and $S_{2,\mathrm{RSD}}$ is that the soft activity in $S_{2,\soft}$ at $R \sim 0.3$ in \figref{fig:jet_spectra_trimsoft_top2} is moved to $S_{2,\mathrm{RSD}}$.
%RSD considers the soft drop condition for each branching in the clustering tree so that the two-point energy correlation related to the soft activity that satisfies the soft drop condition can be included in $S_{2,\mathrm{RSD}}$.
%The discussion of the analytic tractability of those two-point energy correlations are beyond the scope of the paper, but we will leave it for future studies.

\begin{figure}
\begin{center}
\includegraphics[width=0.6\textwidth]{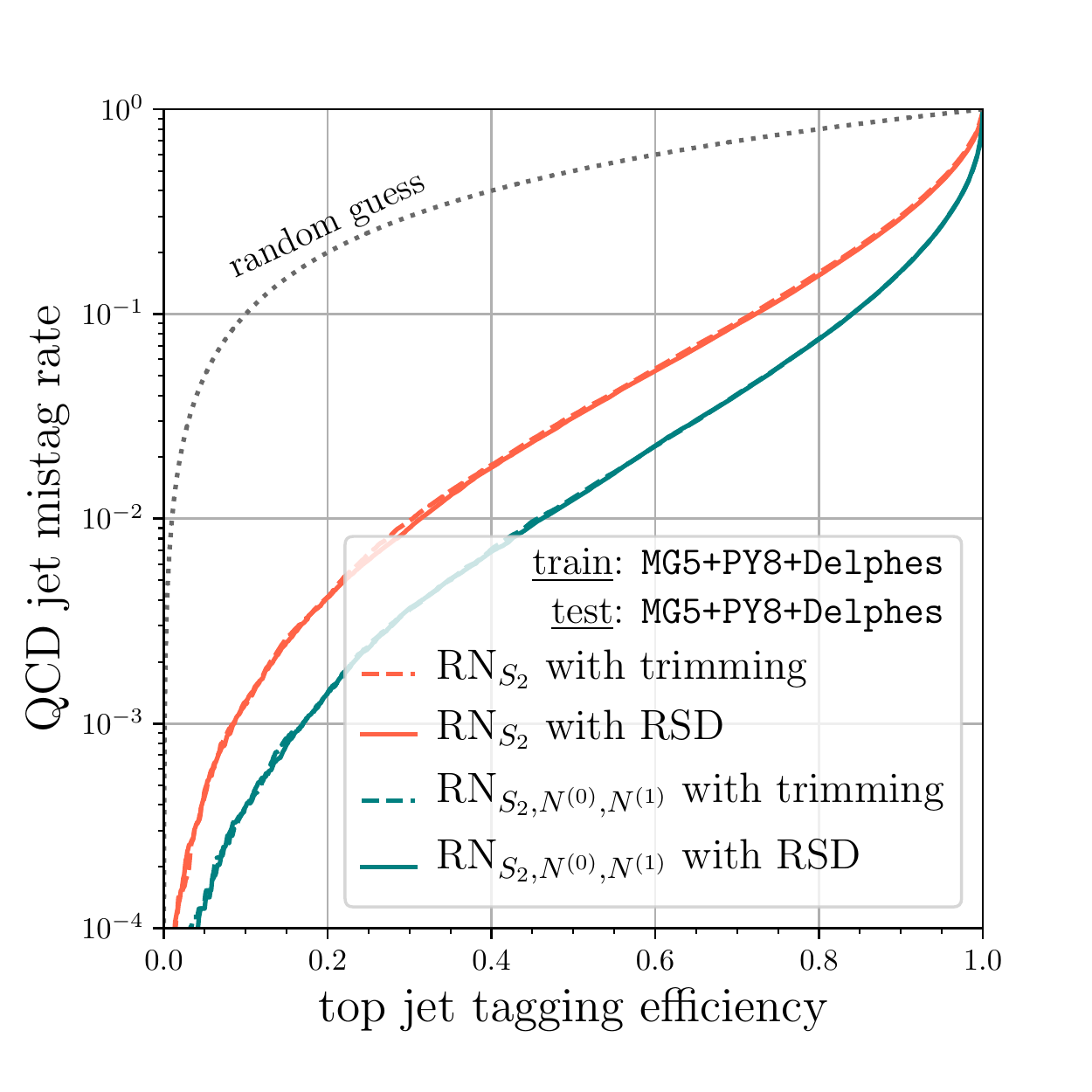}
\end{center}
\caption{
\label{fig:ROC:RN_trim_vs_softdrop}
The ROC curves of $\RN_{S_2}$ and $\RN_{S_2,N^{(0)},N^{(1)}}$ with trimming or RSD.
}
\end{figure}

\Figref{fig:ROC:RN_trim_vs_softdrop} shows the ROC curves of $\RN_{S_2}$ and $\RN_{S_2,N^{(0)},N^{(1)}}$ after replacing inputs
$S_{2,\trim}$, $S_{2,\soft}$, $p_{T,\jet_h}$, and  $m_{\jet_h}$ to
$S_{2,\mathrm{RSD}}$, $S_{2,\mathrm{RSD}^c}$, $p_{T,\jet_r}$, and $m_{\jet_r}$, respectively.
The performance does not change much because the change of inputs is simply a rearrangement of $S_2$ bins related to the soft activity that satisfies the soft drop condition.
Therefore, the impact on the top jet classification performance due to the change of groomer is small.

\subsection{Discussion on other top taggers}
\label{sec:s2_top_tagger:other_cnn}
So far, we compare the performance of the RNs to that of the CNN.
In this subsection, we comment on other top taggers.

\begin{figure}
\begin{center}
\includegraphics[width=0.6\textwidth]{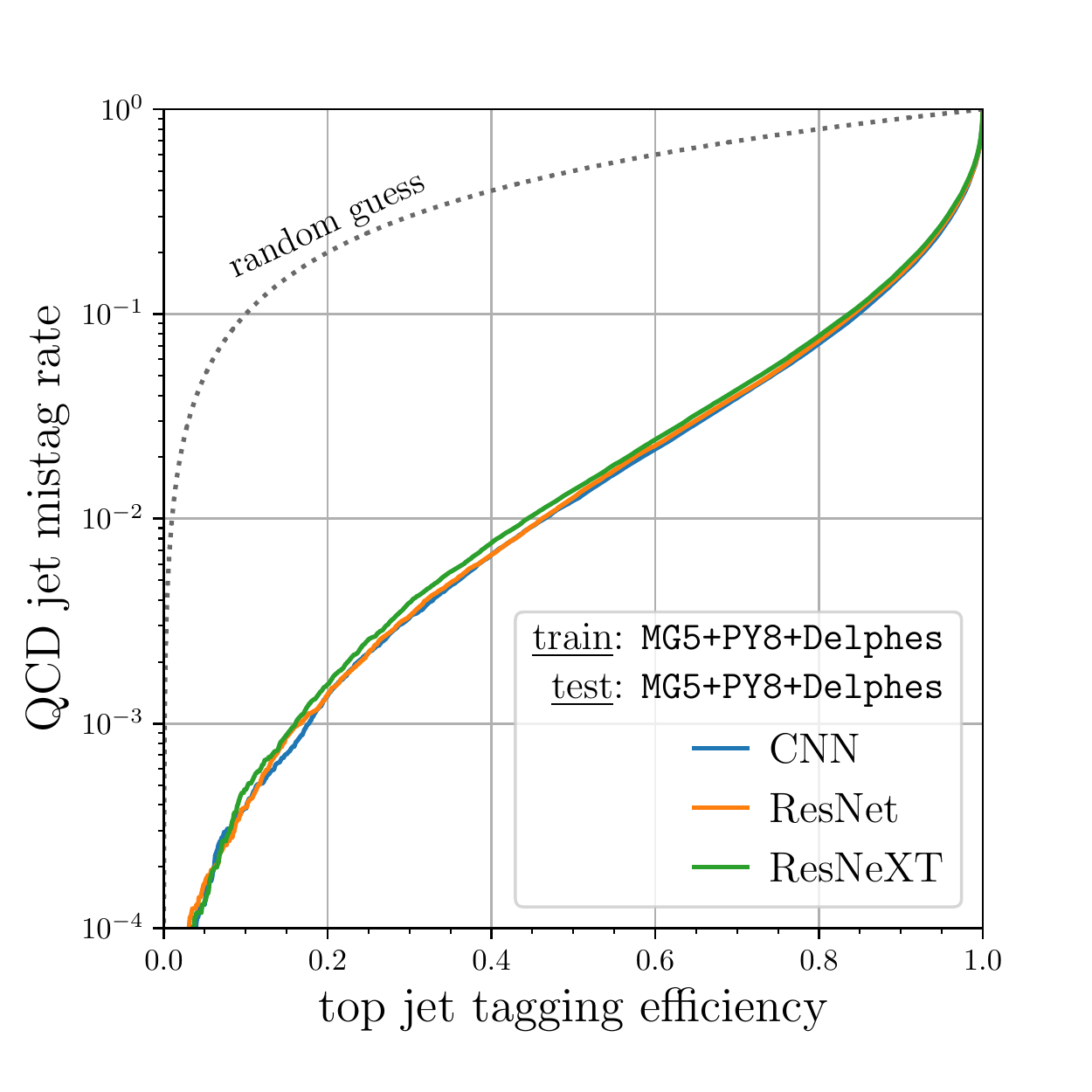}
\end{center}
\caption{
\label{fig:ROC:cnn_variants}
The ROC curves of the CNN and its variants: ResNet and ResNeXt.
}
\end{figure}

In \cite{Kasieczka:2019dbj}, ParticleNet \cite{Qu:2019gqs,10.1145/3326362} and ResNeXt \cite{Xie_2017_CVPR} show a better performance in the top jet classification than the CNN\footnote{Note that this CNN does not take $x_{\kin}$ as inputs and is different from the CNN in this paper.}.
One may wonder if additional features should be included in the RN inputs to reproduce their performance.
However, the networks on figure 5 of \cite{Kasieczka:2019dbj} are not trained on inputs at the same angular resolution. 
It is not clear if the better networks learn additional physical features.
The ResNeXt and CNN in \cite{Kasieczka:2019dbj} use jet images with pixel size 0.025 and 0.04, respectively.
We especially find that the performances of ResNeXt and CNN trained on jet images with pixel size 0.1 are similar.
\Figref{fig:ROC:cnn_variants} shows their ROC curves.

The ResNet \cite{7780459} and ResNeXt in \Figref{fig:ROC:cnn_variants} are the CNN after replacing the chain of the convolutional layers to ResNet \cite{7780459} or ResNeXt modules described in \appendixref{app:implementation:cnn}.
Note that the skip connections in those residual learning networks are for solving the degradation problem \cite{7780459} without deteriorating the universal approximation property of the filter direction of convolutional layers.
If there is no performance degradation due to the depth of the networks, all of those networks should perform similarly.
If we change the pixel size from 0.1 to 0.025, the jet image size changes from $30\times30$ to $120\times120$ and we may need a CNN with more layers or larger filter sizes in order to cover the whole $(\eta,\phi)$ range.
The skip connections may be required to train the network efficiently.
\Figref{fig:ROC:cnn_variants} shows that the CNN is sufficient for the classification in our case.

The ResNeXt in \cite{Kasieczka:2019dbj} shows a similar performance to the ParticleNet.
The ParticleNet is a graph neural network that uses angular coordinates directly, and the angular resolution is not explicitly considered in the inputs.
However, since \texttt{Delphes} provides each constituent's angular position after uniform smearing over corresponding calorimeter bin range \cite{deFavereau:2013fsa}, the inputs of the ParticleNet has implicit angular resolution 0.0174 and 0.1 if the constituent is from electromagnetic and hadronic calorimeter, respectively.
The jet images for the ResNeXt uses pixel width $0.025$, so that the loss of information due to pixelation is small.
%Because of this difference in inputs, the performance improvement due to additionally learned physical features are not convincing.
We leave further investigation between our RNs to those networks in future publications.

\section{Reweighting Distributions of IRC Unsafe Morphological Features}
\label{sec:reweighting}

In \sectionref{sec:s2_top_tagger}, we perform the analysis using sample generated by \py{}, but in this section, we compare the result with the analysis using another event generator and discuss the systematic uncertainties associated with simulations.
Because event generation involves the modeling of soft radiation, the generated events are model-dependent, and the simulator has to be tuned to experimental data.
Describing the distribution of particles in the jet in all circumstances is not trivial. 
Indeed, the simulated distributions of different generators are often significantly different in an extreme kinematic regime, and sometimes neither of them agrees with experimental data. 
The question is how precisely these simulated events should agree with the data. 
For the analysis mainly using high $p_T$ objects, the effects of soft physics are small.
On the other hand, a neural network based jet classifier trained on jet images are capable of utilizing the pattern of soft radiation.
If the agreement between observed and simulated data are ``sufficiently good", we could rely on the simulated data.
In reality, there are yet significant deviations between the MC predictions and experimental data, and the calibrations are necessary. 
Because we know that the less controlled IRC unsafe quantities, such as $N^{(0)}$ and $N^{(1)}$, play an important role in the classification, we focus on calibrating the difference between the experimental and simulated data of those quantities.

To see the systematical error coming from the mismodeling of the parton shower and hadronization, we perform the same classification analysis with different event generators and compare the results.
We choose \hw{} and \py{} for the comparison. 
The two event generators are quite different in modeling of the soft and collinear radiations.
\hw{} uses the angular-ordered shower \cite{Gieseke:2003rz} and the cluster hadronization model. \cite{Webber:1983if,Bahr:2008pv}.\footnote{Dipole shower also can be used, but we do not study the model in this paper.} 
\py{} uses $p_T$ ordered-shower \cite{Sjostrand:2004ef} and the string model of hadronization \cite{Andersson:1983ia,Sjostrand:1984ic}.
The comparison of the radiation pattern of QCD jets is available in various literature \cite{Larkoski:2013eya,Larkoski:2014pca,Gras:2017jty}.
The prediction of the gluon jet distributions differs significantly in each simulator while it more or less agrees with each other for the quark jets.
It is pointed out that prediction is sensitive to the color reconnection modeling.

\begin{figure}
\begin{center}
\begin{subfigure}{0.49\textwidth}
\includegraphics[width=1.0\textwidth]{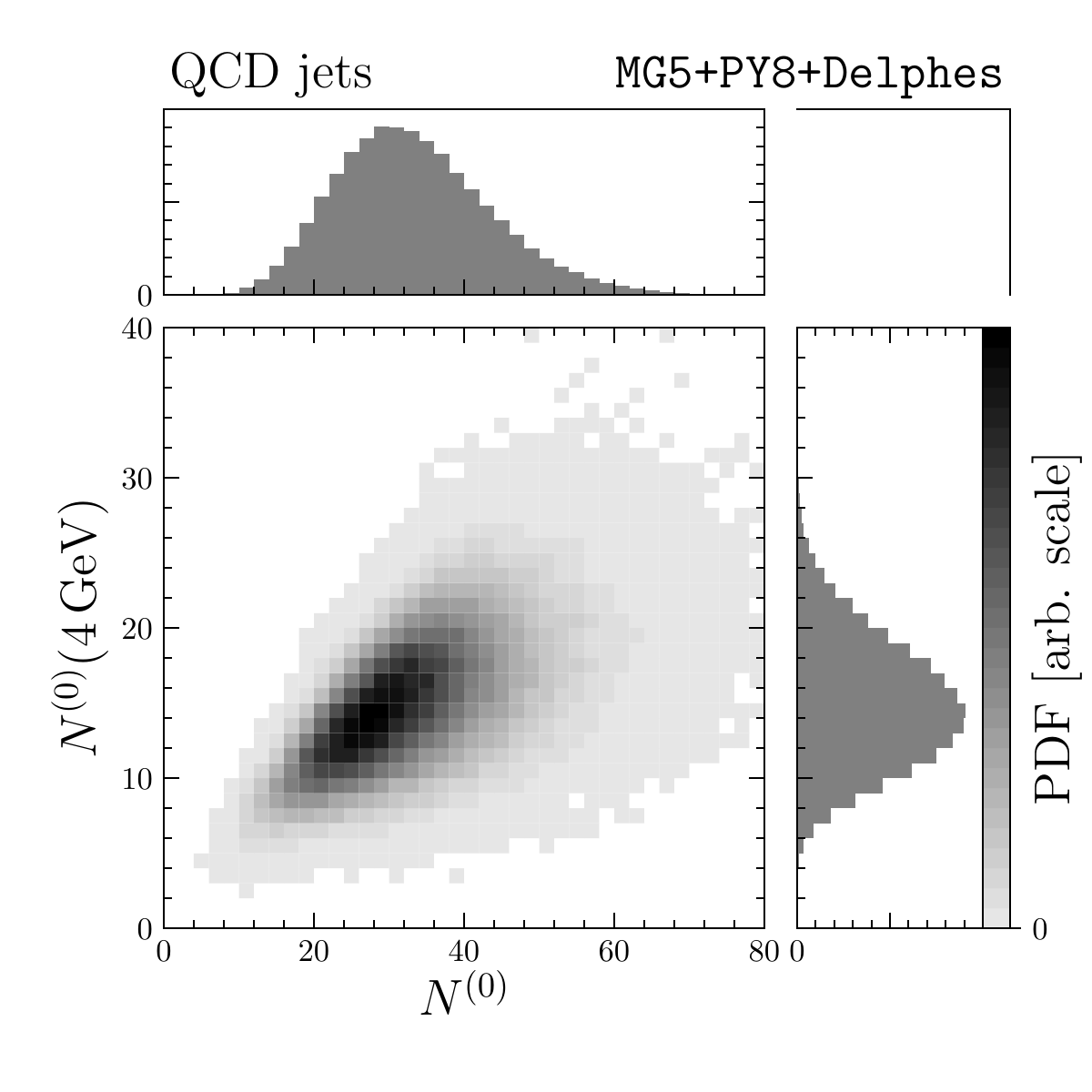}
\caption{}
\end{subfigure}
\begin{subfigure}{0.49\textwidth}
\includegraphics[width=1.0\textwidth]{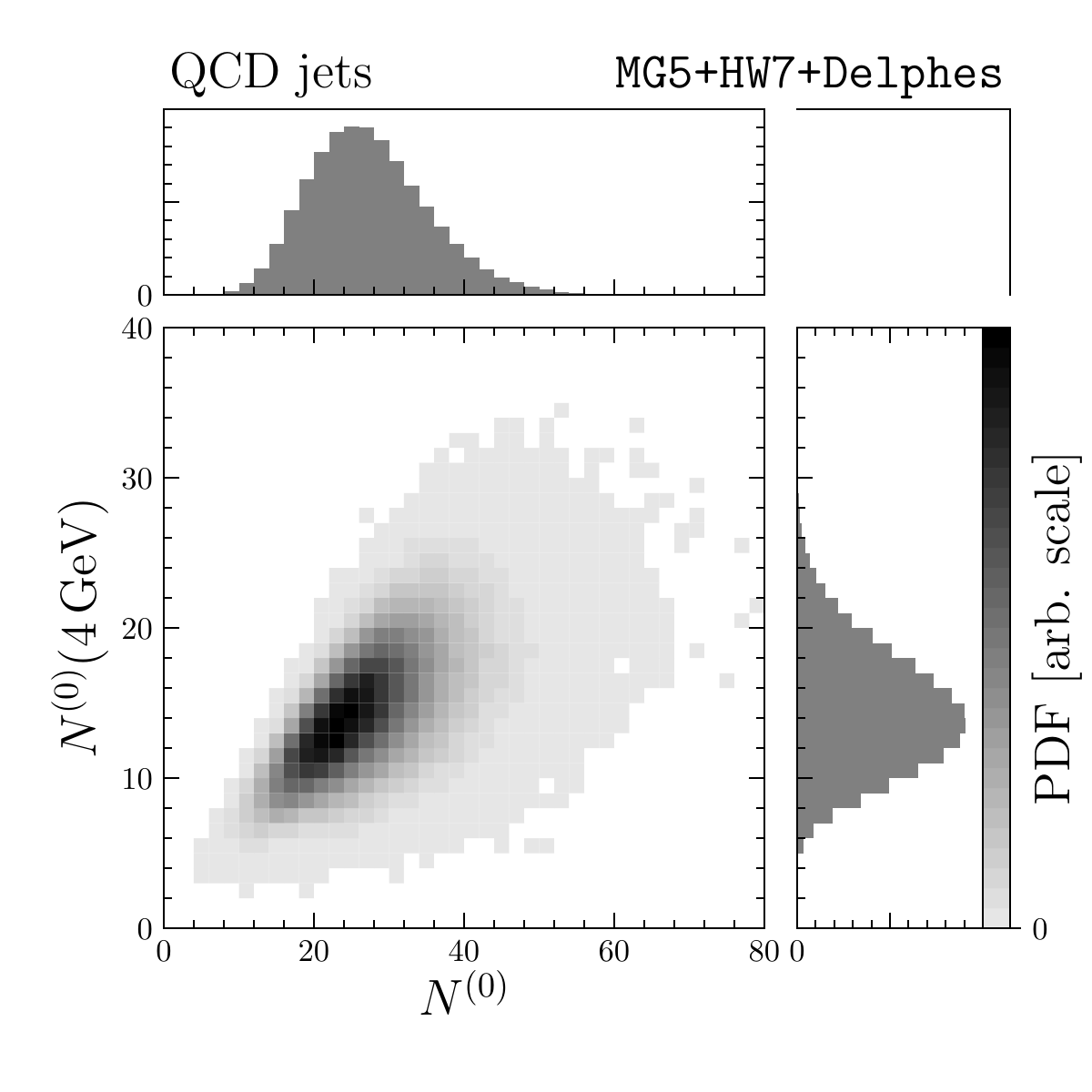}
\caption{}
\end{subfigure}
\end{center}
 \caption{$(N^{(0)},N^{(0)}(4\,\mathrm{GeV}))$ distributions for (a) \py{} and (b) \hw{}} \label{fig:n04}
 \end{figure}

\begin{figure}
\begin{center}
\begin{subfigure}{0.49\textwidth}
\includegraphics[width=1.0\textwidth]{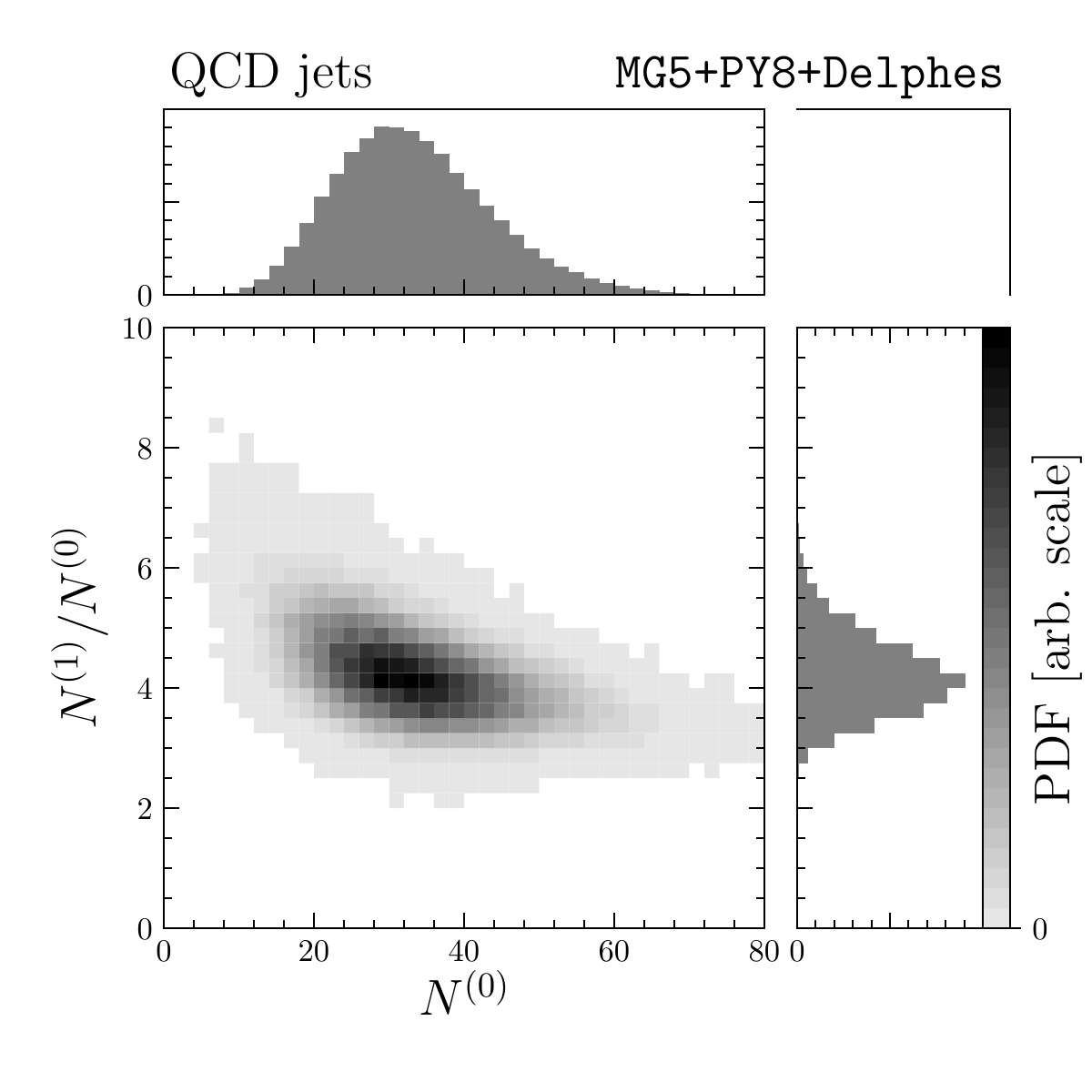}
\caption{}
\end{subfigure}
\begin{subfigure}{0.49\textwidth}
\includegraphics[width=1.0\textwidth]{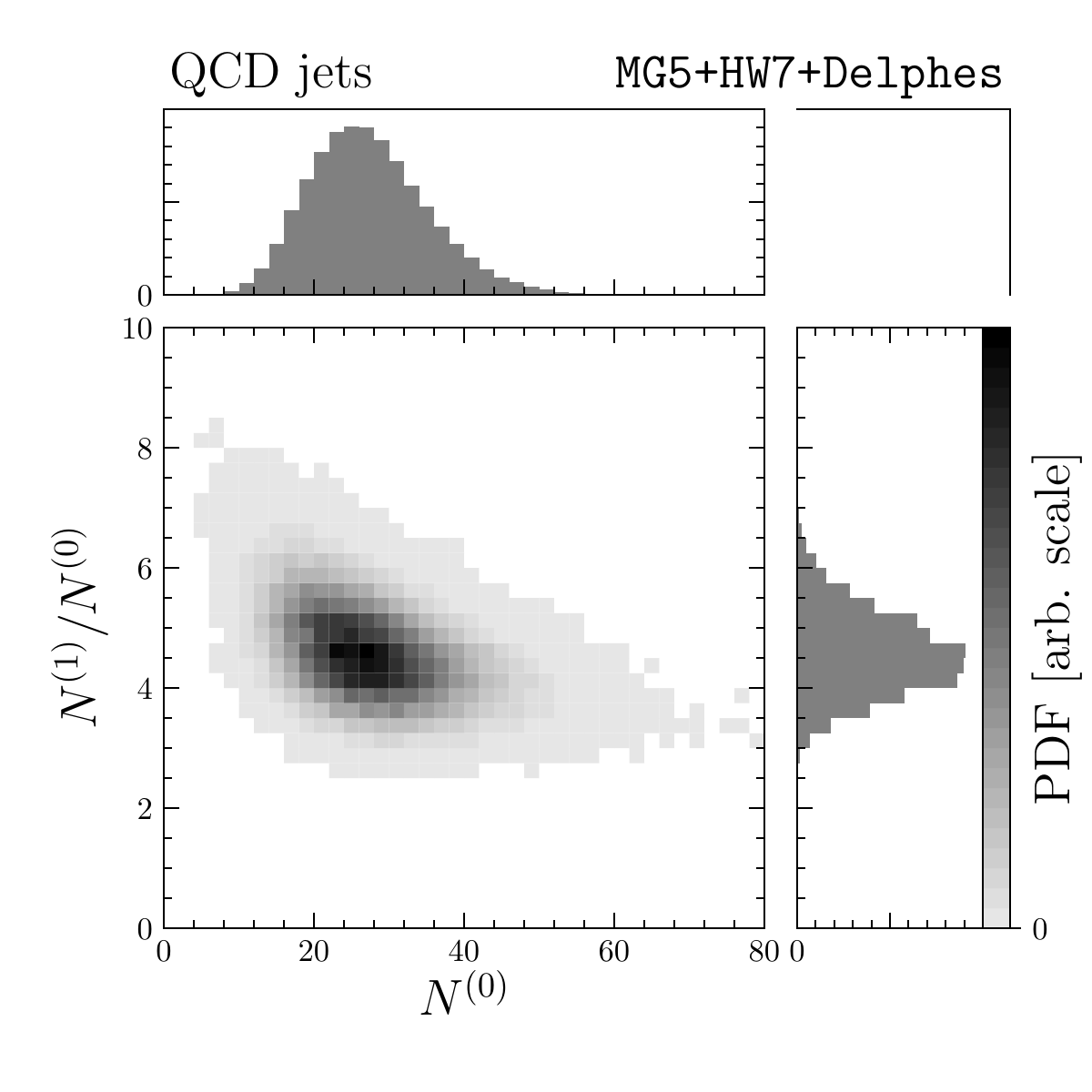}
\caption{}
\end{subfigure}
\end{center}
\caption{
$(N^{(0)},N^{(1)}/N^{(0)})$ distributions for (a) \py{} and (b) \hw{}} \label{fig:n0n1}
\end{figure}

In \figref{fig:n04}, we show the $(N^{(0)}, N^{(0)}(4\,\GeV))$ distributions of the QCD jet samples.  
The $N^{(0)}$ distribution simulated by \py{} is broader than that simulated by \hw{}. 
The tail of the $N^{(0)}$ distribution exceeds 60 for \py{}, but it vanishes at there for \hw{}.
On the other hand, the $N^{(0)}(4\,\GeV)$ distributions of \py{} and \hw{} are similar, as shown in \figref{fig:n_pixel}.
The active pixels with $p_T > 4\,\GeV$ correspond to the particles from high $p_T$ partons in the shower.
Predictions on those partons in the two generators tend to agree, and the predicted $N^{(0)}(4\,\GeV)$ distributions are also similar.
The $N^{(1)}/N^{(0)}$ distributions of \py{} and \hw{} are also similar, as shown in \figref{fig:n0n1}.
Therefore, $N^{(0)}$ should play an important role in the classification.
 
\begin{figure}
  \begin{center}
 \includegraphics[width=0.6\textwidth]{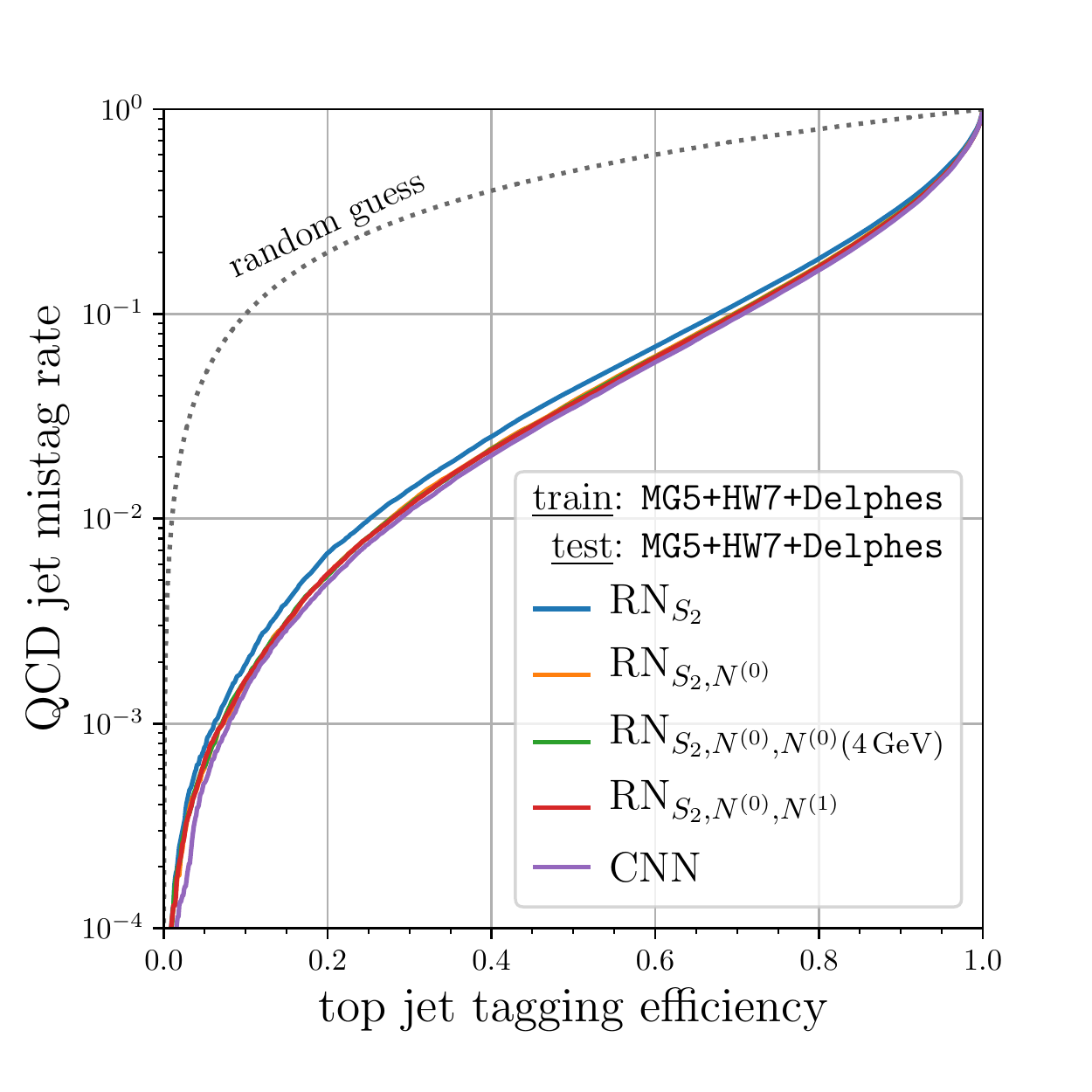}
 \end{center}
 \caption{
The ROC curves of the networks trained on \hw{} samples.  
 }\label{fig:rochw}
 \end{figure}

The separation of the top jets and QCD jets is worse for \hw{}  compared with \py{} discussed in previous sections.
The AUC of the top jet vs.~QCD jet classification predicted by \hw{} is smaller than that predicted by \py{}.
In \figref{fig:rochw}, we show the ROC curves of each classifier trained on \hw{} events. 
The performance of the $\RN_{S_2}$ is similar to that trained on \py{} events.
Once $N^{(0)}$ is additionally considered in the classification, the performance is improved.
However, the improvement from adding $N^{(0)}$ is significantly smaller in \hw{}, because the $N^{(0)}$ distributions of top jets and QCD jets are close, as shown in \figref{fig:n_pixel}.

In the previous analysis, we have shown that inputs $N^{(0)}$, $N^{(0)}(4\,\mathrm{GeV})$,  $N^{(1)}$, and $N^{(1)}(4\,\mathrm{GeV})$ in addition to $S_{2,ab}$ is good enough for building a neural network that fits the CNN output.
At the same time, this indicates that tuning of the event generator focusing on these counting variables can be an efficient way to obtain the simulated data that gives consistent results with the experimental data.

If the difference between the simulated and experimental data is not too large, reweighting simulated events is useful for reducing the difference.
We consider reweighting based on the marginal distribution of interested variables $\bm{x}$.
Let $\rho_{\mathrm{true}}(\bm{x})$ and $\rho_{\mathrm{MC}}(\bm{x})$ be the $\bm{x}$ distributions with true and simulated events, respectively.
The new weight $w_{\mathrm{new}}^{[i_Y]}$ of the event $i_Y$ is given as follows,
\begin{equation}
\label{eqn:reweight_factor}
w_{\mathrm{new}}^{[i_Y]} = \frac{\rho_{\mathrm{true}}(\bm{x}^{[i_Y]})}{\rho_{\mathrm{MC}}(\bm{x}^{[i_Y]})} \cdot w_{\mathrm{old}}^{[i_Y]},
\end{equation}
where $w_{\mathrm{old}}^{[i_Y]}$ is the weight before reweighting.

\begin{figure}[t]
%%%%%%%%%%%%%%
\begin{center}
\begin{subfigure}{0.49\textwidth}
\includegraphics[width=1.0\textwidth]{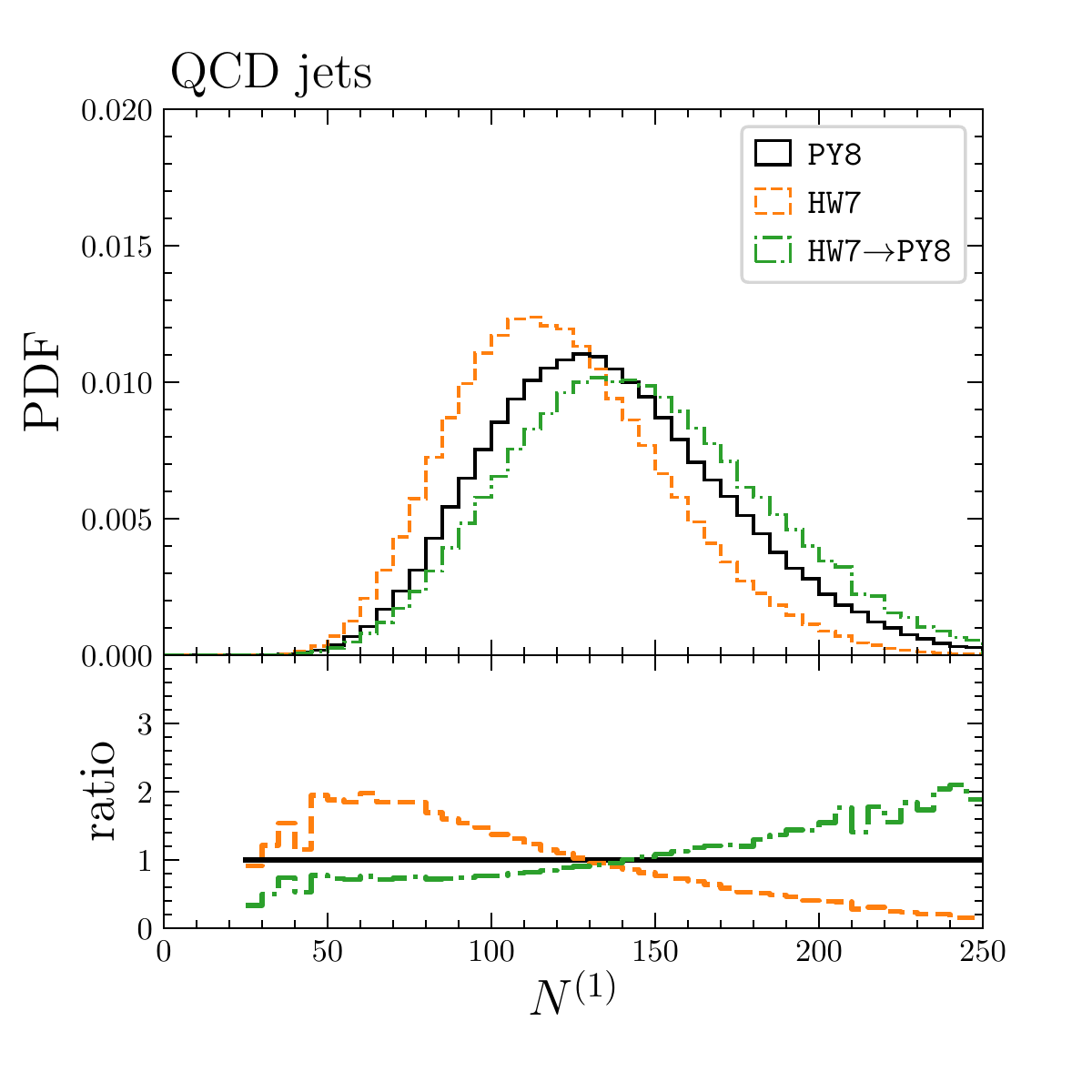}
%\caption{\label{fig:hw2py:a}}
\end{subfigure}
\begin{subfigure}{0.49\textwidth}
\includegraphics[width=1.0\textwidth]{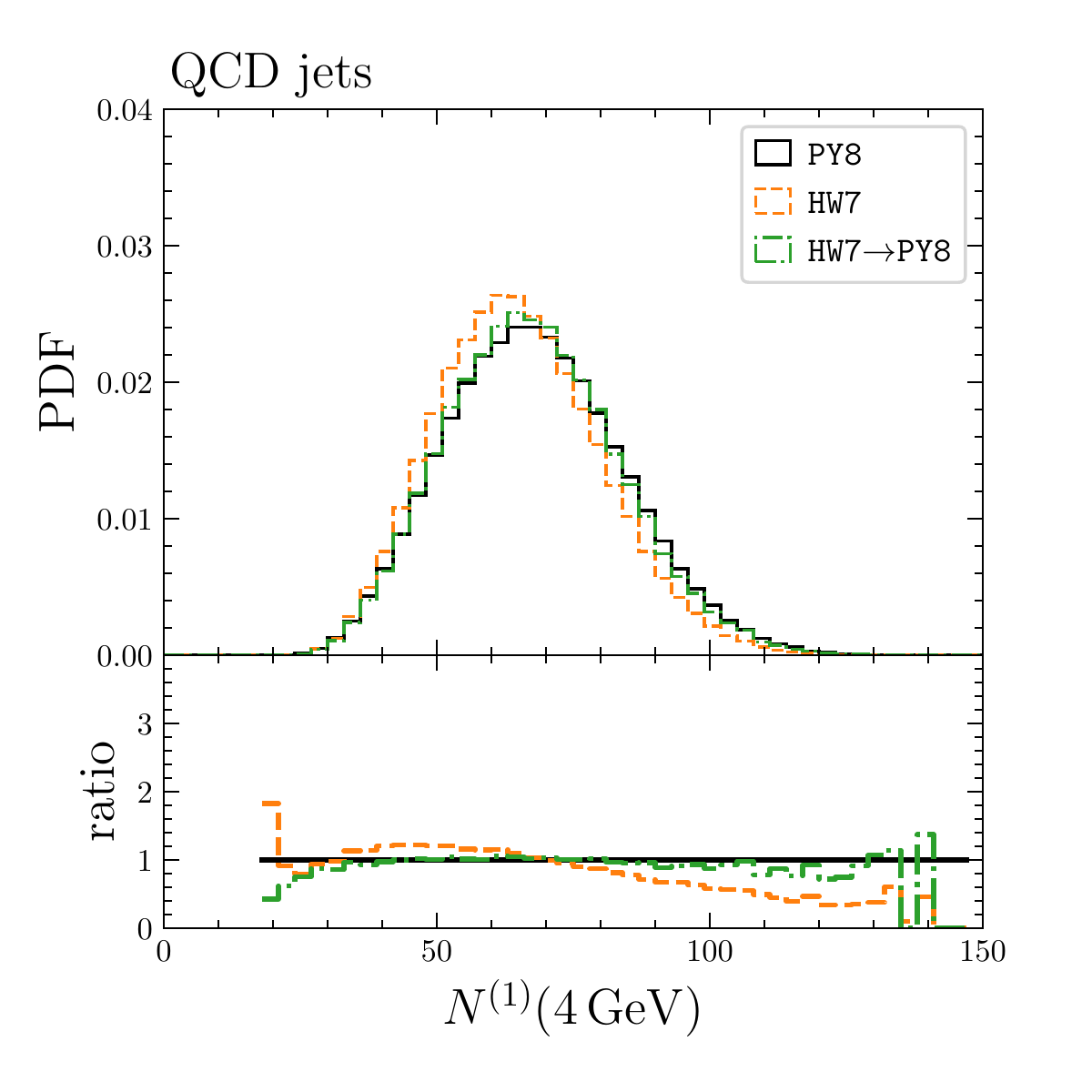}
%\caption{\label{fig:hw2py:b}}
\end{subfigure}
\end{center}
\caption{
\label{fig:reweighted_n1}
The $N^{(1)}$ and $N^{(1)}(4\,\GeV)$ distributions of \py{}, \hw{}, and reweighted \hw{} samples.
}
\end{figure}

Let us perform an exercise to correct $(N^{(0)},N^{(0)}(4\,\GeV))$ distribution, assuming that either one of the distributions $\rho_{\py{}}$ and $\rho_{\hw{}}$ simulated by \py{} and \hw{} is $\rho_{\mathrm{true}}$ while the other is $\rho_{\mathrm{MC}}$. 
We consider the reweighting of these two variables in order to consider a non-trivial case that some of the variables are correlated.
The reweighting factor ${\rho_{\mathrm{true}}(N^{(0)},N^{(0)}(4\,\GeV))} / {\rho_{\mathrm{MC}}(N^{(0)},N^{(0)}(4\,\GeV))}$ is calculated using the normalized histogram of $(N^{(0)},N^{(0)}(4\,\GeV))$, as stated in \appendixref{app:reweighted_dist}.
The $N^{(1)}$ distribution in \figref{fig:reweighted_n1} still disagree after the reweighing, but the deviation is minor.
Because the sample size is limited, we do not attempt to correct all those distribution in this paper.

\begin{figure}
%%%%%%%%%%%%%%
\begin{center}
\begin{subfigure}{0.49\textwidth}
\includegraphics[width=1.0\textwidth]{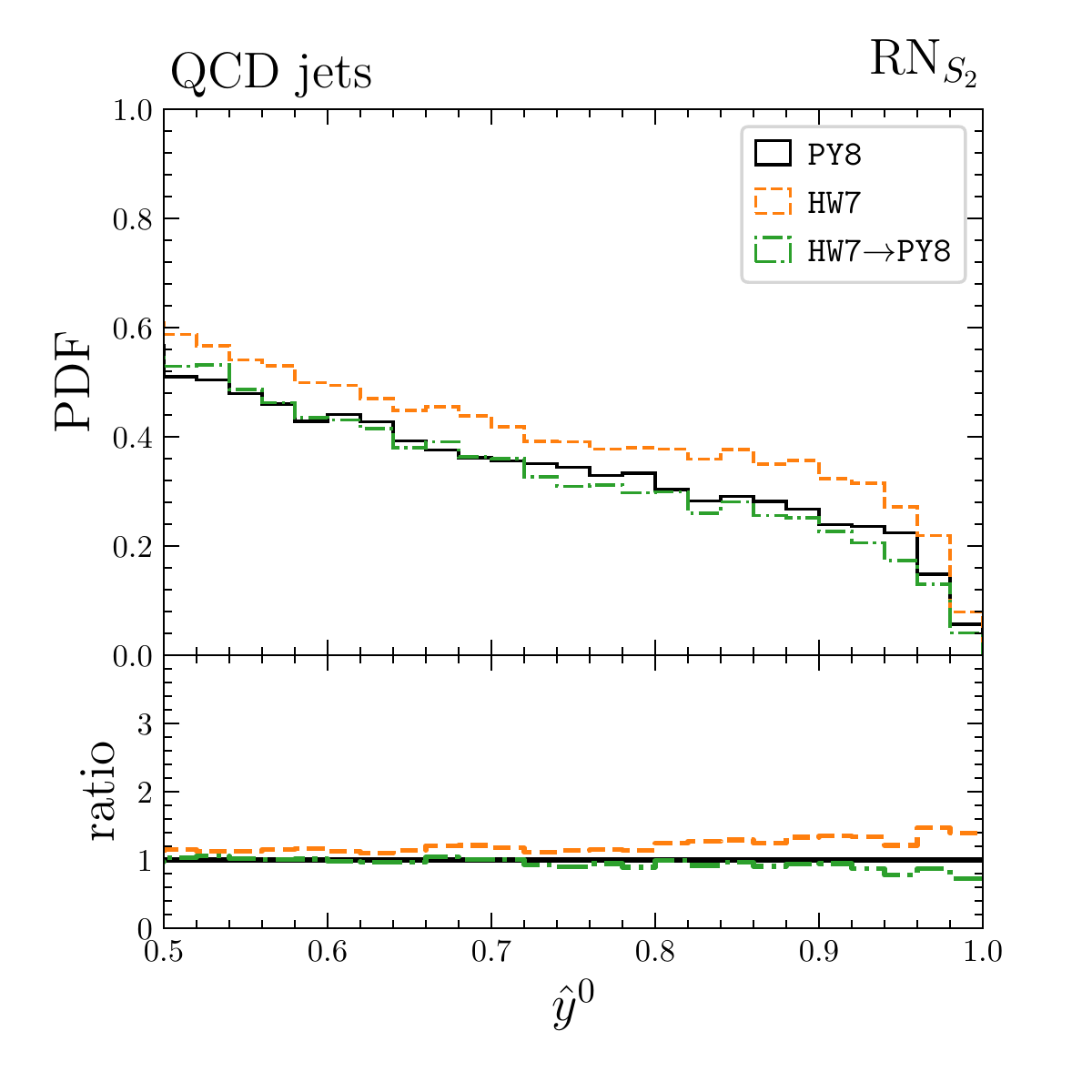}
\caption{\label{fig:hw2py:a}}
\end{subfigure}
\begin{subfigure}{0.49\textwidth}
\includegraphics[width=1.0\textwidth]{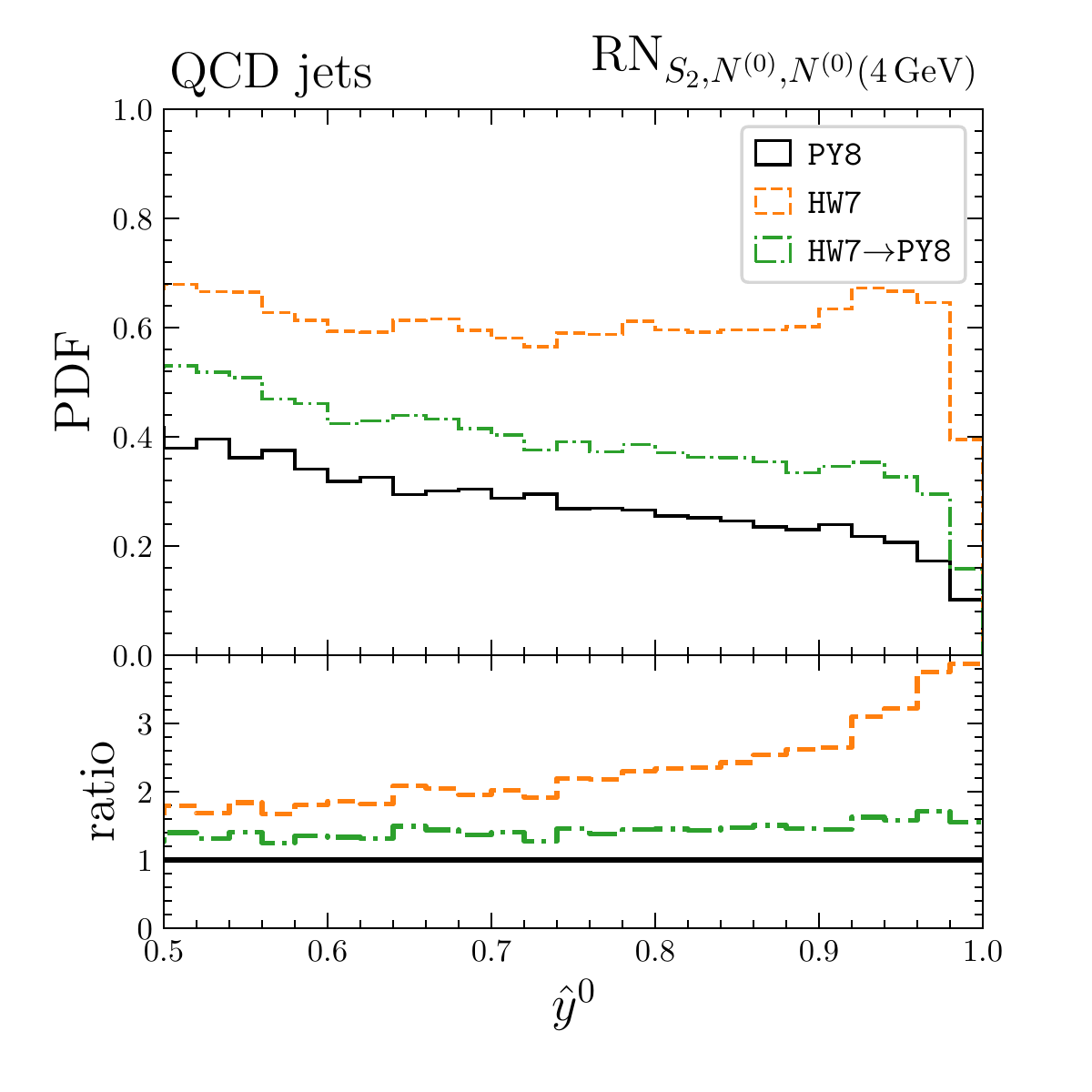}
\caption{\label{fig:hw2py:b}}
\end{subfigure}
\end{center}
%%%%%%%%%%%%%
\begin{center}
\begin{subfigure}{0.49\textwidth}
\includegraphics[width=1.0\textwidth]{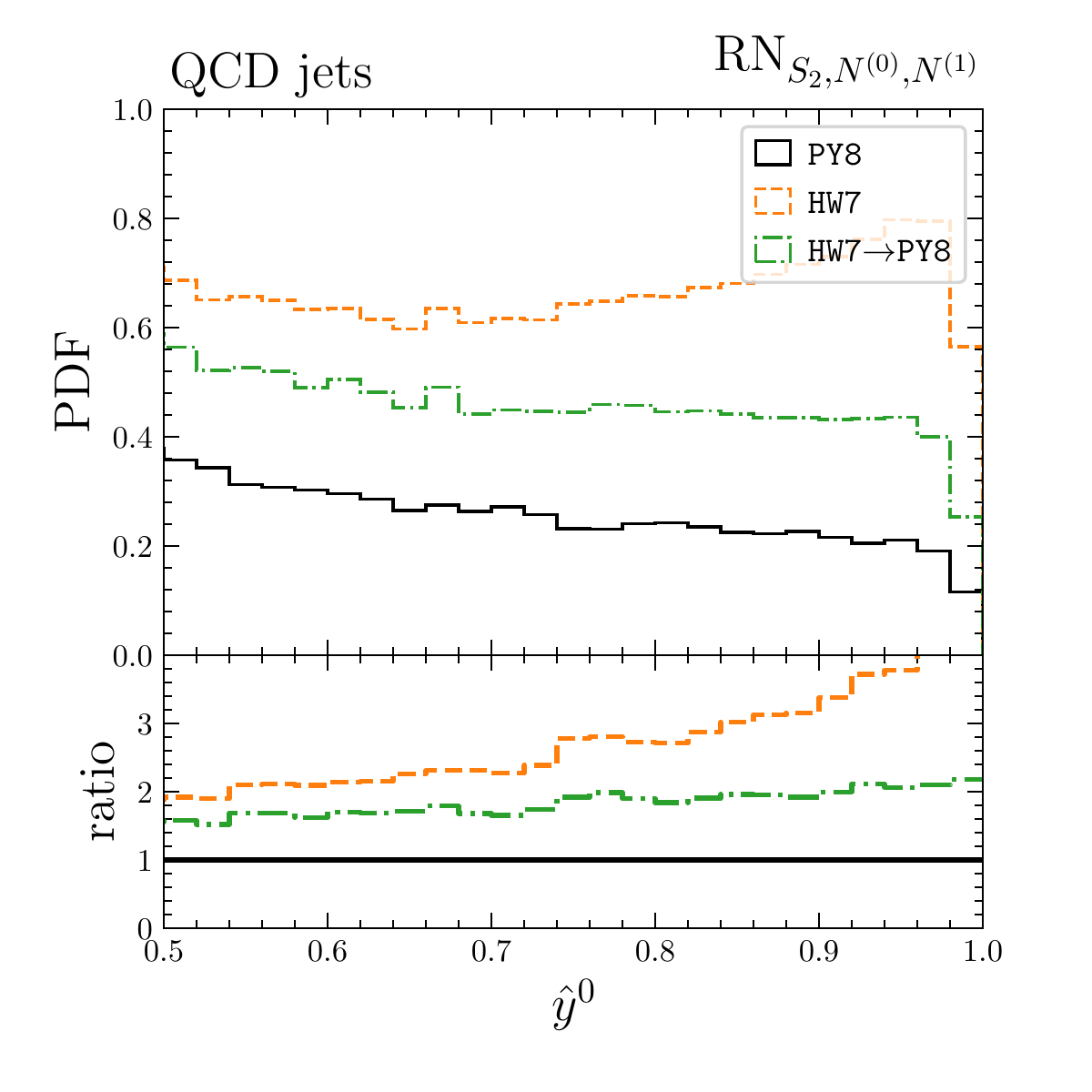}
\caption{\label{fig:hw2py:c}}
\end{subfigure}
\begin{subfigure}{0.49\textwidth}
\includegraphics[width=1.0\textwidth]{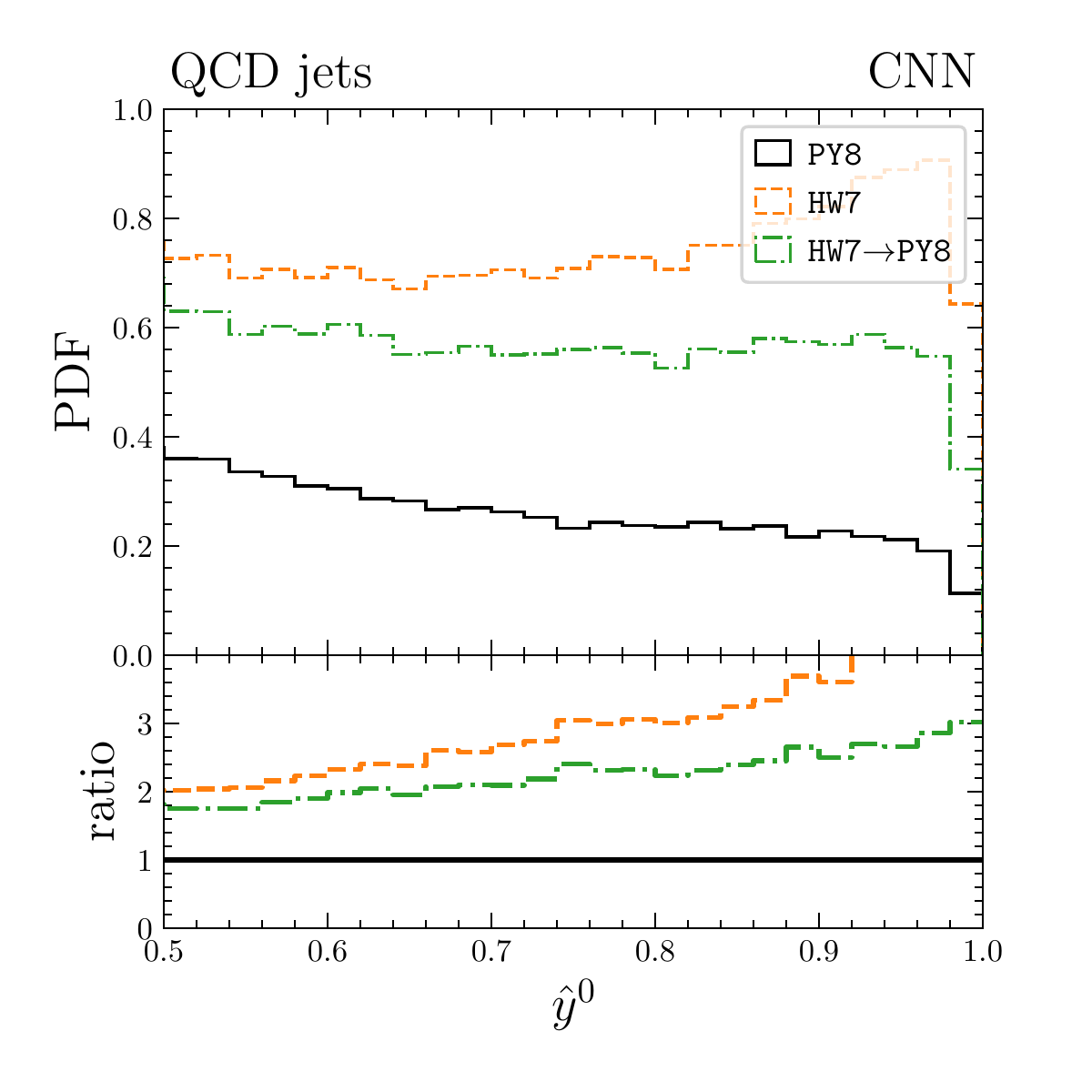}
\caption{\label{fig:hw2py:d}}
\end{subfigure}
\end{center}
\caption{
\label{fig:hw2py}
The $\hat{y}^0$ distributions of \py{} and \hw{} test samples for the model trained on the \py{} events.
The neural networks used in the plots are (a) $\RN_{S_2}$, (b) $\RN_{S_2,N^{(0)},N^{(0)}(4\,\mathrm{GeV})}$, (c) $\RN_{S_2,N^{(0)},N^{(1)}}$, and (d) $\CNN$. 
}
\end{figure}

\Figref{fig:hw2py} shows the $\hat{y}^0$ distributions for QCD jet samples, of the models trained on \py{} samples.
The orange dashed, black solid, and green dot-dashed histograms are the $\hat{y}^{0}$ distributions with \hw{}, \py{}, and reweighted \hw{} samples, respectively.

\Figref{fig:hw2py:a} shows the $\hat{y}^0$ distributions of $\RN_{S_2}$.
This classifier does not use $N^{(0)}$ and $N^{(0)}(4\,\GeV)$ explicitly, but the distribution of reweighted \hw{} samples comes quite close to that of \py{} samples.
The score difference comes from the difference of $S_2$ distribution.
The $S_{2,\soft}$ distribution of \hw{} samples is 10\% smaller than that of \py{} samples.  
The reweighting reduces the difference because $N^{(0)}$ and $S_{2,\soft}$ are correlated.
The Pearson correlation coefficient between $N^{(0)}$ and $S_{2,\soft}^{(i)}$ is 0.3 for both \py{} and \hw{}.\punctfootnote{The correlation between $N^{(0)}$ and $S_{2,\trim}^{(i)}$ is around 0.15 for the bins dominated by the cross-correlation between high $p_T$ constituents.
}
The bin-by-bin ratio of the average $\langle S_{2,\soft}^{(i)} \rangle$ between \hw{} and \py{} samples is about 0.9.
The average $\langle S_{2,\soft}^{(i)} \rangle$ of \hw{} samples increases after the reweighting and the $S_{2,\soft}$ distributions get closer to each other.
On the other hand, the reweighting increases the disagreement of $p_{T,\jet}$ distribution. 
The sum of the weights of the reweighted \hw{} samples with $p_{T,\jet} \sim 500\,\GeV$ is about 20\% larger than that of \py{} samples.
We marginalized $p_{T,\jet}$ during the training so that the impact on the $\hat{y}^0$ distribution is minimal.
Therefore, the agreement seen in \figref{fig:hw2py:a} is mainly due to the correction of $S_{2,ab}$, and it is encouraging.

\Figref{fig:hw2py:b} shows the $\hat{y}^0$ distributions of $\RN_{S_2,N^{(0)},N^{(0)}(4\,\mathrm{GeV})}$.
For $\hat{y}^0 \sim 1$, the ratio of the $\hat{y}^0$ distributions of \hw{} and \py{} exceeds 4, and the distribution of \hw{} samples even peaks near $\hat{y}^0 \sim 1$. 
This means that the model trained on \py{} samples focuses on a particular region in order to get high purity top samples, but the \hw{} samples are still populated in the region.  
In the situation that the \hw{} distribution is  ``true" while \py{} samples are used to build the top jet vs.~QCD jet classifier,  we overestimate the top quark event rate by dijet contamination; adding the  variables whose ``true distributions" are not well understood could cause the problem of this kind.

The ratio between weighted \hw{} and \py{} distributions is constant.
This is nice in order to avoid the systematics along with tightening the cut to reject QCD events.  
On the other hand, the ratio of the reweighted \hw{} samples is much larger than that in \figref{fig:hw2py:a}.
The deviation should come from the mismodeling of the correlation between $N^{(0)}$ and other parameters.  
The difference is even larger if one includes $N^{(1)}$ in the inputs, as shown in \figref{fig:hw2py:c}.
The ratio between the weighted \hw{} and \py{} sample is now nearly a factor of two larger at $y\sim 1$ and even increasing.
This disagreement is not surprising given the very poor sample of \hw{} in the high $N^{(0)}$ region.
Finally, \figref{fig:hw2py:d} is the $\hat{y}^0$ distribution of the CNN model. 
The distributions looks quite similar to those in \figref{fig:hw2py:c} before reweighting, 
but the ratio of the distributions of reweighted \hw{} events and \py{} events is larger than that of $\RN_{S_2,N^{(0)},N^{(1)}}$.

\begin{figure}[h]
 \begin{center}
\includegraphics[width=0.7\textwidth]{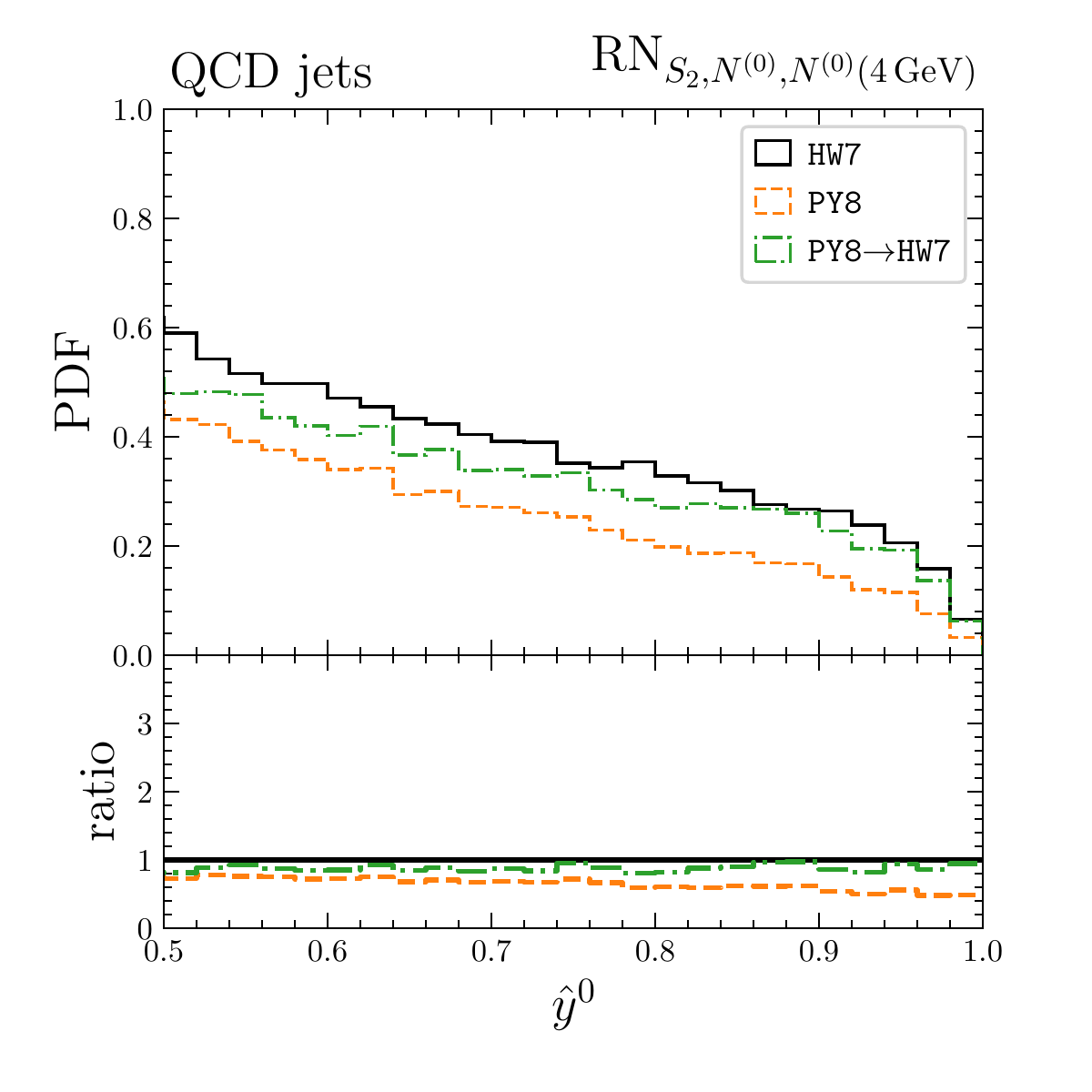}
\end{center}
\caption{The score distribution of \py{} and \hw{} test sample for the model trained by the \hw{} events. }
\label{fig:py2hw}
 \end{figure}

\Figref{fig:py2hw} shows the $\hat{y}^0$ distributions of the model $\RN_{S_2,N^{(0)},N^{(0)}(4\,\mathrm{GeV})}$ trained on \hw{} events. 
Recall that the QCD jets in \py{} samples cover the phase space of the QCD jets simulated by \hw{}, and the reweighting is then effective for transforming the \py{} samples to \hw{} samples.
The opposite is not true because there are QCD jets which are not in \hw{} generated samples. 
The reweighting is not exact because we have only a small number of events in some phase space region, and we see some deviation in $\hat{y}$ distribution, as shown in \figref{fig:hw2py:b}.
If one wishes to describe real data by assigning an appropriate weight for each simulated events, it is better to use a generator setup that covers wider phase space so that we can correct the event distribution by using experimental data afterwords. 
 
The disagreement between \py{} and \hw{} samples remains after the reweighting in this exercise. 
We do not proceed to reweight distributions other than the $(N^{(0)},N^{(0)}(4\,\GeV))$ distribution in this paper because of the statistical limitation.
Neural network-based reweighing \cite{Andreassen:2019nnm} can be helpful for adjusting full phase-space, but it is beyond the scope of this paper.
The difference between the two generators is too large to achieve perfect agreement simply by reweighting. 
Because $N^{(0)}$ and $N^{(1)}$ are important quantities for describing the neural network-based classifier, those generators may be tuned carefully to reproduce the distribution of soft activities in jet images.

\section{Discussions}
\label{sec:conclusion}
In this paper, we have identified essential quantities that the CNN on a jet image is using for the top jet vs.~QCD jet classification. 
The discovered quantities consist of both IRC safe and IRC unsafe observables.
The former includes the IRC safe two-point energy correlation, jet spectrum, as a function of the distance between two jet constituents. 
The latter is an IRC unsafe Minkowski sequence inspired from the Minkowski functionals that describes morphological information on the set of jet constituents. 
It gives a quantitative measure of the area that is occupied by the particles inside jets. 
The first element of the Minkowski sequence is the number of active pixels in the jet image, $N^{(0)}$, and the second element $N^{(1)}$
is the sum of the $N^{(0)}$ and the number of the pixels adjacent to the active pixels. 
These quantities are derivable from a jet image, and the relation network (RN) trained on these quantities (along with kinematic observables) has equivalent performance to the CNN.                           

The IRC safe quantities are theoretically more controlled, especially different event simulators predict consistent distributions. 
On the other hand, the IRC unsafe quantities are described by phenomenological models tuned by the experimental data. 
The classification performance of RN agrees with that of CNN only when we include IRC unsafe Minkowski sequences among the inputs.
The similarity of the performance indicates that the top jet classifier based on CNN uses the geometric information of soft radiation, and we have succeeded in reproducing the CNN predictions using fewer degrees of freedom.

We also point out that the training of the RN is more stable than the CNN.
The stability comes from the fact that the RN classifiers use a restricted set of derived inputs from the jet images, and the loss function of the RN is less complicated than that of CNN.
We measure the variation of the training results by randomly swapping the event orders in the batch training and using a different initial parameter in the networks. 
The variation of the RN output is about factor 3 smaller than that of the CNN output, as we have seen in \tableref{table1}. 

As the IRC safe inputs, we choose the jet spectrum \cite{Lim:2018toa,Chakraborty:2019imr}, which is aggregated two-point energy correlation as the function of the $\Delta R$. 
We introduce the various improvements on the jet spectrum from the previous paper.  In this paper, it is derived from a constrained graph network. A vertex of the graph network corresponds to a jet constituent, while each vertex carries information of the constituent momentum and the subjet ID to which the constituent belongs.  
The edges links between two vertices and represent the two-point correlation between the two constituents. 
For the classification of top jets and QCD jets, we find that the correlation among the trimmed jet and the correlation between the leading subjet and the other constituents, and their geometry are especially useful in the classification. 
We systematically include the three-point structure of the top quark in the two-point energy correlations after removing the leading subjet.
The modularized networks process the two-point correlations separately with global kinematical inputs so that the combined network accepts significantly more inputs without inflating the parameters in the hidden layers.

The classifiers using the IRC unsafe quantities, such as soft pixels of jet image or the Minkowski sequence, could suffer from  systematic uncertainties of the simulation.  
After the identification of the key morphological quantities, we can minimize efforts on calibration by focusing on the $N^{(0)}$ and $N^{(1)}$ distributions. 
The distributions may be corrected relatively easily by reweighting events to calibrate the distributions to the observed data.  
We demonstrate that the reweighting of the simulated events to reproduce the true $N^{(0)}$ distributions greatly reduce the systematic error of the classifiers.
Such tuning of the data reduces the systematic uncertainties in the ML classifications that depend on the simulated events.

In summary, we propose an approach to replace a complex neural network using the low level inputs into a simple network using the processed inputs motivated from a physics point of view.  To this end, we show surprising evidence that the CNN output depends on the geometrical measures expressed by discritized version of the Minkowski functionals.  
These morphological quantities improve jet classification significantly. The study of jet morphology from the data, and comparison to the prediction from event simulation might be an exciting direction to persuade. 
We think the variables may be further extended not only for jet physics but also for the analysis of event geometry or anomaly searches.

\section*{Acknowledgements}
The authors would like to thank Marat Freytsis, Eun-Sol Kim (in kakaobrain), Andrew J. Larkoski, Benjamin Nachman, Maurizio Pierini, David Shih, and Takashi Tsuboi for useful discussions.
This work is supported by the Grant-in-Aid for Scientific Research on Scientific Research B 
(No.~16H03991, 17H02878) and Innovative Areas (16H06492);
World Premier International Research Center Initiative (WPI Initiative), MEXT, Japan;
and the Department of Science and Technology, Government of India, under Grant No.~IFA18-PH 224 (INSPIRE Faculty Award).
M. T. is supported in part by the JSPS Grant-in-Aid for Scientific Research
No.~16H03991, 16H02176, 18K03611, and 19H04613.

\appendix

\section{Setup for Monte-Carlo Event Simulation}
\label{app:mc}

We generate $pp \rightarrow t \bar{t}$ and $pp \rightarrow jj$ events for top jet and QCD jet samples, respectively.
The symbol $j$ represents gluon or (anti-)quark other than the top quark.
The parton level events are generated by \texttt{Madgraph5 2.6.6} \cite{Alwall:2014hca}.
The center of mass energy is 13 TeV.
Produced top quarks are forced to decay into $bW$ and the subsequent $W$ boson decays into two quarks including $b$-quarks.
Since we are only interested in boosted top quarks, we generate events with outgoing partons whose $p_T$ is larger than 450 GeV.
Numbers of the generated $pp \rightarrow t\bar{t}$ and $pp \rightarrow jj$ events with this preselection are 5 million and 10 million, respectively.
The renormalization and factorization scales are set to be $H_T/2$, where $H_T$ is the sum of the transverse energy of each parton, and the parton distribution function is $\texttt{NNPDF23\_lo\_as\_0130\_qed}$ \cite{Ball:2012cx,Ball:2013hta,Carrazza:2013bra,Carrazza:2013wua}.
Two parton shower and hadronization simulations are considered in this paper:
 \texttt{Pythia~8.226}~\cite{Sjostrand:2014zea} with Monash tune~\cite{Skands:2014pea} and \texttt{Herwig~7.1.3} \cite{Bellm:2015jjp,Bahr:2008pv} with default tune \cite{herwigtune,Gieseke:2012ft}. 
The pile-ups are not included but the underlying events and multi-parton interactions are considered.

We use \texttt{Delphes 3.4.1} \cite{deFavereau:2013fsa} for detector simulation with its default ATLAS detector configuration.
Jets are reconstructed from calorimeter towers whose $(\eta,\phi)$ resolutions at electromagnetic and hadronic calorimeters in $|\eta| < 2.5$ are assumed to be (0.0174,$1^\circ$) and (0.1, $10^\circ$), respectively.
Anti-$k_T$ jet clustering algorithm \cite{Cacciari:2008gp} with radius parameter $R_\jet = 1.0$ implemented in \texttt{fastjet~3.3.0}~\cite{Cacciari:2011ma,Cacciari:2005hq} is used to cluster these calorimeter towers into jets.
The leading $p_T$ jets with its transverse momentum $p_{T,\jet} \in [500,600]$ GeV and mass $m_{\jet} \in [150,200]$ GeV are selected for the analysis.
In addition, a top jet sample is required to have quarks from the originating top quark within $R_\jet$ from the jet axis.
After this selection, we have about 950,000 top jets and 350,000 QCD~jets. 
Half of them are used for the training and th others are used for testing.
For jet trimming, we use $k_T$ algorithm \cite{Catani:1993hr,Ellis:1993tq} with radius 0.2 and keep subjets whose energy fraction is larger than 0.05.
The leading $p_T$ subjet $\jet_{1}$ is the highest $p_T$ anti-$k_T$ subjet \cite{Cacciari:2008gp} with radius 0.2.

Note that we have not used matched sample, so that the modeled $p_{T,\jet}$ distribution is not precise beyond the leading order accuracy.
Nevertheless, the changes due to recoiling from extra radiation are not a main interest in this paper, so we use this samples by presuming that the top jets and QCD jets are factorizable.

\section{Kernel Density Estimation of $p_{T,\jet}$ Distribution}
\label{app:kde}

We use the kernel density estimation (KDE) on a finite interval $[p_{T,\jet}^{\min},p_{T,\jet}^{\max}]$ to model the event-by-event weight $f_{p_{T,\jet}}(p_{T,\jet}; Y)$ in \eqref{eqn:loss_cross_entropy}.
First, we transform $p_{T\,jet}$ into a logit $t(p_{T,\jet})$ in order to make the domain unbounded.
\begin{equation}
t(p_{T,\jet}) 
= 
\logit \left( \frac{p_{T,\jet} - p_{T,\jet}^{\min}}{p_{T,\jet}^{\max} - p_{T,\jet}^{\min}} \right) 
= 
- \log \frac{p_{T,\jet} - p_{T,\jet}^{\min}}{p_{T,\jet}^{\max} - p_{T,\jet}}
\end{equation}
The KDE of the sampled logits, $t(p_{T,\jet}^{[i_{N_Y}]})$ is used to estimate the probability density function $f_{P_{T,\jet}} (p_{T,\jet}; Y)$.
\begin{equation}
f_{P_{T,\jet}} (p_{T,\jet}; Y) 
\approx
\frac{t'(p_{T,\jet})}{N_Y} \sum_{i_Y=1}^{N_Y} K_h( t(p_{T,\jet}) - t(p_{T,\jet}^{[i_Y]}) )
, \quad 
t'(p_{T,\jet})
=
\frac{p_{T,\jet}^{\max} - p_{T,\jet}^{\min}}{
(p_{T,\jet}^{\max} - p_{T,\jet})
(p_{T,\jet} - p_{T,\jet}^{\min})
}
\end{equation}
where $K_h$ is a scaled kernel whose bandwith parameter is  $h$.
In particular, a gaussian kernel with bandwith $h=0.25$ is used for the KDE.
\begin{equation}
K_h(x) = \frac{1}{h} K \left(\frac{x}{h} \right)
, \quad 
K(x) = \frac{1}{\sqrt{2\pi}} \exp \left( - \frac{x^2}{2} \right)
\end{equation}

\begin{figure}
\begin{center}
\includegraphics[scale=0.5]{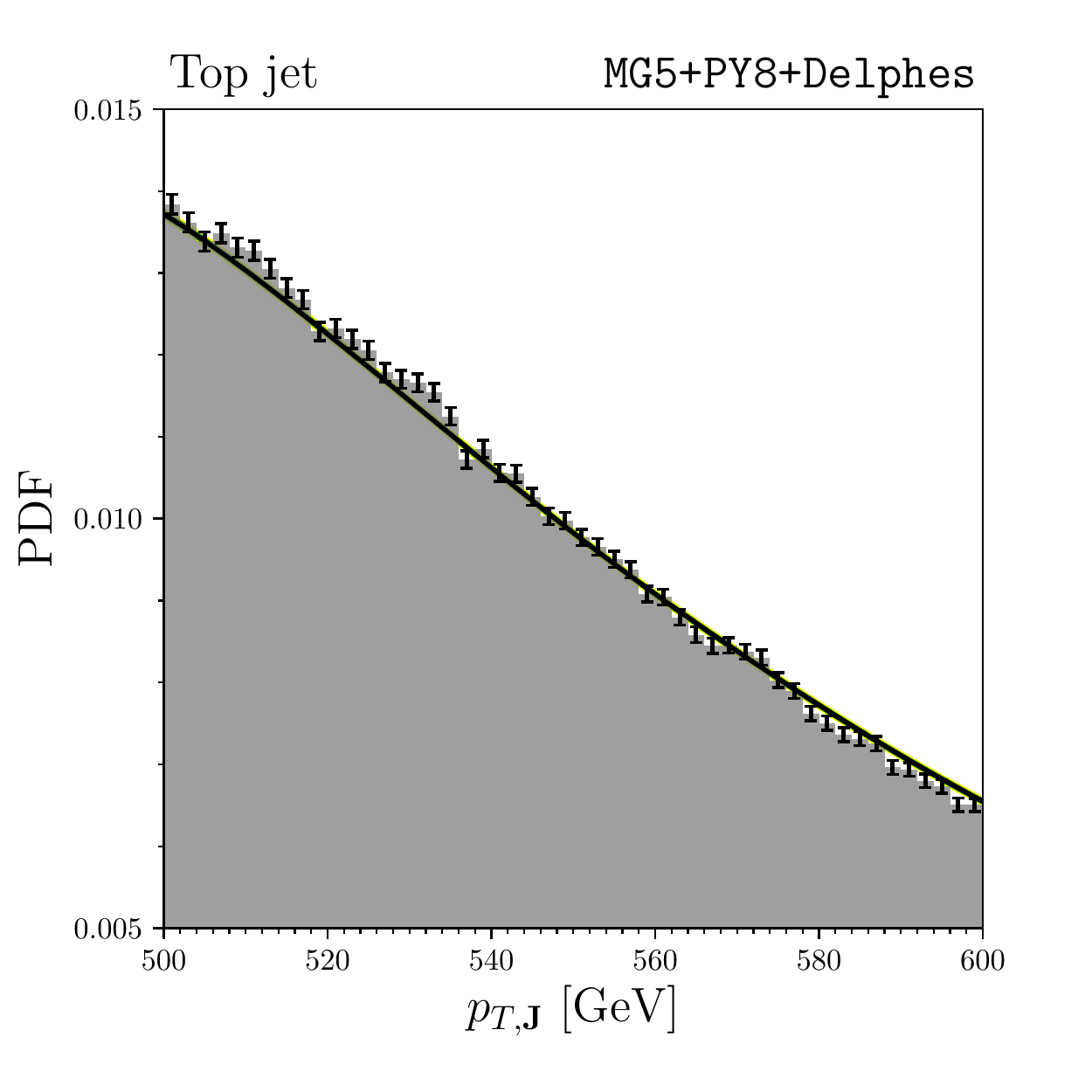}
\includegraphics[scale=0.5]{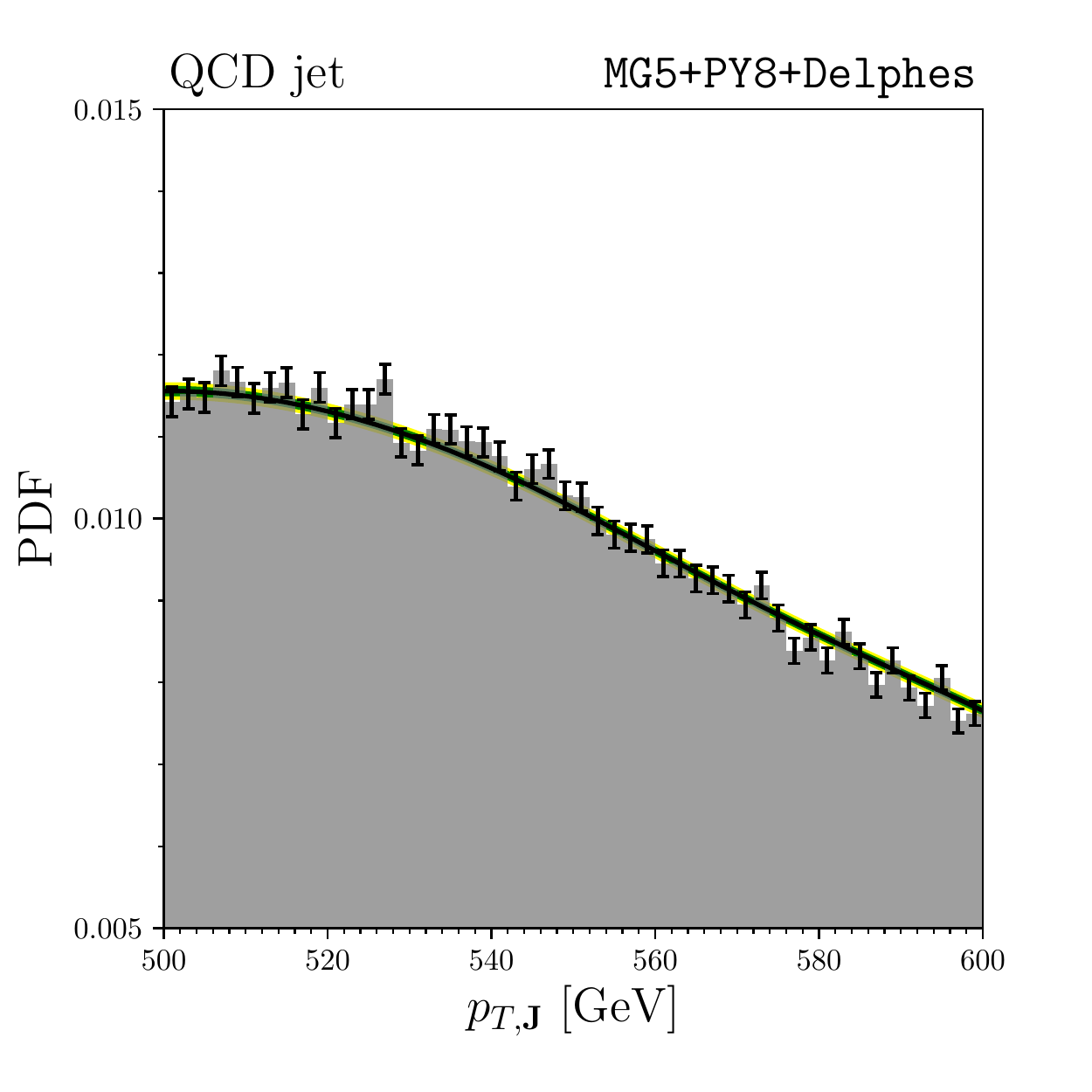}
\end{center}
\caption{
\label{fig:KDE:kin}
The histogram and modeled probability density distribution of $p_{T,\jet}$ for (left) top jet samples and (right) QCD jet samples.
Solid black line is the KDE.
The vertical bar is the statistical uncertainty of each bin.
The green and yellow bands are pointwise statistical uncertainty of the KDE calculated from squared sum of the summands.
}
\end{figure}

However, $t'(p_{T,\jet})$ is singular at $p_{T,\jet}^{\min}$ and $p_{T,\jet}^{\max}$, and the estimation of the probability density near the boundary is less precise.
Instead of using samples after the selection $p_{T,\jet} \in [500, 600]$ GeV, we use a selection with broader $p_{T,\jet}$ range, $[450, 650]$ GeV for KDE only in order to avoid the effects from the singularities.
The KDE is then normalized for $p_{T,\jet} \in [500, 600]$ GeV afterward.
We show the normalized histogram of $p_{T,\jet}$ and the KDE in \figref{fig:KDE:kin}.

\section{Network Implementations}

\subsection{Relation Networks}
\label{app:implementation:rn}
The relation networks used in this paper are implemented as follows.
The module for analyzing the energy correlation with jet trimming, 
$\bm{h}_{\trim} = \MLP_{\trim}(\bm{x}_{\trim},\bm{x}_{\kin})$, consists of two hidden layers,
\begin{eqnarray}
\nonumber
\bm{h}^{(1)}_{\trim}
& = &
\phantombox{
$\FC(\bm{z}_{\trim},\bm{z}_{\kin})$,
}{00000000000000000}
\phantombox{
size: 200,
}{00000000000}
\textrm{activation: ELU}
\\
\nonumber
\bm{h}^{(2)}_{\trim}
& = &
\phantombox{
$\FC(\bm{h}^{(1)}_{\trim})$,
}{00000000000000000}
\phantombox{
size: 200,
}{00000000000}
\textrm{activation: ELU}
\\
\bm{h}_{\trim}
& = &
\phantombox{
$\FC(\bm{h}^{(2)}_{\trim})$,
}{00000000000000000}
\phantombox{
size: 5,
}{00000000000}
\textrm{activation: linear}
\end{eqnarray}
where $\bm{z}_i$ is the standardized inputs of $\bm{x}_i$, and $\FC$ is a fully-connected layer with a given output size and activation function.
Note that we do not apply $L_2$ regularization for the $\FC$s with linear activation. 
The module for analyzing the energy correlation of $\jet_1$ and $\jet \setminus \jet_1$ is as follows.
\begin{eqnarray}
\nonumber
\bm{h}^{(1)}_{\jet_1}
& = &
\phantombox{
$\FC(\bm{z}_{\jet_1},\bm{z}_{\kin})$,
}{00000000000000000}
\phantombox{
size: 200,
}{00000000000}
\textrm{activation: ELU}
\\
\nonumber
\bm{h}^{(2)}_{\jet_1}
& = &
\phantombox{
$\FC(\bm{h}^{(1)}_{\jet_1})$,
}{00000000000000000}
\phantombox{
size: 200,
}{00000000000}
\textrm{activation: ELU}
\\
\bm{h}_{\jet_1}
& = &
\phantombox{
$\FC(\bm{h}^{(2)}_{\jet_1})$,
}{00000000000000000}
\phantombox{
size: 5,
}{00000000000}
\textrm{activation: linear}
\end{eqnarray}
The logits $\bm{u}'$ for the binary classification is implemented as follows.
\begin{eqnarray}
\nonumber
\label{eqn:MLP_logit_layer_first}
\bm{h}^{(1)}_{\logit}
& = &
\phantombox{
$\FC(\bm{h}_{\trim}, \bm{h}_{\jet_1},\bm{z}_{\kin})$,
}{0000000000000000000}
\phantombox{
size: 200,
}{00000000000}
\textrm{activation: ELU}
\\
\nonumber
\bm{h}^{(2)}_{\logit}
& = &
\phantombox{
$\FC(\bm{h}^{(1)}_{\logit})$,
}{0000000000000000000}
\phantombox{
size: 200,
}{00000000000}
\textrm{activation: ELU}
\\
\bm{u}'
& = &
\phantombox{
$\FC(\bm{h}^{(2)}_{\logit})$,
}{0000000000000000000}
\phantombox{
size: 2,
}{00000000000}
\textrm{activation: linear}
\end{eqnarray}
For the relation networks with inputs $\bm{x}_\mathrm{geometry}$, we replace $\bm{h}^{(1)}_{\logit}$ of \eqref{eqn:MLP_logit_layer_first} as follows.
\begin{eqnarray}
\bm{h}^{(1)}_{\logit}
& = &
\phantombox{
$\FC(\bm{h}_{\trim}, \bm{h}_{\jet_1}, \bm{z}_{\mathrm{geometry}})$,
}{000000000000000000000000}
\phantombox{
size: 200,
}{00000000000}
\textrm{activation: ELU},
\end{eqnarray}

\subsection{Convolutional Neural Networks}
\label{app:implementation:cnn}
Our convolutional neural networks are trained on the preprocessed jet images obtained as in \cite{Chakraborty:2019imr}. 
We recluster given jet constituents by $k_T$ algorithm \cite{Catani:1993hr,Ellis:1993tq} with radius parameter $R_\jet = 0.2$ and translate the $(\eta,\phi)$ coordinate so that the leading $p_T$ subjet axis is at $(0,0)$.
If a subleading $p_T$ subjet exists, we rotate the $(\eta,\phi)$ corordinate about the origin so that the subjet is on the positive $y$-axis on the rotated coordinate.
If a third leading $p_T$ subjet exists with a negative $x$ coordinate, we reflect the coordinate to across the $y$ axis so that the third leading $p_T$ subjet always has a positive $x$ coordinate.
The preprocessed jet image $\bm{x}_{\image}$ is a two-dimensional $p_T$-weighted histogram of those regularized constituents on a range $[-1.5,1.5]\otimes[-1.5,1.5]$ with bin size $0.1\times 0.1$.
The energy deposit of each pixel is standardized thereafter.

The vanilla CNN of this paper consists of six convolutional layers with a filter size $3\times3$.
The standardized image $\bm{z}_{\image}$ of $\bm{x}_{\image}$ is fed into a chain of convolutional layers as follows.
\begin{eqnarray}
\nonumber
\bm{h}^{(1)}_{\CNN}
& = &
\phantombox{
$\CONV(\bm{z}_{\image})$,
}{0000000000000000000}
\phantombox{
size: $30 \times 30 \times 16$,
}{000000000000000000}
\phantombox{
filter size: $3 \times 3$,
}{00000000000000000}
\textrm{activation: ELU},
\\
\nonumber
\bm{h}^{(2)}_{\CNN}
& = &
\phantombox{
$\CONV(\bm{h}^{(1)}_{\CNN})$,
}{0000000000000000000}
\phantombox{
size: $30 \times 30 \times 16$,
}{000000000000000000}
\phantombox{
filter size: $3 \times 3$,
}{00000000000000000}
\textrm{activation: ELU},
\\
\nonumber
\bm{h}^{(3)}_{\CNN}
& = &
\phantombox{
$\CONV(\bm{h}^{(2)}_{\CNN})$,
}{0000000000000000000}
\phantombox{
size: $30 \times 30 \times 16$,
}{000000000000000000}
\phantombox{
filter size: $3 \times 3$,
}{00000000000000000}
\textrm{activation: ELU},
\\
\nonumber
\bm{h}^{(3,\POOL)}_{\CNN}
& = &
\phantombox{
$\POOL(\bm{h}^{(3)}_{\CNN})$,
}{0000000000000000000}
\phantombox{
size: $15 \times 15 \times 16$,
}{000000000000000000}
\textrm{
pool size: $2 \times 2$,
}
\\
\nonumber
\bm{h}^{(4)}_{\CNN}
& = &
\phantombox{
$\CONV(\bm{h}^{(3,\POOL)}_{\CNN})$,
}{0000000000000000000}
\phantombox{
size: $15 \times 15 \times 8$,
}{000000000000000000}
\phantombox{
filter size: $3 \times 3$,
}{00000000000000000}
\textrm{activation: ELU},
\\
\nonumber
\bm{h}^{(5)}_{\CNN}
& = &
\phantombox{
$\CONV(\bm{h}^{(4)}_{\CNN})$,
}{0000000000000000000}
\phantombox{
size: $15 \times 15 \times 8$,
}{000000000000000000}
\phantombox{
filter size: $3 \times 3$,
}{00000000000000000}
\textrm{activation: ELU},
\\
\nonumber
\bm{h}^{(6)}_{\CNN}
& = &
\phantombox{
$\CONV(\bm{h}^{(5)}_{\CNN})$,
}{0000000000000000000}
\phantombox{
size: $15 \times 15 \times 8$,
}{000000000000000000}
\phantombox{
filter size: $3 \times 3$,
}{00000000000000000}
\textrm{activation: ELU},
\\
\nonumber
\bm{h}^{(6,\POOL)}_{\CNN}
& = &
\phantombox{
$\POOL(\bm{h}^{(6)}_{\CNN})$,
}{0000000000000000000}
\phantombox{
size: $7 \times 7 \times 8$,
}{000000000000000000}
\textrm{
pool size: $2 \times 2$,
}
\\
\nonumber
\bm{h}^{(7)}_{\CNN}
& = &
\phantombox{
$\FC(\bm{h}^{(6,\POOL)}_{\CNN})$,
}{0000000000000000000}
\phantombox{
size: $200$,
}{000000000000}
\textrm{activation: ELU},
\\
\label{eqn:setup:cnn}
\bm{h}_{\CNN}
& = &
\phantombox{
$\FC(\bm{h}^{(7)}_{\CNN})$,
}{0000000000000000000}
\phantombox{
size: $100$,
}{000000000000}
\textrm{activation: linear},
\end{eqnarray}
where $\CONV$ is a two-dimensional convolutional layer with a given filter size and activation function, and $\POOL$ is a max-pooling layer with a given pool size.
The output size consists of three numbers: the first two numbers represent output image width and height, and the third number is the number of filters.
We simply put $\bm{h}_{\CNN}$ to $\MLP_{\logit}$ by replacing \eqref{eqn:MLP_logit_layer_first} to the following.
\begin{eqnarray}
\bm{h}^{(1)}_{\logit}
& = &
\phantombox{
$\FC(\bm{h}_{\CNN}, \bm{z}_{\kin})$,
}{0000000000000000000}
\phantombox{
size: 200,
}{00000000000}
\textrm{activation: ELU}
\end{eqnarray}

The ResNet in \sectionref{sec:s2_top_tagger:other_cnn} consists of convolutional layers $\bm{h}^{(i+1,\res)}_{\ResNet}$ with skip connection $\bm{h}^{(i+1,\shortcut)}_{\ResNet}$. 
We define a ResNet module of input image $\bm{h}^{(i)}_{\ResNet}$ as follows.
\begin{eqnarray}
\nonumber
\bm{h}^{(i+1,\res)}_{\ResNet}
& = &
\phantombox{
$\CONV\circ\elu\circ\CONV(\bm{h}^{(i)}_{\ResNet})$,
}{0000000000000000}
\\
\nonumber
\bm{h}^{(i+1,\shortcut)}_{\ResNet}
& = &
\begin{cases}
\bm{h}^{(i)}_{\ResNet}
&
\textrm{if $\bm{h}^{(i)}_{\ResNet}$ and $\bm{h}^{(i+1)}_{\ResNet}$ has the same size,
}
\\
\CONV_{1 \times 1}(\bm{h}^{(i)}_{\ResNet}) 
&
\textrm{otherwise,}
\end{cases}
\\
\bm{h}^{(i+1)}_{\ResNet}
& = &
\ResNet(\bm{h}^{(i)}_{\ResNet})
= \elu(
\bm{h}^{(i+1,\res)}_{\ResNet}
+
\bm{h}^{(i+1,\shortcut)}_{\ResNet}
)
\end{eqnarray}
where $\CONV_{1\times 1}$ is a convolutional layer with filter size $1\times1$. 
The hyperparameters of other $\CONV$ will be specified later.
All the convolutional operations above do not have any activation function.
If input image size and output image size are different, we use strided convolution on $\CONV(\bm{h}^{(i)}_{\ResNet})$. 
We build a ResNet by replacing the chain of convolutional layers in \eqref{eqn:setup:cnn} to the following chain of six ResNet modules.
\begin{eqnarray}
\nonumber
\bm{h}^{(1)}_{\ResNet}
& = &
\phantombox{
$\ResNet(\bm{z}_{\image})$,
}{0000000000000000000}
\phantombox{
size: $30 \times 30 \times 16$,
}{000000000000000000}
\phantombox{
filter size: $3 \times 3$,
}{00000000000000000}
\\
\nonumber
\bm{h}^{(2)}_{\ResNet}
& = &
\phantombox{
$\ResNet(\bm{h}^{(1)}_{\ResNet})$,
}{0000000000000000000}
\phantombox{
size: $30 \times 30 \times 16$,
}{000000000000000000}
\phantombox{
filter size: $3 \times 3$,
}{00000000000000000}
\\
\nonumber
\bm{h}^{(3)}_{\ResNet}
& = &
\phantombox{
$\ResNet(\bm{h}^{(2)}_{\ResNet})$,
}{0000000000000000000}
\phantombox{
size: $15 \times 15 \times 8$,
}{000000000000000000}
\phantombox{
filter size: $3 \times 3$,
}{00000000000000000}
\textrm{stride: 2,}
\\
\nonumber
\bm{h}^{(4)}_{\ResNet}
& = &
\phantombox{
$\ResNet(\bm{h}^{(3)}_{\ResNet})$,
}{0000000000000000000}
\phantombox{
size: $15 \times 15 \times 8$,
}{000000000000000000}
\phantombox{
filter size: $3 \times 3$,
}{00000000000000000}
\\
\nonumber
\bm{h}^{(5)}_{\ResNet}
& = &
\phantombox{
$\ResNet(\bm{h}^{(4)}_{\ResNet})$,
}{0000000000000000000}
\phantombox{
size: $8 \times 8 \times 8$,
}{000000000000000000}
\phantombox{
filter size: $3 \times 3$,
}{00000000000000000}
\textrm{stride: 2,}
\\
\nonumber
\bm{h}^{(6)}_{\ResNet}
& = &
\phantombox{
$\ResNet(\bm{h}^{(5)}_{\ResNet})$,
}{0000000000000000000}
\phantombox{
size: $8 \times 8 \times 8$,
}{000000000000000000}
\phantombox{
filter size: $3 \times 3$,
}{00000000000000000}
\\
\bm{h}^{(1)}_{\logit}
& = &
\phantombox{
$\FC(\bm{h}^{(6)}_{\ResNet}, \bm{z}_{\kin})$,
}{0000000000000000000}
\phantombox{
size: 200,
}{00000000000}
\textrm{activation: ELU.}
\end{eqnarray}

The ResNeXt in \sectionref{sec:s2_top_tagger:other_cnn} uses multiple chains of convolutional layers for the residual learning parts $\bm{h}_{\ResNet}^{(i+1,\res)}$ in the ResNet. 
The ResNeXt module with four parallel chains of convolutional layers is defined as follows. 
\begin{eqnarray}
\nonumber
\bm{h}^{(i+1,j)}_{\ResNeXt}
& = &
\elu \circ \CONV\circ\elu\circ\CONV_{1\times 1}(\bm{h}^{(i)}_{\ResNeXt}),
\\
\nonumber
\bm{h}^{(i+1,\res)}_{\ResNeXt}
& = &
\CONV_{1 \times 1}\left(\bigoplus_{j=1}^{4} \bm{h}^{(i+1,j)}_{\ResNeXt}\right),
\\
\nonumber
\bm{h}^{(i+1,\shortcut)}_{\ResNeXt}
& = &
\CONV_{1 \times 1}(\bm{h}^{(i)}_{\ResNeXt}),
\\
\bm{h}^{(i+1)}_{\ResNet}
& = &
\ResNeXt(\bm{h}^{(i)}_{\ResNeXt})
= \elu\left(
\bm{h}^{(i+1,\res)}_{\ResNeXt}
+
\bm{h}^{(i+1,\shortcut)}_{\ResNeXt}
\right),
\end{eqnarray}
where the direct sum of the images represents a stacked image along the filter dimension. 
Since we use many convolutional layers already, we use three of those modules for the image analyzer.
\begin{eqnarray}
\nonumber
\bm{h}^{(1)}_{\ResNeXt}
& = &
\phantombox{
$\ResNeXt(\bm{z}_{\image})$,
}{0000000000000000000}
\phantombox{
size: $30 \times 30 \times 16$,
}{000000000000000000}
\phantombox{
filter size: $3 \times 3$,
}{00000000000000000}
\\
\nonumber
\bm{h}^{(2)}_{\ResNeXt}
& = &
\phantombox{
$\ResNeXt(\bm{h}^{(1)}_{\ResNeXt})$,
}{0000000000000000000}
\phantombox{
size: $30 \times 30 \times 16$,
}{000000000000000000}
\phantombox{
filter size: $3 \times 3$,
}{00000000000000000}
\\
\nonumber
\bm{h}^{(3)}_{\ResNeXt}
& = &
\phantombox{
$\ResNeXt(\bm{h}^{(2)}_{\ResNeXt})$,
}{0000000000000000000}
\phantombox{
size: $15 \times 15 \times 8$,
}{000000000000000000}
\phantombox{
filter size: $3 \times 3$,
}{00000000000000000}
\textrm{stride: 2,}
\\
\nonumber
\bm{h}^{(1)}_{\logit}
& = &
\phantombox{
$\FC(\bm{h}^{(3)}_{\ResNeXt}, \bm{z}_{\kin})$,
}{0000000000000000000}
\phantombox{
size: 200,
}{00000000000}
\textrm{activation: ELU}
\end{eqnarray}

\section{Updating Trainable Parameters with Moving Averages}
\label{app:moving_avg}

The moving average of a network parameter in \sectionref{sec:network_setup} is evaluated as follows.
An updated parameter $\bm{\theta}^{(t)}$ at an epoch $t$ is accumulated into a moving average $\bar{\bm{\theta}}^{(t)}$,
\begin{eqnarray}
\bar{\bm{\theta}}^{(t)} = 
\begin{cases}
0 
& 
t < t_0 
\\
\alpha \bar{\bm{\theta}}^{(t-1)} + (1-\alpha) \bm{\theta}^{(t)}
&
t \geq t_0 
\end{cases}
\end{eqnarray}
where $\alpha = 0.9$.
We accumulate only the updated parameters at the epochs after $t_0 = 50$.
The solution to the recurrence relation is as follows,
\begin{equation}
\bar{\bm{\theta}}^{(t)}
=
\sum_{u=t_0}^{t} \alpha^{t-u} (1-\alpha) \bm{\theta}^{(u)}.
\end{equation} 
As a side effect of the epoch selection, the sum of the weights in the average is not 1. 
As $\bm{\theta}^{(u)}$ approaches its optimum $\bm{\theta}_0$, $\bar{\bm{\theta}}^{(t)}$ approaches to $(1-\alpha^{t-t_0+1}) \bm{\theta}_0$.
The factor $1-\alpha^{t-t_0+1}$ should be corrected to make the moving average also converging to $\bm{\theta}_0$.
We use the following unbiased moving average $\hat{\bm{\theta}}^{(t)}$ of the sequence of $\bm{\theta}^{(t)}$ for the validation and testing,
\begin{equation}
\hat{\bm{\theta}}^{(t)} = \frac{1}{1-\alpha^{t-t_0+1}} \bar{\bm{\theta}}^{(t)} \; \mathrm{for} \; t \geq t_0.
\end{equation}

\section{Evaluation of the Reweighting Factor}
 \label{app:reweighted_dist}
 
\begin{figure}
\begin{center}
\begin{subfigure}{0.49\textwidth}
\includegraphics[width=1.0\textwidth]{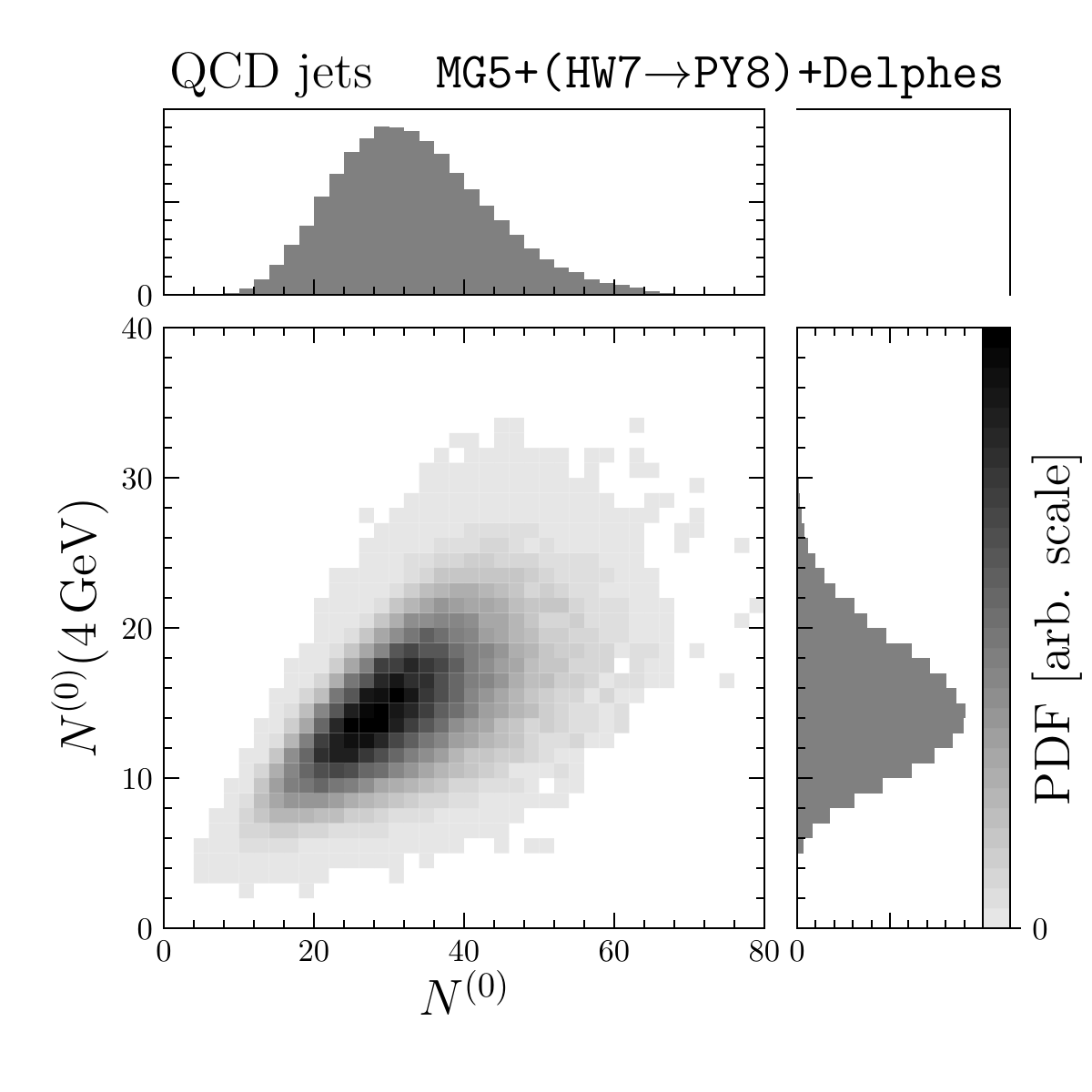}
\caption{}
\end{subfigure}
\begin{subfigure}{0.49\textwidth}
\includegraphics[width=1.0\textwidth]{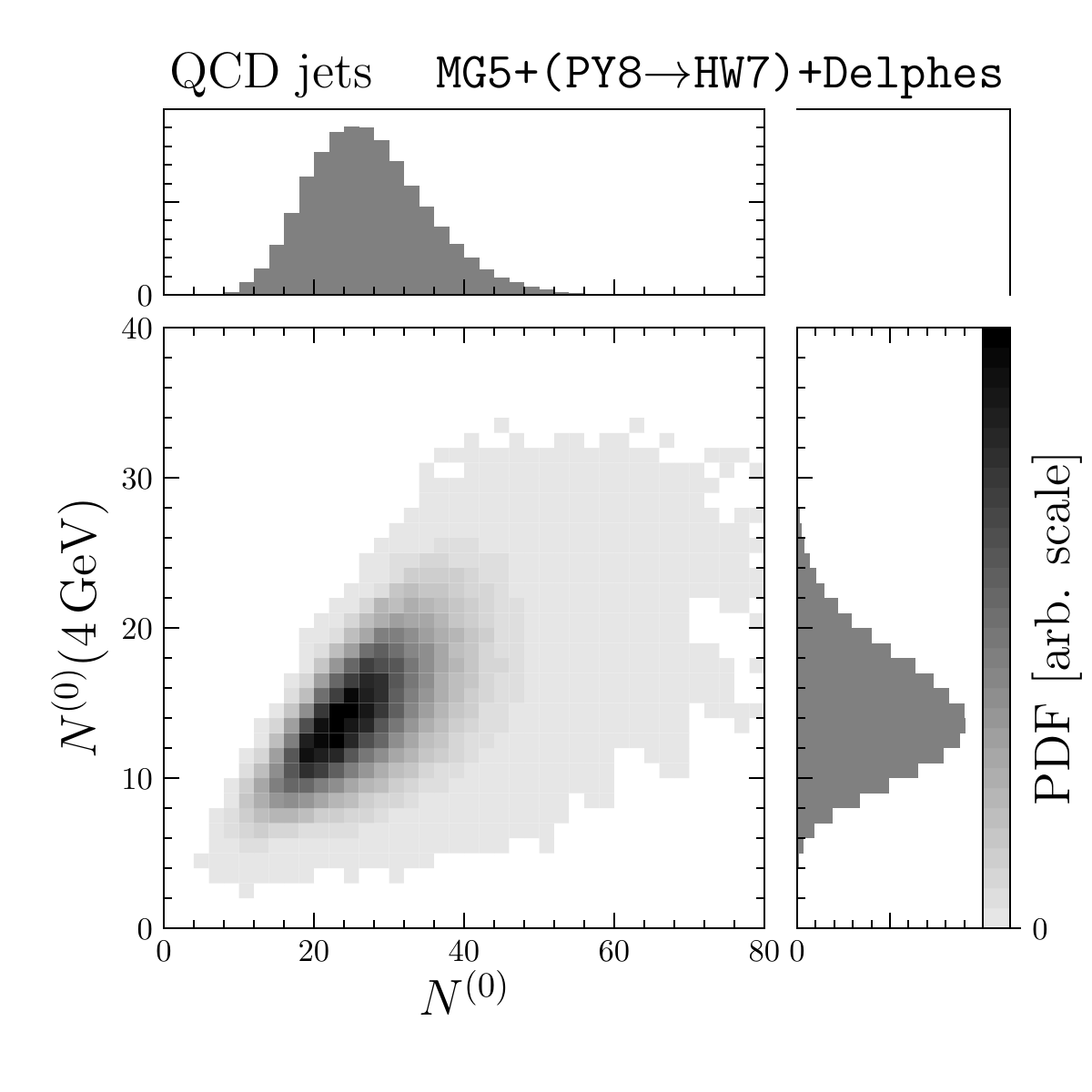}
\caption{}
\end{subfigure}
\end{center}
 \caption{$(N^{(0)},N^{(0)}(4\,\GeV))$ distribution for (a) the weighted \hw{} samples 
 to reproduce \py{} distribution and (b) the weighted \py{} samples  
 to reproduce \hw{} distribution.}\label{fig:n04w}
 \end{figure}

\begin{figure}
\begin{center}
\begin{subfigure}{0.49\textwidth}
\includegraphics[width=1.0\textwidth]{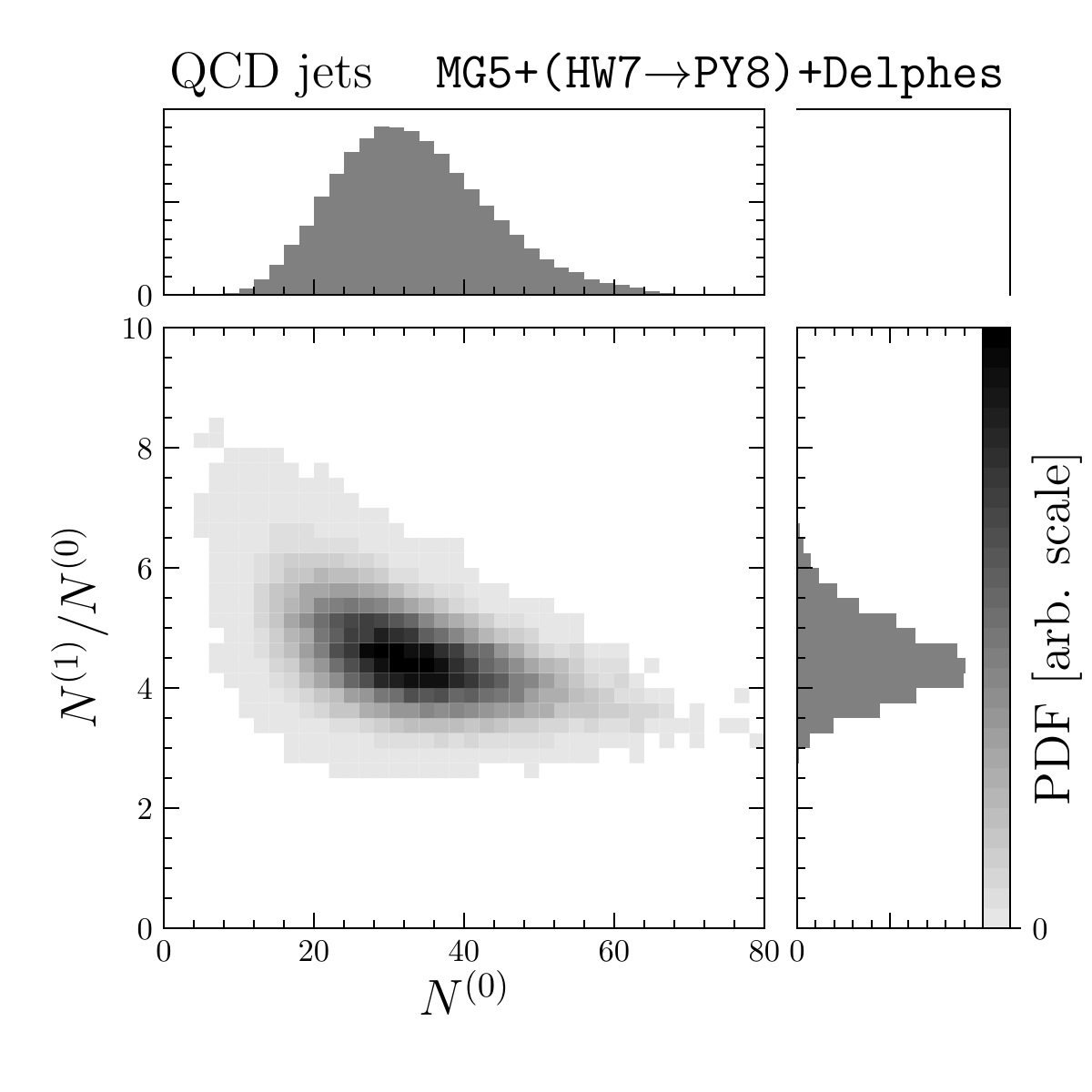}
\caption{}
\end{subfigure}
\begin{subfigure}{0.49\textwidth}
\includegraphics[width=1.0\textwidth]{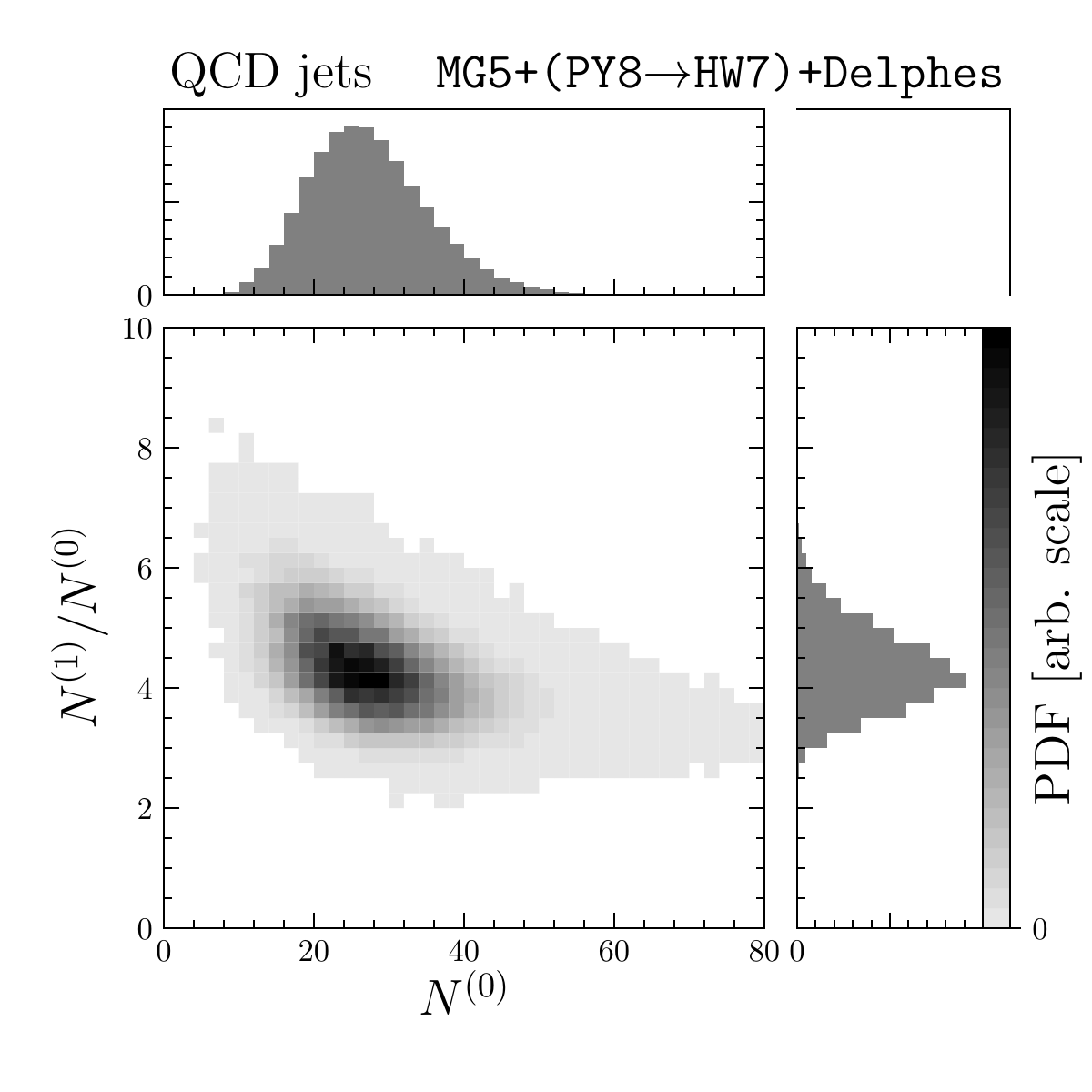}
\caption{}
\end{subfigure}
\end{center}
 \caption{$N_{\mathrm{pixel}}$ vs $N_1/N_{\mathrm{pixel}}$  for (a) the weigthed \hw{} sample 
 to reproduce \py{} distribution and (b) the weighted \py{} sample.  
 to reproduce \hw{} distribution}\label{fig:n0n1w}
 \end{figure}

In \sectionref{sec:reweighting}, we reweight the \hw{} generated events to \py{} generated events by using $(N^{(0)}, N^{(0)}(4\,\mathrm{GeV}))$ distribution.
Since the two numbers are correlated as shown in \figref{fig:n04}, we transform the data first and calculate the reweighting factor using normalized histograms in order to ensure the efficiency of the reweighting.
The transformation of $(N^{(0)}, N^{(0)}(4\,\mathrm{GeV}))$ is defined as follows.
\begin{equation}
\label{eqn:change_of_variable}
(x, y) \rightarrow (x',y') = (x, c_1-c_2 y/x +c_3 x),
\end{equation}
where $c_1=3/2$, $c_2=2$, and $c_3=-1/60$.
For each event, the reweighting factor in \eqref{eqn:reweight_factor} is calculated by the ratio of the corresponding bin values, $\rho_{\py{}}/\rho_{\hw{}}$, where $\rho_{A}$ is the bin value of $(x',y')$ histogram with events generated by $A$.
The reweighting factor for \py{} generated events to obtain distributions of \hw{} generated events can be obtained by a similar procedure. 
The reweighted $(N^{(0)}, N^{(0)}(4\,\mathrm{GeV}))$ distribution and $(N^{(0)}, N^{(1)}/N^{(0)})$ distribution are shown in \figref{fig:n04w} and \figref{fig:n0n1w}, respectively.

%---references ...
\bibliographystyle{JHEP}
\bibliography{JetSubstructureSpectroscopy_Top}

\end{document}